\documentclass[11pt]{article}

\usepackage[margin=1in]{geometry}
\usepackage{setspace}
\usepackage{amsmath,amsfonts,amssymb,mathtools}

\usepackage{amsthm}

\usepackage{times}
\usepackage[varg]{txfonts}
\usepackage{bm}
\usepackage{algorithm}
\usepackage[noend]{algpseudocode}

\usepackage{tikz}
\usetikzlibrary{arrows.meta,positioning,shapes.geometric,shapes.misc,fit,calc}
\usetikzlibrary{shapes,snakes}
\usetikzlibrary{arrows.meta, bending, positioning, patterns}
\usepackage{graphicx}
\usepackage{tikz,pgfplots,pgffor}
\pgfplotsset{compat=1.7}

\usepackage{booktabs}
\usepackage{multirow}
\usepackage{tablefootnote}

\usepackage{enumitem}
\usepackage{xcolor}
\usepackage{caption}
\usepackage{subcaption}
\usepackage[font=small,labelfont=bf]{caption}
\usepackage{listings}
\usepackage{etoolbox}

\usepackage{hyperref}
\usepackage{natbib}
\newtheorem{theorem}{Theorem}
\newtheorem{proposition}{Proposition}

\newtheorem{corollary}{Corollary}
\newtheorem{remark}{Remark}
\newtheorem{definition}{Definition}

\usepackage[disable]{todonotes}
\newcommand{\REV}[1]{}
\newcommand{\AER}[1]{}
\newcommand{\ROne}[1]{}
\newcommand{\RTwo}[1]{}
\newcommand{\TODO}[1]{}
\newcommand{\HWOLD}[1]{}
\newcommand{\HW}[1]{}
\newcommand{\SC}[1]{}
\newcommand{\EB}[1]{}
\newcommand{\RH}[1]{}
\newcommand{\RHdone}[1]{}
\newcommand{\RHreply}[1]{}

\definecolor{cyan}{rgb}{0.0, 1.0, 1.0}

\definecolor{darkgreen}{rgb}{0.0, 0.5, 0.0}
\DeclareMathOperator*{\argmax}{arg\,max}

\newcommand{\Halmos}{\qed}

\newcommand{\conv}{\operatorname{conv}}
\newcommand{\BR}{\mathcal{BR}}
\newcommand{\A}{\mathcal{A}}
\newcommand{\B}{\mathcal{B}}
\newcommand{\I}{\mathcal{I}}
\newcommand{\K}{\mathcal{K}}
\newcommand{\E}{\mathcal{E}}
\newcommand{\X}{\mathcal{X}}
\newcommand{\G}{\mathcal{G}}

\newcommand{\x}{\mathbf{x}}
\newcommand{\y}{\mathbf{y}}
\newcommand{\z}{\mathbf{z}}

\newcommand{\noti}{{-i}}

\newcounter{ecompanion}
\newcommand{\ecompanionsection}[1]{
  \refstepcounter{ecompanion}
  \section*{Appendix \Alph{ecompanion}: #1}
  \addcontentsline{toc}{section}{Appendix \Alph{ecompanion}: #1}
}

\newcommand{\vspaceminus}[1]{}
\newcommand{\barxzero}{{\bf{0}}^N}
\newcommand{\barxswt}{\tilde \x_{\text{sw}}^{\texttt{time}}}
\newcommand{\barxswtime}[1]{\tilde \x_{\text{sw}}^{\texttt{#1}}}
\newcommand{\barxosw}{\bar \x_{\text{sw}}^*}

\begin{document}

\title{\Large\bfseries Random-Restart Best-Response Dynamics for Large-Scale Integer Programming Games and Their Applications
% \\with Applications to Aquatic Invasive Species Prevention
}

\author{
Hyunwoo Lee\thanks{Grado Department of Industrial and Systems Engineering, Virginia Tech. Email: \texttt{hyunwoolee@vt.edu}} \and
Robert Hildebrand\thanks{Grado Department of Industrial and Systems Engineering, Virginia Tech. Email: \texttt{rhil@vt.edu}} \and
Wenbo Cai\thanks{Mechanical and Industrial Engineering, New Jersey Institute of Technology. Email: \texttt{cai@njit.edu}} \and
{\.I.} Esra B{\"u}y{\"u}ktahtak{\i}n\thanks{Grado Department of Industrial and Systems Engineering, Virginia Tech. Email: \texttt{esratoy@vt.edu}}
}

\date{\today}

\maketitle

\begin{abstract}
This paper presents scalable algorithms for computing pure Nash equilibria (PNEs) in large-scale integer programming games (IPGs), where existing exact methods typically handle only small numbers of players. Motivated by a county-level aquatic invasive species (AIS) prevention problem with 84 decision makers, we develop and analyze \emph{random-restart best-response dynamics} (RR-BRD), a randomized search framework for PNEs. For IPGs with finite action sets, we model RR-BRD as a Markov chain on the best-response state graph and show that, whenever a PNE exists and the restart law has positive probability of reaching a PNE within the round cap, RR-BRD finds a PNE almost surely. We also propose a Monte Carlo sampling-and-simulation procedure to estimate success behavior under a fixed round cap, which informs our instance-dependent performance characterization. We then embed RR-BRD as a randomized local-search subroutine within the zero-regret (ZR) framework, yielding \emph{BRD-incorporated zero-regret} (BZR). Using solver callbacks, RR-BRD searches for and supplies PNEs, while ZR separates and adds equilibrium inequalities to tighten the formulation. We introduce \emph{edge-weighted budgeted maximum coverage} (EBMC) games to model AIS prevention and establish PNE existence results for both selfish and locally altruistic utilities. Computational experiments on synthetic EBMC and knapsack problem game instances show that RR-BRD and BZR scale equilibrium computation up to $n \le 30$ players. We further solve a real-world EBMC game derived from the Minnesota AIS dataset with $n = 84$ county players.
\end{abstract}


\bigskip
\noindent\textbf{Funding:} This work was supported in part by the Air Force Office of Scientific Research (AFOSR) and the Minnesota Aquatic Invasive Species Research Center (MAISRC).

\section{Introduction}\label{sec:Intro}
\vspaceminus{3pt}
Integer programming games (IPGs) stand at the crossroads of integer programming and non-cooperative game theory. They study settings in which each player solves an optimization model with discrete decisions, and these decisions influence one another’s utilities. IPGs are non-cooperative, complete-information games: players are self-interested and have complete information about each other’s utilities and strategy sets \citep{carvalho2022computing}. A pure Nash equilibrium (PNE), as defined by \citet{nash1950equilibrium}, is a strategy profile in which no player can improve their utility by unilaterally changing their strategy. Social welfare is the sum of all players’ utilities. In this paper, we develop efficient algorithms for large-scale IPGs to (i) compute a PNE, (ii) compute multiple PNEs, and (iii) identify the PNE with the highest social welfare.

Computing multiple PNEs is an important task in non-cooperative games, as some equilibria may be more desirable than others. In particular, a socially optimal PNE---a PNE that maximizes total social welfare---is a natural target in settings with mediators or central planners who care about system-wide performance. This notion is also central in algorithmic game theory \citep{roughgarden2010algorithmic}, where it underpins standard measures of equilibrium inefficiency relative to socially optimal outcomes. In the IPG literature, two recent lines of work address (i) equilibrium computation and enumeration and (ii) selecting a welfare-maximizing equilibrium: \citet{dragotto2023zero} for PNEs, and \citet{cronert2024equilibrium} for mixed Nash equilibria (MNEs). However, both approaches remain limited to small-scale instances.

Although recent advances in IPG algorithms have improved our ability to compute PNEs, most methods remain limited to small-scale games with only two or three players ($n \le 3$) \citep{dragotto2023zero} or with a smaller number of variables ($m \le 5$) \citep{ sagratella2016computing, schwarze2023branch}. Given the maturity of the field, it is timely to bridge this gap by addressing the large-scale, multi-stakeholder instances required for real-world decision-making. Compelling examples arise in resource allocation decisions across decentralized administrative structures, such as the many states or counties in the United States, naturally forming large-scale IPG settings. Motivated by Minnesota's AIS prevention efforts, which are independently managed by 84 counties, we introduce a new class of non-cooperative IPGs: \textit{edge-weighted budgeted maximum coverage} (EBMC) games. These games model county-level decision-making under budget constraints and inter-county externalities. Solving such problems requires scalable and efficient algorithmic approaches.

To address this challenge, we focus on \textit{random-restart best-response dynamics} (RR-BRD), a randomized variant of best-response dynamics (BRD) \citep{matsui1992best}. We adopt a Markov chain framework and model RR-BRD as a random walk on the best-response (BR) state graph of pure strategy profiles, as introduced by \citet{goemans2005sink}. A basic observation is that the induced random walk eventually enters a recurrent class of the BR state graph; these \textit{sink equilibria} include singleton PNEs as well as non-PNE sink components. RR-BRD mitigates such traps via a restart strategy with a fixed cap on the number of steps per attempt. We show that, whenever a PNE exists and the restart distribution assigns positive probability to its basin of attraction, RR-BRD finds a PNE almost surely, with an instance-dependent bound on the expected number of best-response solves. In practice, we incorporate RR-BRD as 
a randomized local-search subroutine within the zero-regret (ZR) framework of \citet{dragotto2023zero}. This hybrid approach dramatically improves the scalability of PNE computation, increasing the number of solvable players from small-scale cases ($n = 2,3$) to up to 30 players in random datasets and 84 players in real-world datasets.

As a motivating application for large-scale IPGs, we build on optimization-based approaches to invasive species control \citep{buyuktahtakin2018review, chen2023game, kibics2021multistage} by introducing EBMC games, which arise naturally in decentralized, county-level AIS prevention across U.S. lakes. Previous studies have applied the EBMC formulation \citep{caskurlu2014partial} to model AIS inspection at the county \citep{Haight2021} and state levels \citep{haight2023bi}, while others have examined AIS spread across network graphs \citep{escobar2018aquatic}.  In EBMC games, each player controls a subset of vertices in a network graph, and each player maximizes their edge coverage under budget constraints. We define two variants of the utility functions--selfish and locally altruistic--and establish theoretical conditions for the existence of a PNE in these models. Additionally, we validate our IPG framework through experimental results on EBMC games using random and real-world datasets, as well as on knapsack problem games (KPGs) with random datasets. Our findings reveal that IPGs, coupled with the enhanced solvability of the BZR algorithm presented in this paper, offer a powerful and scalable framework for modeling and solving large-scale multi-agent problems.

\subsection{Literature Review} \label{sec:literature}
\vspaceminus{3pt}

Integer programming games (IPGs) are gaining attention for their ability to model strategic interactions beyond traditional normal-form games. Various applications of IPGs have been studied, including facility location and design games \citep{cronert2024equilibrium}, fixed-charge transportation problems \citep{sagratella2020noncooperative}, kidney exchange games \citep{carvalho2017nash}, quadratic IPGs (qIPGs) \citep{sagratella2016computing, schwarze2023branch}, and knapsack problem games (KPGs) \citep{carvalho2021cut,dragotto2023zero}. In these models, each player solves a player-specific integer program (IP), and strategic interactions are typically captured through reciprocally bilinear terms that couple players' decision variables, with either linear objectives (e.g., KPG) or quadratic objectives (e.g., qIPG).

\citet{carvalho2023integer} provide a comprehensive overview of the IPG solution methods, categorizing them into six approaches. Among these, sample generation method (SGM) \citep{carvalho2022computing}, enumerative SGM (eSGM) \citep{cronert2024equilibrium}, and Cut-and-Play (CnP) \citep{carvalho2021cut} compute MNEs, while branching method (BM) \citep{sagratella2016computing}, zero-regret (ZR) \citep{dragotto2023zero}, and branch-and-prune (BnP) \citep{schwarze2023branch} focus on PNEs. Notably, \citet{carvalho2022computing} and \citet{cronert2024equilibrium} use normal-form game methods to compute MNEs in sampled games before verifying them in the original setting. Among exact methods to compute PNEs, the zero-regret (ZR) algorithm \citep{dragotto2023zero} is state-of-the-art for computing socially optimal PNEs via a cutting-plane approach. For approximate NEs, \citet{sankaranarayanan2024best} presents a best-response algorithm to compute approximate PNE and \citet{duguet2025computing} show a piecewise linear approximation to compute approximate MNE in the mixed-integer nonlinear programming cases based on SGM.

In most approaches, equilibrium computation is ultimately reduced to one or more joint optimization problems that couple all (or a subset of) players' strategy sets, either explicitly in normal form or implicitly via extended MIP or complementarity formulations. Consequently, reported computational studies focus on instances with small numbers of players and integer variables per player (typically with $n\cdot m \le 300$, where $n$ is the number of players and $m$ is the number of integer variables per player; see also Section~\ref{sec:RR_BRD_others}). This provides a useful baseline but leaves open the question of how far one can push the tractability of equilibrium computation in IPGs when the number of players is much larger--for example, when $n \cdot m$ is in the thousands.

Best-response dynamics (BRD) \citep{matsui1992best} is a foundational and commonly used method for computing PNEs in normal-form games, where players take turns solving their best-response problems while keeping the strategies of the others fixed. In well-structured classes such as potential games \citep{monderer1996potential} and, more generally, best-response--weakly acyclic games \citep{apt2015classification}, the convergence of BRD to a PNE is well understood. However, for arbitrary normal-form games the best-response state graph may contain directed cycles and non-singleton sink strongly connected components.
\citet{goemans2005sink} interpret the long-run behavior of such dynamics through \textit{sink equilibria}, which capture recurrent behavior of best-response walks; see also the broader dynamics-based perspective of \citet{papadimitriou2019game}. Relatedly, \citet{hakim2024swim} study \emph{limit} (infinite-horizon) absorption over sink SCCs of a (better-or-equal) response graph under a different noisy dynamics; our analysis uses a Markov chain induced by (improving) best-response updates and focuses on finite-horizon PNE-hitting probabilities.

Recent work in algorithmic game theory (AGT) has studied randomized variants of BRD that still follow a single trajectory but randomize either the initial profile or the updating player at each step. For example, \citet{amiet2021pure} and \citet{heinrich2023best} consider games with random payoffs: \citet{amiet2021pure} study BRD in games with binary strategy profiles, and \citet{heinrich2023best} analyze BRD with randomized playing sequences and randomly drawn initial strategy profiles. These works introduce randomness through the play sequence and initial draws from a known or assumed distribution, but they do not employ restart mechanisms. In the IPG context, \citet{sankaranarayanan2024best} analyzes a best-response algorithm with fixed playing sequence for lattice convex-quadratic simultaneous games and shows that, when the sequence of strategy profiles enters a finite trap (detected by a repeated profile), one can extract an approximate MNE from the strategies in that trap. To our knowledge, prior to this work, BRD-type procedures had not been developed as a scalable, restart-based algorithm for computing PNEs in IPGs with finite strategy sets.

\subsection{Contribution}
Our detailed contributions are summarized as follows:

\begin{itemize}

    \item We introduce \emph{random-restart BRD} (RR-BRD) into the IPG literature and analyze it as a randomized search framework for computing PNEs in large-scale games. For IPGs with finite strategy sets, we adopt the best-response (BR) state-graph viewpoint of \citet{goemans2005sink} and model RR-BRD as a Markov chain on this graph. We show that, whenever a PNE exists and the restart distribution assigns positive probability to states from which a PNE is reachable within the round cap, RR-BRD finds a PNE with probability one and admits an instance-dependent bound on the expected number of BR solves. We further contribute to the normal-form game literature by providing a convergence guarantee for BRD under restarts stated directly in terms of BR-state-graph reachability, without assuming random payoffs or restrictive structural conditions (e.g., potential games or BR-weakly acyclic games).

    \item We propose a Monte Carlo sampling-and-simulation procedure, in the spirit of \citet{hakim2024swim}, to estimate per-attempt success under a fixed round cap and to inform an instance-dependent performance characterization of RR-BRD. We also discuss best- and worst-case behavior on general BR state graphs and give a knapsack-style IPG instance illustrating worst-case behavior.
    
    \item Building on this framework, we propose \emph{BRD-incorporated zero-regret} (BZR), which embeds RR-BRD as a randomized local-search subroutine within the ZR \citep{dragotto2023zero}. Using solver callbacks, BZR invokes RR-BRD to search for and supply PNEs, while ZR adds equilibrium inequalities to tighten the formulation. In BZR, integer-feasible incumbents produced by the mixed-integer program (MIP) solver provide randomized initial strategy profiles for RR-BRD, and equilibrium inequalities are generated from the best-response profiles encountered along RR-BRD trajectories. We also discuss how RR-BRD can serve as a modular local-search subroutine within other IPG algorithms.

    \item We introduce edge-weighted budgeted maximum coverage (EBMC) games, a novel class of non-cooperative integer programming games where players maximize edge coverage under budget constraints on a network. We examine two variants of the players' utility functions--locally altruistic and selfish--and prove the existence of a PNE in locally altruistic EBMC games via potential game arguments, as well as provide sufficient conditions for the existence of a PNE in selfish EBMC games.

    \item We conduct extensive experiments to assess the scalability and effectiveness of RR-BRD and BZR on synthetic EBMC and KPG instances with up to 30 players. Across 135 test cases (27 per variant for the two EBMC and three KPG types), RR-BRD finds a PNE in 128 instances. When embedded in ZR, BZR substantially improves performance: it finds at least one PNE in 128 instances (vs.\ 33 for ZR) and identifies 179 PNEs for EBMC (vs.\ 35) and 3160 for KPGs (vs.\ 38). Overall, RR-BRD is effective both as a standalone algorithm and as a local-search subroutine that greatly improves ZR’s scalability.

    \item We implement our algorithms on a real-world EBMC instance with 84 county planners in Minnesota's AIS inspection programs. For the selfish EBMC game, RR-BRD finds a socially optimal PNE; without this algorithm, it would be unclear whether a PNE exists, making PNE-based benchmarks difficult to use. For the locally altruistic EBMC game, we identify a socially optimal solution that is also a PNE. Comparing these equilibria, we show how the behavioral specification (selfish vs.\ locally altruistic) changes both system-wide performance and the distribution of individual utilities, with locally altruistic equilibria often yielding higher individual utilities.
\end{itemize}
This paper is organized as follows. Section~\ref{sec:preliminaries} provides preliminaries on IPGs, normal-form games, and the best-response state graph. Section~\ref{sec:convergence-brd} presents our RR-BRD framework and its probabilistic convergence guarantees. Section~\ref{sec:BZR} shows how RR-BRD can be incorporated into the ZR framework and other algorithms. Section~\ref{sec:game-theoretic} formally defines EBMC games as an illustrative IPG, highlighting variations in utility functions across players. Section~\ref{sec:existence} establishes conditions for the existence of PNEs in EBMC games. Section~\ref{sec:computational} reports computational experiments that demonstrate the effectiveness and generalizability of RR-BRD and BZR on both random and real-world datasets, including random knapsack game datasets representing broader classes of resource-constrained IPGs. Finally, Section~\ref{sec:Conclusion} offers concluding remarks. For quick reference on acronyms and notation used throughout the paper, see Appendix~\ref{appendix:acronyms}.

\section{Preliminaries}\label{sec:preliminaries}
\vspaceminus{3pt}
This section provides the modeling and state-graph foundations used throughout the paper.
Section~\ref{sec:IPG_PNE} defines IPGs and PNE, and Section~\ref{sec:normal_br_state} introduces the normal-form best-response state graph (and its Markov-chain reduction) used to describe and analyze our RR-BRD in Section~\ref{sec:convergence-brd}.

\subsection{Integer Programming Games and Pure Nash Equilibria} \label{sec:IPG_PNE}
We introduce the fundamental definitions and concepts related to IPGs and PNEs. Following \citet{koppe2011rational}, an IPG is defined as a tuple
\(
\G = \bigl(N, (\mathcal{X}_i)_{i \in N}, (u_i)_{i \in N}\bigr),
\)
where \(N := \{1, \dots, n\}\) denotes the set of players, 
\(\mathcal{X}_i := \{\x^i \in \mathbb{Z}^{m_i}: A^i \x^i \leq \mathbf{b}^i \}\) represents the pure-integer strategy set for player \(i\) with rational matrix \(A^i\), vector \(\mathbf{b}^i\), and dimension \(m_i\), and
\(u_i (\x^i,\x^{\noti}): \mathcal{X} \rightarrow \mathbb{R}\) is the utility function for player \(i\), parametrized by the strategies \(\x^{\noti}\) of all other players. Here, \(\mathcal{X}:=\mathcal{X}_1 \times \dots \times  \mathcal{X}_n\) denotes the joint strategy space.

Throughout, we assume that each \(\mathcal{X}_i\) is nonempty and \textit{polyhedrally bounded}, so that each player has a finite number of pure strategies and the joint strategy space \(\mathcal{X}\) is finite. This assumption is used when relating IPGs to finite normal-form games in the sequel.

The utility function of player $i$ is often expressed as
\vspaceminus{5pt}
\begin{equation} \label{eq:IPG_obj}
    u_i (\x^i,\x^{\noti})
    = (\x^i)^\top Q^i \x^i 
      + (\mathbf{d}^i)^\top \x^i
      + \smashoperator[lr]{ \sum_{p \in N \setminus \{i\} }} (\x^{p})^\top Q_{p}^i \x^i, 
\end{equation}
where \(Q^i\) is a symmetric matrix (not necessarily positive semidefinite), \(\mathbf{d}^i\) is a linear coefficient vector, and the terms \((\x^{p})^\top Q_{p}^i \x^i\) capture inter-player interactions through \textit{reciprocally bilinear} terms. 

In an IPG, players act non-cooperatively, each solving an individual integer program (IP) to maximize their utility. The best-response problem for player \(i\) is
\begin{equation} \label{eq:BR_subproblem}
    \max_{\x^i} \bigl\{ u_i (\x^i,\x^{\noti}) : \x^i \in \mathcal{X}_i \bigr\}.
    \vspaceminus{15pt}
\end{equation}
We assume that the best-response problems \eqref{eq:BR_subproblem} are tractable and solved to optimality.

A \textit{strategy profile} \(\x = (\x^1, \dots, \x^n)\) is a collection of all players' strategies. A profile \(\hat\x\) is a \textit{pure Nash equilibrium (PNE)} if, for each player \(i \in N\), the strategy \(\hat\x^i\) is a best response (BR) to the others' strategies \(\hat{\x}^{\noti}\), i.e.,
\(
u_i(\hat{\x}^i, \hat{\x}^{\noti}) 
\geq u_i(\x^i, \hat\x^{\noti}),
\, \forall\, \x^i \in \mathcal{X}_i.
\)
\citet{carvalho2018existence} show that deciding whether a general IPG admits a PNE is a \(\Sigma_2^p\)-complete problem and that PNEs may fail to exist even in simple two-player IPGs.

We define the social welfare function as
\(
\phi(\x) := \sum_{i \in N} u_i (\x^i,\x^{\noti}),
\)
and refer to a \textit{socially optimal} strategy profile (or \textit{optimal social-welfare} (OSW) solution) as
\(
\barxosw \in \arg \max_{\x \in \mathcal{X}} \phi (\x).
\)
Let \(\mathcal{S}_{\text{pne}}(\G)\) denote the set of all PNEs of the IPG instance $\G$. The \textit{best PNE} is then defined as
\(
\hat{\x}^*_{\text{pne}} \in \arg \max_{\x \in \mathcal{X} \cap \mathcal{S}_{\text{pne}}(\G)} \phi (\x).
\)
The \textit{price of stability} (POS) \citep{roughgarden2010algorithmic}, a common measure of inefficiency in game theory, is defined as the ratio
\(
\text{POS} \;=\; \frac{\phi(\barxosw)}{\phi(\hat{\x}_\text{pne}^*)}.
\)
A larger POS indicates a larger gap in social welfare between the best PNE and the socially optimal solution. 
% We summarize the main symbols and abbreviations in Table~\ref{tab:abbreviations}.

\paragraph*{IPGs as finite normal-form games.}
When each player's pure-integer feasible set \(\mathcal{X}_i\) is finite and polyhedrally bounded, an IPG can be viewed as a finite normal-form game by setting \(S_i := \mathcal{X}_i\) as the pure strategy set of player \(i\) and defining utilities \(u_i\) at each joint profile to equal the IP objective values. This representation preserves payoffs and best responses and induces the same best-response state graph and Markov chain as those defined below. We emphasize that this is a conceptual viewpoint—explicit enumeration of \(\prod_{i\in N}\mathcal{X}_i\) is generally intractable—and our methods avoid it by computing BRs via player-specific IPs.

\subsection{Normal-Form Games, Best-Response State Graph, and Markov-Chain Reduction} \label{sec:normal_br_state}
Let \(N=\{1,\dots,n\}\) denote the players. For each \(i\in N\), let \(S_i\) be a \textit{finite} set of pure strategies and set \(S:=\prod_{i\in N} S_i\). Utilities \(u_i:S\to\mathbb{R}\) are arbitrary but deterministic. 
For each $i\in N$ and profile $s=(s_i)_{i\in N}\in S$, define the set of strictly improving best responses (BRs) as
\(
\BR_i^{+}(s)
:=\left\{\, t_i \in \arg\max_{r_i\in S_i} u_i(r_i,s_{-i}) \;:\; u_i(t_i,s_{-i}) > u_i(s_i,s_{-i}) \right\}.
\)
We say that player \(i\) has an \textit{improving BR move} at \(s\) if \(\BR_i^{+}(s)\neq\emptyset\); otherwise, player \(i\) is already best-responding at \(s\) (possibly with ties), and we regard this as a \textit{no-move} situation for player \(i\).

\paragraph*{Best-response state graph and a time-homogeneous Markov chain.}

The \textit{best-response (BR) state graph} is \(\mathcal{U}=(S,E)\) with a directed edge
\(
s \;\to\; (t_i,s_{-i})
\)
for each \(s\in S\), \(i\in N\), and \(t_i\in \BR_i^{+}(s)\); that is, edges correspond to strictly improving BR moves. We also interpret no-move updates as self-loops \(s\to s\). A profile \(s^*\) is a PNE iff \(\BR_i^{+}(s^*)=\emptyset\) for all \(i\in N\)--equivalently, \(s^*\) has no outgoing strictly improving BR edges, so all BR updates reduce to self-loops.

Fix a scheduler \(\pi\) with \(\pi_i>0\) for all \(i\in N\) (e.g., uniform \(\pi_i=1/n\)). For each player \(i\in N\) and state \(s\in S\), we specify a \textit{tie-breaking kernel} \(\lambda_i(\cdot\mid s)\), a probability distribution on \(S\), such that:
\begin{itemize}
    \item if \(\BR_i^{+}(s)\neq\emptyset\), then \(\lambda_i(\cdot\mid s)\) is supported on the improving BR successors
    \(
    \{\, (t_i,s_{-i}) : t_i\in\BR_i^{+}(s) \,\},
    \)
    and each such successor has positive probability (for instance, uniform choice over \(\BR_i^{+}(s)\));
    \item if \(\BR_i^{+}(s)=\emptyset\), then \(\lambda_i(s\mid s)=1\), i.e., player \(i\) does not move when already best-responding (even if there are other tied best responses).
\end{itemize}
We then define transition probabilities
\begin{equation}\label{eq:transition}
P(s,s')\ :=\ \sum_{i\in N}\pi_i\,\lambda_i(s'\mid s),\qquad s,s'\in S.
\end{equation}
Thus \((X_t)_{t\ge 0}\) with \(X_{t+1}\sim P(X_t,\cdot)\) is a finite, time-homogeneous Markov chain whose positive-probability transitions are exactly the edges of \(\mathcal{U}\) (including self-loops). By the Markov property, the next state depends on the current state and a fresh draw \(I_t\sim\pi\) together with the tie-breaking kernels. At a PNE \(s^*\), \(\BR_i^{+}(s^*)=\emptyset\) for all \(i\), hence \(\lambda_i(s^*\mid s^*)=1\) and
\(
P(s^*,s^*)=1
\),
so \(\{s^*\}\) is an absorbing class. Standard notions such as communicating classes, recurrent and transient states, hitting probabilities, and expected hitting times apply.

\section{Convergence of Best-Response Dynamics in Finite Normal-Form Games}
\label{sec:convergence-brd}

We establish convergence guarantees for random-restart best-response dynamics (RR-BRD) in IPGs via the following steps: (i) cast an IPG with finite pure-integer feasible sets as a finite normal-form game; (ii) construct the best-response (BR) state graph; (iii) compute its strongly connected components (SCCs) and the associated condensation directed acyclic graph (DAG); and (iv) identify sink SCCs, distinguishing singleton \textit{PNE sinks} from multi-node \textit{non-PNE sinks}. Steps (i)--(ii) are described in Section~\ref{sec:preliminaries}. The baseline fact is that a random walk on the BR state graph eventually enters a sink SCC and remains there; thus it converges either to a singleton PNE sink or to a non-PNE sink (trap) \citep{goemans2005sink}. Our addition is to introduce \textit{random restarts} aimed at reaching singleton PNE sinks, yielding a Las Vegas--style guarantee: if a PNE exists and the restart distribution places positive mass on any state from which some PNE sink is reachable, then RR-BRD finds (and certifies) a PNE almost surely.

\subsection{SCCs, the Condensation DAG, and Recurrent Classes}\label{sec:SCC}

Let $\mathrm{SCC}(\mathcal{U})$ denote the set of strongly connected components of $\mathcal{U}$, and let $\mathcal{D}$ be the \textit{condensation DAG} of $\mathcal{U}$, obtained by contracting each SCC (so $\mathcal{D}$ is finite and acyclic). An SCC $S$ is a \textit{sink SCC} if and only if no BR edge leaves $S$ (equivalently, the node corresponding to $S$ is a sink in $\mathcal{D}$). We distinguish: (i) a \textit{PNE sink}, i.e., a singleton sink SCC $\{s^*\}$ (necessarily a PNE), and (ii) a \textit{non-PNE sink} (or \textit{trap}), i.e., a sink SCC $S$ with $|S|>1$. This structure is illustrated in Figure~\ref{fig:connected-condensation-dag}.

\begin{figure}[h!]
\centering
\resizebox{0.8\textwidth}{!}{%
% \begin{tikzpicture}[
%   >=Latex,
%   node distance=10mm and 14mm,
%   scc/.style={draw, rounded corners, minimum width=16mm, minimum height=9mm, align=center},
%   sink/.style={draw, double, circle, minimum size=10mm, align=center, thick},
%   arr/.style={->, line width=0.6pt},
%   note/.style={draw=none, font=\footnotesize, align=center}
% ]

\begin{tikzpicture}[
  >=Latex,
  node distance=10mm and 14mm,
  label distance=2pt,
  every label/.style={font=\footnotesize},
  scc/.style={draw, rounded corners, minimum width=16mm, minimum height=9mm, align=center},
  sink/.style={draw, double, circle, minimum size=10mm, align=center, thick},
  arr/.style={->, line width=0.6pt},
  note/.style={draw=none, font=\footnotesize, align=center}
]

% --- Root SCC ---
\node[scc, label=below:{SCC $A$}] (A) {};

% --- Upper branch toward P ---
\node[scc, right=28mm of A, label=below:{SCC $B$}] (B) {};
\node[scc, right=24mm of B, label=below:{SCC $C$}] (C) {};
\node[sink, right=22mm of C, label=below:{PNE sink $P$}] (P) {};

% --- Lower branch toward C ---
\node[scc, below=18mm of B, label=above:{SCC $D$}] (D) {};
\node[scc, right=24mm of D, label=above:{SCC $E$}] (E) {};
\node[sink, right=22mm of E, label=above:{non\mbox{-}PNE sink $M$}] (S) {};

% % --- Root SCC (multi-node) ---
% \node[scc] (A) {SCC $A$};

% % --- Upper branch toward P (mix of singleton/multi SCCs) ---
% \node[scc, right=28mm of A] (B) {SCC $B$};       % singleton
% \node[scc, right=24mm of B] (U1) {SCC $U$};    % multi
% \node[sink, right=22mm of U1, label=right:{\footnotesize PNE sink}] (P) {$P$};

% % --- Lower branch toward C (mix of multi/singleton SCCs) ---
% \node[scc, below=18mm of B] (D) {SCC $D$};       % multi
% \node[scc, right=24mm of D] (E) {SCC $E$};       % singleton
% \node[sink, right=22mm of E, label=right:{\footnotesize non\mbox{-}PNE sink}] (C) {$C$};

% --- Edges (DAG; overall connected) ---
\draw[arr] (A) -- (B);
\draw[arr] (A) -- (D);
\draw[arr] (B) -- (C);
\draw[arr] (C) -- (P);
\draw[arr] (D) -- (E);
\draw[arr] (E) -- (S);

% Optional cross-edge (keeps acyclic, reinforces connectedness)
\draw[arr] (B) |- (E);

% --- Internal dots to show SCC cardinalities ---
% Helper macro-ish placements via calc
% A: multi (three dots)
\fill ($(A.center)+(-4pt,3pt)$) circle (1.2pt) coordinate (a1);
\fill ($(A.center)+(4pt,3pt)$)  circle (1.2pt) coordinate (a2);
\fill ($(A.center)+(0pt,-3pt)$) circle (1.2pt) coordinate (a3);
\draw[->, shorten >=1pt, shorten <=1pt] (a1) .. controls +(-6pt,6pt) and +(-6pt,6pt) .. (a2);
\draw[->, shorten >=1pt, shorten <=1pt] (a2) .. controls +(6pt,-8pt) and +(6pt,-8pt) .. (a3);
\draw[->, shorten >=1pt, shorten <=1pt] (a3) .. controls +(-8pt,-6pt) and +(-8pt,-6pt) .. (a1);

% B: singleton (one dot)
\fill ($(B.center)+(0pt,0pt)$) circle (1.2pt);

% U1: multi (three dots)
\fill ($(C.center)+(-4pt,3pt)$) circle (1.2pt) coordinate (u1);
\fill ($(C.center)+(4pt,3pt)$)  circle (1.2pt) coordinate (u2);
\fill ($(C.center)+(0pt,-3pt)$) circle (1.2pt) coordinate (u3);
\draw[->, shorten >=1pt, shorten <=1pt] (u1) .. controls +(-6pt,6pt) and +(-6pt,6pt) .. (u2);
\draw[->, shorten >=1pt, shorten <=1pt] (u2) .. controls +(6pt,-8pt) and +(6pt,-8pt) .. (u3);
\draw[->, shorten >=1pt, shorten <=1pt] (u3) .. controls +(-8pt,-6pt) and +(-8pt,-6pt) .. (u1);

% D: multi (two dots)
\fill ($(D.center)+(-3pt,0pt)$) circle (1.2pt) coordinate (d1);
\fill ($(D.center)+(3pt,0pt)$)  circle (1.2pt) coordinate (d2);
\draw[->, shorten >=1pt, shorten <=1pt] (d1) .. controls +(-6pt,6pt) and +(-6pt,6pt) .. (d2);
\draw[->, shorten >=1pt, shorten <=1pt] (d2) .. controls +(6pt,-6pt) and +(6pt,-6pt) .. (d1);

% E: singleton
\fill ($(E.center)+(0pt,0pt)$) circle (1.2pt);

% S: multi sink with an internal cycle (three dots + cycle arrows)
\fill ($(S.center)+(-4pt,3pt)$) circle (1.2pt) coordinate (c1);
\fill ($(S.center)+(4pt,3pt)$)  circle (1.2pt) coordinate (c2);
\fill ($(S.center)+(0pt,-4pt)$) circle (1.2pt) coordinate (c3);
\draw[->, shorten >=1pt, shorten <=1pt] (c1) .. controls +(-6pt,6pt) and +(-6pt,6pt) .. (c2);
\draw[->, shorten >=1pt, shorten <=1pt] (c2) .. controls +(6pt,-8pt) and +(6pt,-8pt) .. (c3);
\draw[->, shorten >=1pt, shorten <=1pt] (c3) .. controls +(-8pt,-6pt) and +(-8pt,-6pt) .. (c1);

% P: singleton sink (single dot)
\fill ($(P.center)+(0pt,0pt)$) circle (1.2pt);

\end{tikzpicture}%
}
\caption{Illustration of a Connected Condensation DAG with Two Terminal Sinks: a singleton \emph{PNE sink} $P$ and a multi-node \emph{non-PNE sink} $M$. Boxes and circles are SCCs; internal dots depict the number of profiles each SCC contains (singletons vs.\ multi-node SCCs). Arrows are edges of the acyclic condensation graph. SCC labels (A, B, C, D, E, P, M) are arbitrary and used solely for illustration. In potential or BR-weakly acyclic games, non-PNE sinks do not occur; all sink SCCs are singleton PNEs, so the branch labeled $M$ is absent.}
\label{fig:connected-condensation-dag}
\end{figure}

Figure~\ref{fig:connected-condensation-dag} illustrates that best-response dynamics may converge either to a singleton PNE sink or to a non-PNE sink, from which escape is impossible---motivating the use of random restarts developed in Section~\ref{subsec:rrrbrd}. Because \eqref{eq:transition} assigns positive probability exactly to BR edges, the Markov chain's communicating classes coincide with the SCCs of $\mathcal{U}$; in particular, its \textit{recurrent classes} are precisely the \textit{sink SCCs}. Thus, every PNE is a singleton recurrent (absorbing) class, while any non-PNE sink is a closed, non-singleton trap. BRD as a random walk (i.e., equivalently, with randomized playing sequences) samples \(I_t \in N\) i.i.d.\ with \(\Pr(I_t=i)=\pi_i>0\) and, conditional on \(I_t=i\), updates \(s_{t+1}\) by applying one improving best-response move for player \(i\) according to \(\lambda_i(\cdot\mid s_t)\). This produces exactly the Markov chain with transition kernel \(P\) in \eqref{eq:transition}.

Randomized playing sequences \citep{heinrich2023best} can help escape \textit{non-sink} cycles (for example, cycles in an SCC with at least one outgoing edge, such as the SCCs labeled $A$, $C$ and $D$ in Figure~\ref{fig:connected-condensation-dag}), whereas a fixed update order can get stuck there (i.e., the fixed update order can prevent the escaping BR move). However, using a randomized playing sequence is not enough; once the chain enters a non-PNE trap, it cannot leave. In particular, when non-PNE traps exist, a random playing sequence alone does \textit{not} ensure reaching a PNE, as emphasized in the sink-equilibrium framework of \citet{goemans2005sink}.

\subsection{Deterministic Convergence in Structured Subclasses}
\paragraph*{Potential games and best-response weak acyclicity.}
(compare Fig. \ref{fig:connected-condensation-dag}: non-PNE sink is absent in these subclasses)
A finite game is a (weighted) \textit{potential game} if some $\psi:S\!\to\!\mathbb{R}$ tracks unilateral improvements: for all $i$ and $t_i\in S_i$,
\(
u_i(t_i,s_{-i})-u_i(s_i,s_{-i})=w_i\bigl(\psi(t_i,s_{-i})-\psi(s_i,s_{-i})\bigr)
\)
with $w_i>0$.
By the finite improvement property, every BR path strictly increases $\psi$ and therefore reaches a PNE in finitely many steps \citep{monderer1996potential}; classic subclasses include congestion games \citep{rosenthal1973class} and player-specific congestion games \citep{milchtaich1996congestion}.
A game is \textit{BR-weakly acyclic} if every sink SCC in the BR state graph is a singleton PNE (equivalently, from every start, there exists a finite BR path to a PNE). In such games, \textit{every} BR trajectory (under any scheduler) reaches a PNE in finite time because infinite trajectories would have to reside in a non-PNE sink, which is excluded by definition; see, e.g., \citet{apt2015classification} for systematic treatments and related characterizations.

\medskip
\noindent\textit{Bridge to the general case.}
Outside these structured subclasses, the BR state graph may contain non-PNE trap components, which can trap BRD even under randomized playing sequences (see the sink-equilibrium viewpoint of \citealt{goemans2005sink}). The next subsection shows that introducing \textit{random restarts} yields Las Vegas--style guarantees and instance-dependent bounds in arbitrary finite games.

\subsection{Round Random-Restart Best-Response Dynamics (RRR--BRD)}
\label{subsec:rrrbrd}
RRR--BRD is a round-based variant of BRD that incorporates \textit{random restarts}. A \textit{restart}
means that if the dynamics fails to reach a PNE within a prescribed cap, we abandon the current
trajectory, reinitialize the strategy profile (according to a restart distribution), and repeat. Instead of updating one player at a time, RRR--BRD \textit{batches} updates into \textit{rounds}:
in each round, every player plays one improving BR move, in a uniformly random order.
This makes local PNE certification simple: if no player changes in a round (i.e., no improving BR exists when each player is selected), then the current profile is a PNE. Hereafter, we use RRR--BRD to denote the implemented algorithm; we retain RR--BRD as a conceptual umbrella term.

The guarantees we develop are \textit{Las Vegas} rather than deterministic: randomness affects the trajectory and runtime, but the algorithm never returns an incorrect solution---whenever it returns, it outputs a \emph{certified} PNE. If a PNE exists and a capped attempt succeeds with positive probability,
then independent unbounded restarts imply almost-sure convergence to a PNE. We now present RRR--BRD (Algorithm~\ref{alg:rrrbrd}).

\begin{algorithm}[H]
\caption{Round Random-Restart BRD (RRR--BRD)}
\label{alg:rrrbrd}
\begin{algorithmic}[1]
\small
\Require restart law $\mu$ on $S$; round cap $R\in\mathbb{N}$
\While{true}
  \State sample $s \sim \mu$   \Comment{fixed-cap restart: if no PNE within $R$ rounds, restart}
 \label{alg:rrrbrd_sampler}
  \For{$r=1,2,\dots,R$}
     \State draw a random permutation $\Pi$ of $\{1,\dots,n\}$ \Comment{uniform}
     \State $s^{\mathrm{start}} \gets s$
     \State \textbf{for} $i$ in order $\Pi$ \textbf{do}
        apply one improving best-response update for player $i$
     \State \textbf{if} $s = s^{\mathrm{start}}$
\textbf{then} \Return $s$ \Comment{PNE certified by end-of-round no-change}
  \EndFor
\EndWhile
\end{algorithmic}
\end{algorithm}

\paragraph*{Non-PNE sinks and the avoidance of trap detection.}
The BR state graph may contain non-singleton SCCs and, in particular, non-PNE traps. A natural question is whether Algorithm~\ref{alg:rrrbrd} could implement a ``stop-on-trap'' rule---e.g., restart immediately (Line~\ref{alg:rrrbrd_sampler}) once the
trajectory enters a non-PNE trap. While appealing in principle, this requires detecting sink membership, which is computationally intractable in general for succinctly represented games. To the best of our knowledge, there is no general complexity classification of the \textsc{In-a-Sink} problem for arbitrary succinct normal-form representations; existing hardness results are established for specific classes. For example, \textsc{In-a-Sink} is PSPACE-complete for graphical games/BGP~\citep{fabrikant2008complexity}, and across several natural game classes (e.g., anonymous, player-specific or weighted congestion, valid-utility, and two-sided markets) sink membership and the existence of non-singleton sinks are PSPACE-complete, while even PNE existence can be NP-hard~\citep{mirrokni2009complexity}.
Accordingly, our algorithm avoids trap detection entirely: 
Algorithm~\ref{alg:rrrbrd} uses only inexpensive \emph{local} checks for PNE certification and fixed-cap restarts, and never requires constructing or reasoning over the full BR state graph.

\paragraph*{Round-based (macro) Markov chain and hitting times.}
For a permutation $\pi$ of $\{1,\dots,n\}$, let $F_{\pi}$ denote the (random)
operator that applies exactly one improving BR update for each player
in the order $\pi$, using the tie-breaking kernels $(\lambda_i)_{i\in N}$ from
Section~\ref{sec:normal_br_state}. If $(\Pi_r)_{r\ge 0}$ are i.i.d.\ uniform
permutations, then the round-boundary process
\(
S_{r+1} \;=\; F_{\Pi_r}(S_r), \, r=0,1,2,\dots,
\)
is a time-homogeneous Markov chain on the finite state space $S$.
PNEs are still absorbing states. Moreover, any sink SCC of the stepwise BR state
graph remains closed under round transitions, so the recurrent classes are unchanged. For any $C\subseteq S$, define the (round) hitting time
$\tau_C := \inf\{r\ge 0:\ S_r \in C\}$ and write $\tau_{\mathrm{PNE}}$ for the hitting time of the set of PNE.

\paragraph*{Restarts and per-attempt success probability.}
A restart law $\mu$ is any probability distribution on $S$.
For a fixed round cap $R\ge 1$, define the per-attempt success probability
\begin{equation}\label{eq:q_mu_R}
q_\mu[R]
~:=~
\mathbb{E}_{s\sim\mu}\!\big[\Pr_s(\tau_{\mathrm{PNE}}\le R)\big],
\end{equation}
i.e., the probability that a single RRR--BRD attempt started from $s\sim\mu$
reaches (and certifies) a PNE within at most $R$ rounds. Note that $q_\mu[R]>0$
whenever $\mu$ assigns positive probability to at least one state from which a
PNE is reachable via improving BR moves within $R$ rounds.
(In contrast, \citet{hakim2024swim} study \emph{limit} (infinite-horizon)
absorption probabilities into \emph{sink SCCs} under a different noisy dynamics.)

We now establish the following result; proof is given in Appendix
\ref{appendix:proofs}.

\begin{theorem}
[RRR--BRD: almost surely success and expected complexity]
\label{thm:rrr-one}
Assume a PNE exists and the restart law $\mu$ assigns positive mass to states from which a PNE is reachable via improving best-response moves. If $q_\mu[R]>0$, then:
\textup{(i)} RRR--BRD finds a PNE with probability $1$ and \textup{(ii)} The expected total number of best-response solves is bounded by \(
\displaystyle
\mathbb{E}[\#\text{BR solves}] \ \le\ \frac{nR}{q_\mu[R]}.
\)
\end{theorem}

\begin{remark}\label{rmk:full_support_q_mu}
If a PNE exists and $\mu$ has full support on $S$, then $q_\mu[R]>0$ for all
$R\ge 1$ because $\mu$ assigns positive probability to every state, including
each PNE state (and thus a PNE can be reached within one capped attempt).
Therefore, the condition $q_\mu[R]>0$ in Theorem~\ref{thm:rrr-one} holds under
any full-support restart law and more generally whenever $\mu$ assigns positive
mass to the attraction basin of some PNE. 
\end{remark}

Theorem \ref{thm:rrr-one} concerns the idealized unbounded-restart variant; in computation we cap the number of restarts at $L$ and report failure if no PNE is reached. If each BR solve takes time at most $T_{\BR}$, then the expected time until discovery
\(
\mathbb{E}[\text{time}] \le \mathbb{E}[\#\text{BR solves}]\,T_{\BR}
\le \frac{nR}{q_\mu[R]}T_{\BR}.
\)
Here $T_{\BR}$ depends on the BR implementation and may be difficult to characterize.

\paragraph*{RRR-BRD in IPGs.}
We now specialize RRR--BRD to IPGs. The scheme is analogous to Algorithm~\ref{alg:rrrbrd}, but states are represented by integer-feasible strategy profiles $\x = (\x^1,\dots,\x^n)$, and BR updates are implemented by solving player-specific IPs. The round-based random permutation is encoded in Line~\ref{alg:RRR_BRD_IPG_line3}, and random restarts are encoded in Line~\ref{alg:RRR_BRD_IPG_line11}.

\begin{algorithm}[h!]
\caption{RRR-BRD Algorithm for Integer Programming Games}
\begin{algorithmic}[1]
\small
\Require  IPG instance, initial strategy profile $\bar{\mathbf{x}}_{\text{init}}$, max rounds $R$, max restarts $L$
\State Initialize current profile $\bar{\mathbf{x}} \gets \bar{\mathbf{x}}_{\text{init}}$ (optional) \textbf{or} $\bar{\mathbf{x}} \leftarrow$ \textsc{Random\_Generation}($\mathcal{X}$) \label{alg:RRR_BRD_IPG_line0} 
\For{$l = 1$ to $L$} \label{alg:RRR_BRD_IPG_line1}
    \For{$r = 1$ to $R$} \label{alg:RRR_BRD_IPG_line2}
        \State draw a random permutation of players $N \gets \textsc{RandomPermutation}(N)$  \label{alg:RRR_BRD_IPG_line3}
        \For{each player $i$ in $N$} \label{alg:RRR_BRD_IPG_line4}
        \State If an improving best response exists, compute $\hat{\x}^i \in \arg\max_{\x^i} u_i(\x^i,\bar{\x}^{\noti})$ and update $\bar{\x}\gets(\hat{\x}^i,\bar{\x}^{\noti})$.
        \EndFor \label{alg:RRR_BRD_IPG_line6}
        \If{consecutive profiles are identical} \label{alg:RRR_BRD_IPG_line7}
            \Return $\bar{\x}$ \Comment{PNE certified by end-of-round no-change}  \label{alg:RRR_BRD_IPG_line8}
        \EndIf \label{alg:RRR_BRD_IPG_line9}
    \EndFor \label{alg:RRR_BRD_IPG_line10}
    \State $\bar{\mathbf{x}} \leftarrow$ \textsc{Random\_Generation}($\mathcal{X}$) \Comment{restart: generate a new initial strategy profile}\label{alg:RRR_BRD_IPG_line11}
\EndFor \label{alg:RRR_BRD_IPG_line12}
\State \Return \texttt{None} \Comment{no PNE found within caps $R,L$}  \label{alg:RRR_BRD_IPG_line13}
\end{algorithmic}
\label{alg:RRR_BRD_IPG}
\end{algorithm}

We now present the following corollary; proof is given in Appendix \ref{appendix:proofs}.

\begin{corollary}[RRR--BRD for IPGs]\label{cor:rrr-ipg}
Consider an IPG $\G=(N,(\mathcal{X}_i)_{i\in N},(u_i)_{i\in N})$ with
polyhedrally bounded feasible sets
$\mathcal{X}_i=\{\x^i\in\mathbb{Z}^{m_i}:A^i\x^i\le b^i\}$, so that
$\mathcal{X}=\prod_{i\in N}\mathcal{X}_i$ is finite and can be viewed as the
state space $S$ of a finite normal-form game. Let $\mu$ be the restart law
induced by \textsc{Random\_Generation}$(\mathcal{X})$ in
Algorithm~\ref{alg:RRR_BRD_IPG}. If $\G$ admits a PNE and $q_\mu[R]>0$, then
Algorithm~\ref{alg:RRR_BRD_IPG} with unbounded restarts ($L\to\infty$) finds a
PNE with probability $1$ and satisfies
\(
\ \mathbb{E}[\#\text{BR solves}] \le \frac{nR}{q_\mu[R]}.  
\)
In particular, if \textsc{Random\_Generation} has full support on $\mathcal{X}$
and a PNE exists, then $q_\mu[R]>0$ for every $R\ge 1$ (Remark~\ref{rmk:full_support_q_mu}).
\end{corollary}

Any feasible-solution generator with broad coverage over $\mathcal{X}$ can be used to implement $\mu$. We provide explicit random feasible-solution generators---both maximal and full-support variants---for KPG and EBMC in Appendix~\ref{appendix:feas_generator}. We alternate these variants across restarts; the resulting mixture retains full support on $\mathcal{X}$. We also cap $R$ and $L$, since we do not know \emph{a priori} whether a PNE exists: if no PNE exists, the conceptual RRR--BRD would never terminate. Moreover, when integrating with a MIP solver, the solver supplies natural initial profiles, so imposing finite caps is practical.

In terms of runtime, if each BR solve takes at most $T_{\BR}$, then the expected runtime bound follows immediately by multiplying the BR-solve bound by $T_{\BR}$. In IPGs, each BR solve is implemented by solving a player-specific IP to optimality, so $T_{\BR}$ can be difficult to characterize theoretically. In practice, one may treat $T_{\BR}$ as an empirical per-solve runtime estimate under the chosen solver settings.

\subsection{Complexity Analysis for RRR--BRD}

The practical question is how to select the round cap \(R\) in RRR--BRD. As \(R\) increases, \(q_\mu[R]\) \eqref{eq:q_mu_R} is nondecreasing, creating a trade-off in expected time. Taking \(R\) to be very large is impractical in a failed attempt, and the idealized ``stop-on-trap'' behavior is unavailable. Throughout our study, we therefore fix a modest cap \(R \in [10,20]\), which empirically balances per-attempt cost and \(q_\mu[R]\).

Because the BR state graph can contain SCCs of size \(>1\) and non-PNE traps, performance can be poor even when a PNE exists: if \(q_\mu[R]\) is extremely small, then after \(r\) restarts the total work \(nRr\) can exceed crude enumeration scales with non-negligible probability. In particular, as a \textit{worst-case event}, more than \(r\) restarts are required (equivalently, the algorithm fails for \(r\) consecutive attempts), which occurs with probability \(\Pr(R^\ast>r)=(1-q_\mu[R])^r\). See Appendix~\ref{appendix:q_mu_R_small} for a schematic and a knapsack-style example where most restarts are absorbed by a non-PNE trap and \(q_\mu[R]\) is exponentially small. As a \textit{best-case event}, the first attempt can succeed, which occurs with probability \(q_\mu[R]\). Very roughly, \(q_\mu[R]\) captures instance difficulty for RRR--BRD under \(\mu\) and \(R\). Our guarantees are Las Vegas--style: success occurs almost surely when \(q_\mu[R]>0\), while efficiency is governed by these instance-dependent success probabilities.

Since \(q_\mu[R]\) is not directly computable in large IPGs, we estimate it
empirically via Monte Carlo simulation over restart states. This aligns with the
sampling-and-simulation approach used to estimate absorption behavior on
response graphs \citep{hakim2024swim}. While their work defines transition via weakly improving deviations and studies infinite-horizon absorption into sink SCCs under noisy dynamics, we estimate a finite-horizon probability of hitting the PNE set within \(\le R\) rounds under improving best-response updates.

Given a restart law \(\mu\), we estimate
\(
\tilde q_\mu[R]\ :=\ \mathbb{E}_{s\sim\mu}\!\big[\mathbf{1}\{\text{RRR--BRD from $s$ reaches a PNE within } R\text{ rounds}\}\big]
\)
using Monte Carlo by averaging the indicator over a finite set of starts. We report \(\text{avg\_rounds}\), i.e., the mean number of rounds over all attempts (using \(R\) for failures and the realized round count for successes). Furthermore, based on $\tilde q_\mu[R]$, we report the expected step bound (ESB) for each instance. 

We examine RRR--BRD in two settings:
\begin{enumerate}[leftmargin=1.25em]
\item[(i)] \textbf{Normal-form reductions of small IPGs.}
We consider instances of the form
\(
\max_{\x^i \in \{0,1\}^m}\ \big\{ \mathbf{d}_i^\top \x^i + \sum_{p \in N:\,p \neq i} (\x^{p})^\top Q_{p}^i \x^i  \big\}, \text{ where }
\mathbf{d}_i\in\mathbb{Z}^p, \text{ and }\ Q_p^i \in\mathbb{Z}^{m \times m}, \text{ with entries in }[-100,100].
\)
For $n \in \{2,3\}$ and $m \in \{3,4\}$, we sample 10 random games per $(n,m)$, retaining only instances that admit at least one PNE. For each instance, \(\mu\) is taken from a uniform distribution over all pure profiles \(S\). Since \(|S|\) is modest, we set \(R=10\) and estimate $\tilde q_\mu[R]$ by testing every starting state.

\item[(ii)] \textbf{Knapsack problem games (KPG).} (see Section~\ref{sec:7.4} for the full definition).
We implement RRR--BRD for IPGs (Algorithm~\ref{alg:RRR_BRD_IPG}) with a round cap \(R=20\) and a restart cap \(L=100\). \(\mu\) is taken from the empirical distribution over 100 randomly generated feasible starts per instance using a feasible-solution generator (Appendix \ref{appendix:feas_generator}); \(\tilde q_\mu[R]\) is then estimated from these starts. We use the KPG instances from \citet{dragotto2023zero} for which at least one PNE has been verified in their experiments.
\end{enumerate}

The summaries for (i) and (ii) appear in Tables~\ref{tab:RRR_BRD_normal_form} and \ref{tab:RRR_BRD_KPG_random}. Full results are in Appendix~\ref{appendix:full_simulation}. In all simulated cases \(\tilde q_\mu[R]>0\), and the average \(\tilde q_\mu[R]\) value often exceeds 0.5, implying high success probability with a modest number of restarts. While $|S|$ can be counted exactly for (i), the number of integer-feasible profiles for (ii) cannot be counted exactly; nevertheless, it is expected to be substantially larger than in (i) due to the many binary decisions, even under knapsack constraints. Note that the average ESB is a conservative upper bound; the realized number of best-response solves is often much smaller.

\begin{table}[h!]
\caption{Normal-Form Reductions: Empirical Success Within $R=10$ Rounds Under Uniform Restarts.}
\centering
\scalebox{0.9}{
\begin{tabular}{ccccccc}
\toprule
$n$ & $k$ & $|S|$ & \# inst.\ with $\tilde q_\mu[R]>0$ & avg.\ $\tilde q_\mu[R]$ & avg\_rounds & avg\_ESB ($\frac{nR}{q_\mu[R]}$) \\
\midrule
2 & 3 & $64$   & $10/10$ & $1.00$ & $2.6$ & 20 \\
2 & 4 & $256$  & $10/10$ & $0.925$ & $3.2$ & 26 \\
3 & 3 & $512$  & $10/10$ & $0.58$ & $6.4$ & 94.4 \\
3 & 4 & $4096$ & $10/10$ & $0.72$ & $5.7$ & 62.4 \\
\bottomrule
\end{tabular}
}
\label{tab:RRR_BRD_normal_form}
\end{table}

\begin{table}[h!]
\caption{KPGs: Empirical Success Within $R=20$ Rounds Under Random Restarts.}
\label{tab:RRR_BRD_KPG_random}
\centering
\scalebox{0.9}{
\begin{tabular}{ccccc}
\toprule
type & \# inst.\ with $\tilde q_\mu[R]>0$ & avg.\ $\tilde q_\mu[R]$ & avg\_rounds & avg\_ESB ($\frac{nR}{q_\mu[R]}$) \\
\midrule
A & 23/23 & 1   & 3.6 & 49.6 \\
B & 20/20 & 1  & 3.4  & 48 \\
C & 7/7 & 0.60 & 10.1  & 2135.2\\
\bottomrule
\end{tabular}
}

\vspace{0.1cm}
\footnotesize{
% The detailed explanation about types (A), (B), (C) is explained in Section~\ref{sec:7.4}.
% For any player $i$, 
Types (A) and (B) draw the interaction coefficients equally and uniformly from the set in $[1,100]$ and independently and uniformly from the set in $[1,100]$, respectively. Type (C) draws them from a uniform distribution over $[-100,100]$. 
}
\end{table}

This Monte Carlo exercise is used only to contextualize complexity, not as a prerequisite for implementation. Note that the estimate $\tilde q_\mu[R]$ is inherently \emph{algorithm- and parameter-dependent} outside structured classes (e.g., potential games or BR-weakly acyclic games). Accordingly, we do not propose estimating $\tilde q_\mu[R]$ in practice; rather, we run RRR-BRD directly and treat $\tilde q_\mu[R]$ as an \emph{ex post} diagnostic. In particular, even a \emph{single} successful run that returns a PNE immediately certifies $q_\mu[R]>0$ for that $(\mu,R)$ (and yields a PNE to report), whereas failure to hit a PNE within a finite budget does not rule out $q_\mu[R]>0$.

\section{Best-response Dynamics Incorporated Zero-regret Algorithm} \label{sec:BZR}
This section develops our algorithmic framework for computing PNEs in large-scale IPGs.
We first review the lifted formulation and equilibrium inequalities (EIs) used in ZR
(Section~\ref{sec:lifted}), and then present our best-response-incorporated variant and
implementation details (Section~\ref{sec:algorithm_details}). We also show how RRR--BRD
can be used as a randomized local-search subroutine (Section~\ref{sec:RR_BRD_others}) and, if it terminates
without identifying any PNE, 
how it can be adapted to provide an approximate-equilibrium certificate (Section~\ref{ss:approx_pne}).

\subsection{Lifted Space and Perfect Equilibrium Formulation} \label{sec:lifted}
Following \citet{dragotto2023zero}, we work in a lifted space in which all players' utilities are linearized. Let $\x := (\x^1,\dots,\x^n)$ collect the players' decision vectors and $\z$ denote auxiliary variables used to linearize nonlinear terms (e.g., bilinear products). Let $\K$ be the lifted feasible set,
\(
\K \;:=\; \{(\x,\z) : \x^i \in \X_i \ \forall i\in N,\ \text{$(\x,\z)$ satisfies all linearization constraints and bounds}\},
\)
and let $\conv(\K)$ be its convex hull. Any integer point $(\x,\z)\in \K$ corresponds to a unique strategy profile $\x$; conversely, every pure-strategy profile has at least one lifted representative in $\K$.

Let $\mathcal{S}_{\text{pne}}(\G)$ denote the set of all PNE profiles of the IPG instance $\G$. Define the \textit{perfect equilibrium formulation} as
\(
\E \;:=\; \{(\x,\z)\in \conv(\K) : \x \in \conv(\mathcal{S}_{\text{pne}}(\G)).\}
\) Let $\BR_i( \x^{\noti})$ denote the set of BRs of player $i$ to a profile $\x^{\noti} \in \prod_{k \in N, k\neq i} \mathcal{X}_{k}$. 
%\SC{Has $\mathcal{X}_{k}$ been defined?}
%\HW{See Section~\ref{sec:IPG_PNE}.}
ZR is a cutting-plane algorithm using EIs, that is, linear inequalities valid for $\E$. Such inequalities are obtained from BRs: for any player $i$ and any BR $\hat \x^i \in \BR_i(\x^{-i})$, the inequality
\begin{equation} \label{eq:equilibrium_ineqs}
    u_{i}(\hat{\x}^i,\x^{\noti}) \;\leq\; u_{i}(\x^i,\x^{\noti}),
    \vspaceminus{18pt}
\end{equation}
is valid for all $(\x,\z)\in \E$ (with utilities evaluated in the lifted space). \citet{dragotto2023zero} show that collecting all such inequalities over all players and all BRs yields an \textit{equilibrium closure} that coincides with $\E$ (Theorem~1 in \cite{dragotto2023zero}).

Given a candidate point $(\bar \x,\bar \z)\in\conv(\K)$, the \emph{equilibrium separation oracle} solves, for each player $i$,
\(
\hat \x^i \in \argmax_{\x^i\in \X_i} \bigl\{u_i(\x^i, \bar \x^{-i})\bigr\}
\). If $u_i(\hat \x^i,\bar \x^{-i})>u_i(\bar \x^i,\bar \x^{-i})$, the associated EI \eqref{eq:equilibrium_ineqs} is violated at $(\bar \x,\bar \z)$ and is added as a cut; if no such violation occurs for any $i$, then $\bar \x$ is a PNE. 

The original ZR algorithm iteratively solves a master problem of the form
\begin{equation}\label{eq:joint_optimization}
\max\{ \phi(\x,\z) : (\x,\z)\in \K,\ (\x,\z) \text{ satisfies all cuts in }\Omega\},
\end{equation}
where $\Omega$ is a growing family of EIs. At each iteration, ZR solves \eqref{eq:joint_optimization} to optimality in the lifted space, queries the equilibrium separation oracle at the optimal solution $(\bar \x,\bar \z)$, and either (i) certifies $\bar \x$ as a welfare-maximizing PNE or (ii) adds violated EIs to $\Omega$ and repeats. This yields a cutting-plane method that is provably correct and finitely terminating whenever a PNE exists. We note that regardless of the choice of the objective function in \eqref{eq:joint_optimization}, the capability of computing a PNE via the separation procedure remains valid. 

\paragraph*{Motivation for the BZR algorithm.}
The BZR algorithm is motivated by the complementary strengths and weaknesses of ZR and RRR-BRD.

\emph{(i) Limitations of ZR and how RRR-BRD complements it.}
For large-scale IPGs, the ZR approach faces several practical challenges. Solving the joint optimization problem \eqref{eq:joint_optimization} to optimality can be very expensive when $\mathcal{X}$ encodes many players and variables. Even when the POS is small, many cuts may be needed to move from the socially optimal solution $\barxosw$ to $\hat{\x}^*_{\text{pne}}$. Moreover, the original implementation only adds EIs for BRs that are currently violated, even though additional BRs could, in principle, be used to cut off larger portions of the non-equilibrium region. RRR-BRD (Algorithm~\ref{alg:RRR_BRD_IPG}) addresses these issues by rapidly finding PNEs at scale and by generating many BRs along its trajectories. 
Embedding RRR-BRD as a randomized local-search subroutine inside ZR supplies PNEs in instances where plain ZR struggles to find any PNE within the time limit, and provides a rich pool of best responses that can be recycled as additional EIs.

\emph{(ii) Limitations of RRR-BRD and how ZR complements it.}
RRR-BRD itself has only probabilistic guarantees: it can efficiently find PNEs when they exist but cannot certify that no PNE exists. When integrated with the lifted ZR formulation, RRR-BRD benefits from diverse, welfare-oriented starting points produced by the MIP solver through callbacks, which inject both randomness and structure into the initial profiles. At the same time, the combined BZR algorithm inherits ZR's structural capabilities: given sufficient time, ZR can certify ``no PNE'' by proving infeasibility of the equilibrium formulation and can be configured to enumerate PNEs (e.g., via Hamming-distance cuts). Thus, ZR supplies the certification and enumeration machinery that RRR-BRD lacks, while RRR-BRD supplies the scalable search capability that ZR lacks.

\subsection{Algorithm Details} \label{sec:algorithm_details}
We now present the BZR algorithm (Algorithm~\ref{alg:BZR}) and illustrate its behavior in Figure~\ref{fig:BZR_illustration}, which depicts the joint strategy space $\mathcal{X}$ for two players. The BZR algorithm can be viewed as a combination of a randomized local-search subroutine and a cutting-plane method with delayed (lazy) separation of EIs.

\textbf{Callback function and separation procedure:}
We use the MIP solver’s callback mechanism that triggers on integer-feasible solutions (e.g., \texttt{MIPSOL} in Gurobi \citep{gurobi}). While many types of intervention in the branch-and-bound tree are possible (e.g., custom branching rules), we restrict our use of callbacks to (i) collect integer-feasible points and (ii) inject EIs. Each incumbent integer solution $\bar\x$ returned by the callback is passed to the separation procedure (Lines~\ref{alg:oracle_1}--\ref{alg:oracle_6}). The oracle checks whether $\bar\x$ is a PNE; if not, it adds violated EIs. In Figure~\ref{fig:BZR_illustration}, these cuts are shown in green.

\textbf{RRR-BRD as a randomized local-search subroutine:} 
If $\bar\x$ is not a PNE, it is used as an initial strategy profile for RRR-BRD (Line~\ref{alg:BZR_3}). Along each RRR-BRD trajectory, we store the BRs encountered; for any BR $\hat\x^i$ whose associated EI has not yet been added, we may inject it into $\Omega$ as an additional cut (Lines~\ref{alg:additional_1}--\ref{alg:additional_3}). In Figure~\ref{fig:BZR_illustration}, the RRR-BRD trajectory is illustrated by the blue arrow, and these additional EIs are drawn in red. Note that this step is valid regardless of whether RRR-BRD finds a PNE in that run: every trajectory consists of valid BRs, hence all associated EIs are valid. If a newly found PNE improves the current best value, we update the lower bound and set $\hat\x_{\text{pne}}^* \gets \bar\x$ (Line~\ref{alg:BZR_5}). If RRR-BRD fails to find a PNE within the caps $(R,L)$, no primal update occurs and the algorithm simply proceeds to the next callback solution.

\textbf{Termination:} 
The dual (upper) bound $UB$ is updated by solving
\(
UB \;=\; \max \left\{ \phi(\x, \z) \;\middle|\; \x \in \K_{\text{relax}} \cap \Omega \right\},
\)
where $\K_{\text{relax}}$ denotes the current relaxation of $\K$ and $\Omega$ is the set of accumulated EIs. The BZR algorithm terminates once $UB - LB \le \texttt{tol}$ (Line~\ref{alg:BZR_7}--\ref{alg:BZR_8}), where $\texttt{tol}$ is the solver's default optimality tolerance. This ensures that either the best PNE has been found or that the PNE formulation is infeasible (no PNE exists).

\begin{remark} \label{rmk:X_relax}
The set $\K_{\text{relax}}$ consists of the LP relaxation of $\K$, augmented with general-purpose cutting planes generated by the solver. In our implementation, this dual bound $UB$ may remain above $\barxosw$ because BZR does \emph{not} require solving 
\(
\max \{\phi(\x,\z) \mid (\x,\z) \in \K \cap \Omega\}
\)
to optimality at each iteration. Since $UB$ is shaped jointly by general cuts and a subset of all possible EIs, we do not attempt to add every available EI. Instead, we add only a limited number per callback (e.g., up to $n$), and only for players whose BRs have not yet been used to generate cuts. 
\end{remark}
    
\begin{algorithm}[h!]
\begin{algorithmic}[1]
\vskip4pt
\small
\Require IPG instance, parameters max rounds $R$, max restarts $L$, tolerance \texttt{tol} 
\Ensure Best pure Nash Equilibrium $\hat{\x}_{\text{pne}}^*$ or \texttt{None}
\State Initialize: $LB \gets -\infty$, \, $UB \gets \max \{ \phi(\x, \z) \mid (\x,\z) \in \K_{\text{relax}} \}$, \, $\Omega \gets \emptyset$, $\mathcal{H} \gets \emptyset$ ($\mathcal{H}$ stores best responses) \label{alg:BZR_1}

\While{$UB - LB > \texttt{tol}$ and time limit not exceeded}
    \State $\bar{\x} \gets$ incumbent integer solution from \texttt{MIPSOL} of $\max \{ \phi(\x,\z) \mid \x \in \K \cap \Omega \}$ \Comment{Callback Function}\label{alg:callback}

    \For{$i \in N$} \label{alg:oracle_1}
        \State Compute a best response: $\hat{\x}^i \in \arg\max_{\x^i \in \mathcal X_i} \left\{ u_i(\x^i, \bar{\x}^{\noti}) \right\}$  \label{alg:oracle_2}
        \If{$u_i(\hat{\x}^i, \bar{\x}^{\noti}) > u_i(\bar{\x}^i, \bar{\x}^{\noti})$} \label{alg:oracle_3}
            \State Add $u_i(\hat{\x}^i, \x^{\noti}) \leq u_i(\x^i, \x^{\noti})$ to $\Omega$ and add $\hat \x^i$ to $\mathcal{H}$  \Comment{Add violated EIs} \label{alg:oracle_4}
        \EndIf \label{alg:oracle_5}
    \EndFor \label{alg:oracle_6}

    \If{any violation is found ($\bar \x$ is not a PNE)}  \label{alg:BZR_2}
    \State $\bar{\x} \gets$ \texttt{RRR-BRD}($\bar{\x}, R, L$) (Algorithm \ref{alg:RRR_BRD_IPG}) and store the best responses found in $\mathcal{H}$ \label{alg:BZR_3} \Comment{Apply RRR-BRD}
    \For{(a limited number of) best responses $\hat\x^i \in \mathcal{H}$, $i \in N$} \label{alg:additional_1}
    \If{the EI for $\hat{\x}^i$ has not yet been added}\label{alg:additional_2}
    \State Add $u_i(\hat{\x}^i, \x^{\noti}) \leq u_i(\x^i, \x^{\noti})$ to $\Omega$   \Comment{Add additional EIs} \label{alg:additional_3}
    \EndIf
    \EndFor
    % \EndIf
    \EndIf
    \If{ $\bar{\x} \text{ is a PNE and } \phi(\bar{\x}) \geq LB$} $\hat{\x}_{\text{pne}}^* \gets \bar{\x}$ and $LB \gets \phi(\bar{\x})$ \label{alg:BZR_5}
    \EndIf
    \State $UB \gets \max \{ \phi(\x, \z) \mid (\x,\z) \in \K_{\text{relax}} \cap \Omega \}$ \label{alg:BZR_7}
\EndWhile

\If{$\hat{\x}_{\text{pne}}^* \neq \texttt{None}$} \Return $\hat{\x}_{\text{pne}}^*$ \textbf{else} \Return \texttt{None} \Comment{Return best PNE found or No PNE found} \EndIf\label{alg:BZR_8}

\caption{Best-Response Dynamics Incorporated Zero-regret (BZR) Algorithm}
\label{alg:BZR}
\end{algorithmic}
\end{algorithm}

\begin{figure}[h!]
\noindent
\begin{minipage}[t]{0.60\textwidth}
\vspace{0pt}
\centering
\resizebox{\textwidth}{!}{
  \begin{tikzpicture}[auto]
    % Axes
    \draw[->] (0,0) -- (8,0) node[right] {\( u_1(\mathbf{x}^1, \mathbf{x}^2) \)};
    \draw[->] (0,0) -- (0,8) node[above] {\( u_2(\mathbf{x}^1, \mathbf{x}^2) \)};

    % Define shifted PNE coordinates
    \coordinate (A) at (2.5,2.3);
    \coordinate (B) at (5.0,2.2);
    \coordinate (C) at (5.5,4.7);
    \coordinate (E) at (2.1,4.5);
    % \coordinate (F) at (3.9,3.9);
    \coordinate (G) at (3.7,5.7);

    % Define shifted non-PNE coordinates
    \coordinate (P) at (3.6,6.9);
    \coordinate (L) at (1.3,5.6);
    \coordinate (H) at (0.7,3.2);
    \coordinate (I) at (2.5,1.0);
    \coordinate (O) at (6.7,1.6);
    \coordinate (D) at (6.6,3.6);
    \coordinate (Q) at (6.4,6.0);
    \coordinate (R) at (6.0,5.6);
    
    % Define shifted non-PNE coordinates (inside)
    \coordinate (J) at (2.9,5.8);
    \coordinate (K) at (6.0,2.2);
    \coordinate (M) at (6.0,3.5);
    \coordinate (N) at (4.6,5.75);
    \coordinate (S) at (2.2,5.2);
    \coordinate (T) at (3.8,1.8);

    % Define non-PNE coordinates in the BR path (midpoints of blue arrows)
    \coordinate (AA) at (5.8,1.9);   % midpoint of O--B
    \coordinate (BB) at (3.65,6.3);  % midpoint of P--G
    \coordinate (CC) at (5.75,5.15); % midpoint of R--C

    % Shading for convex hull of feasible integer points (yellow)
    \fill[yellow!30, opacity=0.5, draw=black] (P) -- (L) -- (H) -- (I) -- (O) -- (D) -- (Q) -- cycle;
    
    % Fill the convex hull of PNEs (blue)
    \fill[blue!30, opacity=0.6, draw=black] (E) -- (A) -- (B) -- (C) -- (G) -- cycle;

    % Plot PNE points automatically without labels
    \foreach \point in {A,B,C,E,G} {
        \fill[black] (\point) circle (2pt);
    }

    % Plot non-PNE points automatically without labels
    \foreach \point in {D,H,I,J,K,L,M,N,O,P,Q,R,S,T,AA,BB,CC} {
        \draw[black] (\point) circle (2pt);
    }

    % Sparse contour lines for x + y = c
    \foreach \c in {2,4,6,8} {
        \draw[gray, dashed, opacity=0.5] (0,\c) -- (\c,0);
    }
    \draw[gray, dashed, opacity=0.5] (2,8) -- (8,2);
    \draw[gray, dashed, opacity=0.5] (4,8) -- (8,4);
    \draw[gray, dashed, opacity=0.5] (6,8) -- (8,6);

    % Blue arrowed line from node O to node B
    \draw[blue, thick, ->, line width=1.0pt, >=latex] (O) -- (B);

    \coordinate (B) at (5.0,2.2);
    
    %  Equilibrium Inequalities cutting off O
    \draw[green] (6.3,0.8) -- (7.0,3.2);

    %  Additional EIs in the O--B region (cutting off AA)
    \draw[red] (5.2,0.9) -- (6.3,2.7);
    \draw[red] (5.4,0.7) -- (6.2,2.9);
    \draw[red] (3.5,1.5) -- (6.5,2.9);
    \draw[red] (3.5,1.1) -- (6.7,3.4);

    % Blue arrowed line from node P to node G
    \draw[blue, thick, ->, line width=1.0pt, >=latex] (P) -- (G);
    
    \coordinate (G) at (3.7,5.7);
    %  Equilibrium Inequalities cutting off P
    \draw[green] (1.5,6.5) -- (5.5,7.0);
    \draw[green] (1.4,6.8) -- (5.5,6.8);

    %  Additional EIs in the P--G region (cutting off BB)
    \draw[red] (2.5,6.0) -- (4.8,6.5);
    \draw[red] (2.5,6.3) -- (4.8,6.3);
    \draw[red] (2.3,5.3) -- (5.0,6.1);
    \draw[red] (2.3,5.6) -- (5.2,5.8);

    % Blue arrowed line from node R to node C
    \draw[blue, thick, ->, line width=1.0pt, >=latex] (R) -- (C);

    %  Equilibrium Inequalities cutting off R
    \draw[green] (4.8,6.2) -- (7.0,4.8);
    \draw[green] (4.8,6.5) -- (7.0,4.4);

    %  Additional EIs in the R--C region (cutting off CC)
    \draw[red] (4.6,5.6) -- (6.8,4.8);
    \draw[red] (4.6,6.1) -- (6.8,4.2);
    \draw[red] (4.3,6.1) -- (6.4,3.6);
    \draw[red] (4.6,6.4) -- (6.4,3.0);
    
    % Arrow in (1,1) direction moved further up and left
    \draw[gray, ->, dashed, thick] (0.5,6.5) -- (1.5,7.5) node[midway, below right] {$\textrm{max}( \phi({\bf x}))$};

    % Legend
    \begin{scope}[shift={(8.0,5.5)}]
        % Non-PNE
        \draw[black] (0,0) circle (2pt) node[right=5pt] {Non-PNEs};
        % PNE
        \fill[black] (0,-0.6) circle (2pt) node[right=5pt] {PNEs};
        % Yellow region (feasible integer points)
        \fill[yellow!30, opacity=0.5, draw=black] (-0.2,-1.1) rectangle (0.2,-1.5) ;
        \node[right=5pt] at (0,-1.3) {Conv($\mathcal{X}$)};
        % Blue region (convex hull of PNE)
        \fill[blue!30, opacity=0.6, draw=black] (-0.2,-2.0) rectangle (0.2,-2.4);
        \node[right=5pt] at (0,-2.2) {Conv($\mathcal{X}\cap \mathcal{S}_{pne}$)};
        % Equilibrium Inequalities
        \draw[green] (-0.2,-3.0) -- (0.2,-3.0);
        \node[right=5pt] at (0,-3.0) {Equilibrium Inequalities};
        % Additional EIs
        \draw[red] (-0.2,-3.7) -- (0.2,-3.7);
        \node[right=5pt] at (0,-3.7) {Additional EIs};
        % BR path
        \draw[blue, thick] (-0.2,-4.4) -- (0.2,-4.4);
        \node[right=5pt] at (0,-4.4) {RRR-BRD trajectory};
    \end{scope}
\end{tikzpicture}
}
\caption{Illustration of the BZR Algorithm in a Two-Player Game.}
\label{fig:BZR_illustration}
\end{minipage}
\hfill
\begin{minipage}[t]{0.40\textwidth}
\vspace{0pt}
\normalfont\normalsize
\begin{remark}
\vspace{0.3cm}
Figure~\ref{fig:BZR_illustration} depicts many EIs, with some appearing ``deeper'' (e.g., those that are binding at PNEs). However, we do not establish any general dominance relation among EIs. For general IPGs, we are not aware of a principled way to identify a strictly stronger subclass of EIs. A baseline observation is that including EIs associated with \textit{all} BRs yields the perfect-equilibrium formulation $\mathcal{E}$. Accordingly, BZR focuses on generating a practically useful subset of EIs, rather than attempting a full characterization.
\end{remark}
\end{minipage}
\end{figure}

\noindent An illustrative two-player example that works explicitly through the BZR computations is provided in the Appendix~\ref{appendix:BZR_example}. To complement this small-scale walkthrough, we also include an exponential-separation example between ZR and BZR in Appendix~\ref{appendix:exponential-gap}.

\subsection{RRR-BRD as a Randomized Local-Search Subroutine for Other Algorithms} 
\label{sec:RR_BRD_others}

\citet{carvalho2023integer} organize IPG algorithms around three building blocks: an \textit{approximate} phase, which constructs an inner or outer approximation of each player's strategy set, yielding sets $\tilde{\mathcal{X}}_i$; a \textit{play} phase, which computes a tentative equilibrium \textit{with respect to} this approximated game; and an \textit{improve} phase, which verifies the tentative solution against the original game and refines the approximation as needed. In existing methods, the play phase is typically implemented via one or more \textit{joint} optimization problems that couple all (or a subset of) players' decisions, either explicitly in normal form or implicitly through extended MIP and complementarity-based formulations. RRR-BRD provides a complementary play-phase mechanism: rather than solving a global joint problem over a potentially very large feasible region, it performs a randomized walk on the BR state graph, repeatedly solving BR problems without ever constructing the full normal form.

Table~\ref{tab:algo_overview} summarizes the main exact algorithms for computing Nash equilibria in IPGs, adapted from \citet{carvalho2023integer} and extended with RRR-BRD. The first six methods are based on \textit{deterministic} existence and correctness proofs for their targeted equilibrium notions (MNE or PNE) in the classes considered. By contrast, RRR-BRD provides a \textit{probabilistic} (Las Vegas) guarantee. This trades deterministic worst-case guarantees for a random-walk--based mechanism that scales well in practice.

\vspaceminus{6pt}
\begin{table}[h!]
\caption{Summary of Algorithms for Computing Nash equilibria in IPGs.}
\small
\centering
\scalebox{0.9}{
\begin{tabular}{llllllll}
\toprule
\textbf{Method}   & \textbf{NE Type} & \textbf{Approx} & \textbf{Play}  
& \textbf{Alg Feas} 
& \textbf{Largest IPG Instances Tested}  & \textbf{Reference} \\ 
\midrule
SGM              & Mixed      & Inner      & FP   
& Deter.  
& KPG: \(n \leq 3\), \(m \leq 100\) & \citep{carvalho2022computing}   \\ \hline
eSGM             & Mixed      & Inner      & MIP  
& Deter.  
& KPG: \(n \leq 3\), \(m \leq 60\)  & \citep{cronert2024equilibrium}  \\ \hline
CnP              & Mixed      & Outer      & CP   
& Deter.  
& KPG: \(n \leq 3\), \(m \leq 100\) & \citep{carvalho2021cut}  \\ \hline
BM               & Pure       & Outer      & CP   
& Deter.  
& qIPG: \(n \leq 6\), \(m \leq 3\)  & \citep{sagratella2016computing}   \\ \hline
ZR               & Pure       & Outer      & MIP  
& Deter.  
& KPG: \(n \leq 3\), \(m \leq 100\)  & \citep{dragotto2023zero} \\ \hline
BnP              & Pure       & Outer      & MIP  
& Deter.  
& qIPG: \(n \leq 3\), \(m \leq 5\)  & \citep{schwarze2023branch}  \\ \hline
\textbf{RR-BRD}  & \textbf{Pure} & \textbf{Inner} & \textbf{BRD} 
& \textbf{Prob.} 
& \textbf{KPG:} $\boldsymbol{n \leq 30, m \leq 100}$, \textbf{EBMC:} $\boldsymbol{n \leq 84}$ & \textbf{This paper} \\
\bottomrule
\end{tabular}
}

\vspace{3pt}
\footnotesize{\textit{Note.} 
This table is adapted from Table~1 in \cite{carvalho2023integer} and extended to include RR-BRD. \\
`Approx' = approximate phase, `Play' = play phase. 
`Alg Feas' indicates whether the algorithm has a \textit{deterministic} (Deter.) or \textit{probabilistic} (Prob.) theoretical guarantee to find its targeted equilibrium whenever one exists in the considered class of IPGs.\\[-3pt]
FP = feasibility problem, MIP = mixed-integer problem, CP = complementarity problem,  KPG = knapsack problem games,\\[-3pt]
qIPG = quadratic IPG. \(n\): number of players; \(m = m_i, i \in N\): number of integer variables per player.}
\label{tab:algo_overview}
\end{table}
\vspaceminus{6pt}

Conceptually, RRR--BRD can be plugged into any IPG algorithm that occasionally produces an integer-feasible joint strategy profile $\x \in \mathcal{X}$. Once such a profile is available, RRR--BRD can be launched from $\x$ (on the induced finite game) to search locally for a PNE. If a PNE is found, it can be recorded as a feasible solution without altering the logical structure or exactness guarantees of the base algorithm; if not, the base algorithm simply continues unaffected. In this sense, RRR--BRD acts as a modular, PNE-oriented local-search layer that can be invoked whenever integer-feasible joint strategies become available.

\paragraph*{Branching-type methods (BM and BnP).}
In the \textit{branching method} (BM) of \citet{sagratella2016computing} and the \textit{branch-and-prune} (BnP) algorithm of \citet{schwarze2023branch}, the search for equilibria is organized as a tree whose nodes correspond to restricted subgames defined by additional bounds or fixings on players' variables. At each node, a continuous auxiliary problem (e.g., a relaxed NEP or complementarity system) is solved, pruning tests may discard parts of the search region, and branching refines players' feasible sets. Integer-feasible joint profiles arise, for example, from rounding, local search, or exhaustive enumeration within a node's box. Whenever such a profile is available, RRR--BRD can be run on the corresponding restricted subgame, using that profile as a starting point to search for a PNE within the node.

\paragraph*{Sampling-based and outer-approximation methods (SGM, eSGM, CnP).}
In the sample generation method (SGM) \citep{carvalho2022computing}, its enumeration variant eSGM \citep{cronert2024equilibrium}, and the Cut-and-Play (CnP) algorithm \citep{carvalho2021cut}, an IPG is approximated by a finite normal-form game built from sampled or externally generated pure strategies. At intermediate iterations, these methods maintain (i) finite strategy subsets $\tilde S_i \subseteq S_i$ for each player and (ii) one or more integer-feasible joint profiles in $\tilde S := \prod_i \tilde S_i$ (e.g., candidate equilibria or best-response combinations in the sampled game). RRR--BRD can then be applied to the current finite game restricted to $\tilde S$, using any such profile as a starting point. If RRR--BRD reaches a PNE of the sampled game, then (by construction) this profile corresponds to a degenerate mixed Nash equilibrium of the original IPG and can be stored as a candidate solution or benchmark.

\subsection{Approximate PNE Certificates upon Early Termination}
\label{ss:approx_pne}
RRR--BRD can also return an \textit{approximate-equilibrium certificate} when it does not reach an exact PNE within the allotted budget. For any profile $\bar{\x}$, let $\hat{\x}^i\in\arg\max_{\x^i\in\mathcal{X}_i} u_i(\x^i,\bar{\x}^{-i})$ be a best response for player $i$, and assume $u_i(\x^i,\x^{-i})>0$ for all feasible profiles. Define
\begin{equation}
\alpha_i(\bar{\x})
~:=~
\frac{u_i(\hat{\x}^i,\bar{\x}^{-i})}{u_i(\bar{\x}^i,\bar{\x}^{-i})},
\qquad
\alpha(\bar{\x}) ~:=~ \max_{i\in N} \alpha_i(\bar{\x}).
\label{eq:alpha_def}
\end{equation}
Then $\bar{\x}$ is an $\alpha(\bar{\x})$-approximate PNE in the multiplicative sense, and $\alpha(\bar{\x})=1$ if and only if $\bar{\x}$ is an exact PNE. Implementation and reporting details are provided in Appendix~\ref{appendix:approximate_PNE}.

\section{Edge-weighted Budgeted Maximum Coverage (EBMC) Games} \label{sec:game-theoretic}
\vspaceminus{3pt}
This section presents the edge-weighted budgeted maximum coverage (EBMC) game as a motivating example of large-scale IPGs. Section~\ref{sec:EBMCP} introduces the EBMC problem, and Section~\ref{sec:EBMC_games} develops two EBMC games with different utility functions. Section~\ref{sec:EBMC_types} characterizes the role of types and network structure, and Section~\ref{sec:EBMC_AIS} connects EBMC to AIS-prevention resource allocation.
\vspaceminus{3pt}
\subsection{The Edge-weighted Budgeted Maximum Coverage (EBMC) Problem}
\label{sec:EBMCP}
\vspaceminus{3pt}

The budgeted maximum coverage (BMC) problem \citep{chekuri2004maximum} aims to maximize element coverage by selecting subsets within a limited budget. The EBMC problem is a specialized case of BMC, where edges represent elements, and vertices represent subsets. Both problems are NP-hard \citep{chekuri2004maximum}, and \cite{caskurlu2014partial} further demonstrate that EBMC remains NP-hard even when restricted to bipartite graphs.

We model the underlying network as a directed graph $(\I, \A)$, where $\I$ is the set of vertices and $\A$ is the set of arcs. The players are indexed by $N := \{1,\dots,n\}$. Each player $i \in N$ controls a subset of vertices $\I_i \subseteq \I$, and these subsets form a partition of the vertex set, i.e., $\bigsqcup_{i \in N} \I_i = \I$. We write $\A_i \subseteq \A$ for the set of arcs whose endpoints both lie in $\I_i$, and use
$\delta_+(\I_i)$ (resp.\ $\delta_-(\I_i)$) for the set of arcs with tail (resp.\ head) in $\I_i$ and head (resp.\ tail) in $\I\setminus \I_i$. Each arc $(j,k) \in \A$ has a nonnegative weight $w_{jk}$, and each player $i \in N$ has a budget $\B_i$. The decision variables are
$x_j \in \{0,1\}$, indicating whether vertex $j \in \I$ is selected, and
$y_{jk} \in \{0,1\}$, indicating whether arc $(j,k) \in \A$ is \textit{covered} (i.e., at least one of its endpoints is selected).

The Social Welfare (SW) model (\eqref{eq:SW_1} --\eqref{eq:SW_3}) represents the centralized planning perspective. The objective function \eqref{eq:SW_1} maximizes total edge coverage over the arc set $\A$. Constraint \eqref{eq:SW_2} enforces that an edge $(j,k)$ can be covered only if at least one of its adjacent vertices, $j$ or $k$, is selected. Constraint \eqref{eq:SW_3} imposes the individual budget limit for each player $i \in N$.

\noindent \textbf{The Social Welfare (SW) Model }
\begin{subequations}
\label{modelSW}
\begin{align}
\max_{\substack{
x_j \in \{0,1\} \;\; \forall j \in \mathcal{I} \\
y_{jk} \in \{0,1\} \; \; \forall (j,k) \in \mathcal{A}
}}\quad & \sum_{(j,k) \in \mathcal{A}} w_{jk} y_{jk} \label{eq:SW_1} \\
\text{s.t.} \quad
& y_{jk} \leq x_j + x_k &   \forall (j,k)  \in \mathcal{A} \label{eq:SW_2} \\
& \sum_{j \in \mathcal{I}_i} x_j \leq \mathcal{B}_i & \forall i \in N. \label{eq:SW_3}
\end{align} \label{eq:SW}
\end{subequations}
The optimal solution to the SW model is the OSW solution ($\barxosw$), as introduced in Section \ref{sec:preliminaries}.

\subsection{A Non-cooperative EBMC Game Among $N$ Players}
\label{sec:EBMC_games}
\vspaceminus{3pt}
Although the SW model in \eqref{eq:SW} allocates budgets to each player, it primarily represents the objectives of the central planner, not the players. Consequently, the optimal solutions of this model may not align with the individual goals of the players. We start by defining a player-specific optimization problem that is contingent on the decisions of other players. Subsequently, we introduce two types of utility functions to model player interactions: a locally altruistic function and a selfish function, based on the different sets of arcs.

For each player $i \in N$, let $\I_{\noti}$ denote the set of vertices not controlled by player $i$, and let $\bar{\x}^{\noti} \in \{0,1\}^{\I_{\noti}}$ be the corresponding decisions of all other players. The optimization model for a single player is given by:

\noindent\textbf{The Single-Player (SP) Model}
\begin{subequations}
\begin{align}
\max \quad & \Tilde{u}_i(\y) \label{eq:CO0} \\
\text{s.t.}\ \
&  y_{jk} \leq x_j + x_k & \forall \, (j,k) \in \A \label{eq:CO1} \\
&  \sum_{j \in \I_i} x_j \leq \B_i \label{eq:CO2} \\
& x_j = \bar x^{\noti}_j & \forall \, j \in \I_{\noti} \label{eq:CO3} \\
& x_j \in \{0,1\} & \forall \, j \in \I_{i} \label{eq:CO4} \\
& y_{jk} \in \{0,1\} & \forall \, (j,k) \in \A. \label{eq:CO5}
\end{align}
\label{eq:CO}
\end{subequations}
The SP model \eqref{eq:CO}, defined by \eqref{eq:CO0}--\eqref{eq:CO5}, describes player $i$'s problem parameterized by $\bar{\x}^{\noti}$. When an edge connects a vertex $j \in \I_i$ of player $i$ to a vertex $k \in \I_{\noti}$ of some other player, a positive decision $x_k = 1$ allows edge $(j,k)$ to be covered even if $x_j = 0$. Thus, the actions of other players can have a positive external effect on player $i$'s objective. The objective function $\Tilde{u}_i(\y)$ in \eqref{eq:CO0} is left generic so that it can be tailored to the specific interests of player $i$. 
Since each \(y_{jk}\) can be written as \(y_{jk} = \max(x_j, x_k) = x_j + x_k - x_j x_k\), the objective can equivalently be expressed as a function of $\x$, denoted by $u_i(\x)$. This equivalence enables interchangeable use of \(\Tilde{u}_i(\y)\) and \(u_i(\x)\) under the condition that \(y_{jk} = \max(x_j, x_k)\). While using \(\y\)-variables emphasizes edge-level outcomes influenced by other players, expressing the utility directly in terms of \(\x\)-variables reveals player-level interactions via bilinear terms, which is standard in many IPG formulations.

Including all edges in Constraints \eqref{eq:CO1} and \eqref{eq:CO5} is computationally inefficient, especially for large graphs. Instead, arc sets can be restricted to relevant edges using the refined definition introduced below.

\begin{definition}[Induced Arc Sets]
 \label{def:Induced Subgraph}
Let   $\mathcal{D} = (\I, \mathcal{A})$ be a directed graph and let the subset of vertices $\I_i, i \in N$ be a partition of $\I$, i.e., $\I:=\bigsqcup_{i \in N} \I_i$.
The \textit{induced neighborhood arc set} $\A[\I_i]$ consists of all arcs that have at least one endpoint in $\I_i$. The \textit{induced inbound arc set} $\A^-[\I_i]$ consists of all arcs that have their terminal points in $\I_i$. Formally,
$\A[\I_i] := \{(j, k) \in \A_i \cup \delta_-({\I}_i) \cup \delta_+({\I}_i) \}$ and  $\A^-[\I_i] := \{(j, k) \in \A_i \cup \delta_-({\I}_i) \}$.
\end{definition}

Regardless of the choice of induced arc sets in Definition~\ref{def:Induced Subgraph}, the joint constraints of the SP model~\eqref{eq:CO}--interpreted by treating the parameterized variables $\bar{\x}^{\noti}$ as decision variables in Constraint~\eqref{eq:CO3}--correspond to those of the SW model~\eqref{eq:SW_2}--\eqref{eq:SW_3}. 
Based on these arc-set definitions, we consider two variants of the EBMC game, using different utility functions : (i) \textit{locally altruistic} utilities, where player $i$ maximizes edge coverage over $\A[\I_i]$, i.e., all edges incident to $\I_i$ (including outgoing arcs); and (ii) \textit{selfish} utilities, where player $i$ maximizes edge coverage over $\A^-[\I_i]$, i.e., only edges whose head lies in $\I_i$ (incoming arcs). Interpreting an edge $(j,k)$ as a risky or critical edge (e.g., for environmental risk, security, or regulatory responsibility), Case (i) models a player who shares responsibility for any risk connected to their nodes, whereas Case (ii) models a player who is concerned only with risks incident to its own vertices $\I_i$.

\vspace{3pt}
\noindent\paragraph*{\textbf{Locally Altruistic Game}}
In this game, the utility function of the player $i$ is the coverage of the edges using the induced neighborhood arcs.
Each player $i$ solves the following optimization problem:

\noindent \textbf{Locally Altruistic Single-Player (LASP) Model}
\begin{subequations}
\begin{align}
\max \quad & \tilde{u}_i^{\text{Alt}}(\y) := \ \  \smashoperator[lr]{\sum_{(j, k) \in \A[\I_i]}} \ \ w_{jk} y_{jk}  \label{eq:LASP_0} \\
\text{s.t.}\ \
&  y_{jk} \leq x_j + x_k \ \ & \forall (j,k) \in \A[\I_i] \label{eq:LASP_1} \\
&  \eqref{eq:CO2}-\eqref{eq:CO4} \label{eq:LASP_2} \\
& y_{jk} \in \{0,1\} \ \  & \forall (j,k) \in \A[\I_i]. \label{eq:LASP_3}
\end{align}
\label{eq:LASP}
\end{subequations}
Note that the arc sets $\A$ in Constraints \eqref{eq:CO1} and \eqref{eq:CO5} are replaced by $\A[\I_i]$. The locally altruistic game, denoted by $\mathcal{G}^{\text{Alt}}$, is an EBMC game with $N$ players, where each player solves a LASP model \eqref{eq:LASP}, defined by \eqref{eq:LASP_0}--\eqref{eq:LASP_3}. 
The locally altruistic game is not fully altruistic: player $i$'s utility depends only on arcs incident to $\I_i$ not on the entire network.

\vspace{0.2cm}
\noindent \paragraph*{\textbf{Selfish Game}}
In this game, the utility function of player $i$ is the edge coverage of the induced inbound arc set, which contains incoming arcs controlled by player $i$. Then, each player $i$ solves the following problem:

\noindent \textbf{Selfish Single-Player (SSP) Model}
\begin{subequations}
\begin{align}
\max \quad & \tilde{u}_i^{\text{Self}}(\y) := \ \ 
\smashoperator[lr]{\sum_{(j,k) \in \A^-[\I_i]}} \ \ w_{jk} y_{jk} \label{eq:SSP_0} \\
\text{s.t.}\ \
&  y_{jk} \leq x_j + x_k & \forall \, (j,k) \in \A^-[\I_i] \label{eq:SSP_1} \\
&  \eqref{eq:CO2}-\eqref{eq:CO4} \label{eq:SSP_2} \\
& y_{jk} \in \{0,1\} & \forall \, (j,k) \in \A^-[\I_i]. \label{eq:SSP_3}
\end{align}
\label{eq:SSP}
\end{subequations}
Note that the arc sets $\A$ in Constraints \eqref{eq:CO1} and \eqref{eq:CO5} are replaced by $\A^-[\I_i]$. The selfish game, denoted by $\mathcal{G}^{\text{Self}}$, is an EBMC game with $N$ players, where each player solves their SSP model \eqref{eq:SSP}, defined by \eqref{eq:SSP_0}--\eqref{eq:SSP_3}. We provide an illustrative two-player example of selfish EBMC interactions and the resulting BRD trajectory in the Appendix \ref{appendix:two-player}.

\paragraph*{Selfish Non-Game}
In this scenario, each player $i$ solves the SSP model \eqref{eq:SSP} independently, without engaging in any strategic interaction--that is, each player assumes that no other players will select any vertices. Consequently, player $i$ determines its strategy by solving
\(
\bar\x^i_{\text{ng}} \in \argmax_{\x^i \in \{0,1\}^{\I_i}} u_i(\x^i,{\bf 0}^{\noti}),
\)
where ${\bf 0}^{\noti}$ denotes the zero vector for all players other than $i$. The combination of these individually optimal strategies forms the \textit{selfish non-game} strategy profile, denoted by $\bar\x_{\text{ng}}$.

\paragraph*{Equilibrium Inequality for Selfish EBMC Games}
We will show in Section~\ref{sec:alt_proof} that the OSW solution $\barxosw$ is a PNE for locally altruistic games. Thus, $\barxosw$ will be treated as the best PNE in this setting, and no further algorithmic comparisons are needed. In contrast, for selfish games, we derive the following equilibrium inequality (EI). For any player $i \in N$ and any best response $\hat{\x}^i$ to a profile $\bar{\x}^{\noti}$, we have
$$
\smashoperator[lr]{\sum_{(j,k) \in \A_i}} w_{jk} \bigl(\hat{x}_j+\hat{x}_k -\hat{x}_j \hat{x}_k\bigr) + \smashoperator[lr]{\sum_{(j,k) \in \delta_{-}(\I_i)}}  w_{jk}\bigl(x_j+\hat{x}_k -x_j \hat{x}_k\bigr) \leq  \smashoperator[lr]{\sum_{(j,k) \in \A_i \cup \delta_{-}(\I_i)}} w_{jk}y_{jk}, \\
$$
for all $x_j, x_k, y_{jk} \in \{0,1\} $ satisfying Constraints \eqref{eq:SSP_1}--\eqref{eq:SSP_3}. 
Note that only the best-response variables $\hat{x}_j$ with $j\in \I_i$ can be fixed in the EI. Moreover, in EBMC the $\y$-variables serve as auxiliary variables in the lifted formulation, so when constructing EIs we can substitute $y_{jk}=x_j+x_k-x_jx_k$ and fix the appropriate $\x$-variables as implied by the relevant arc-set definitions.

\paragraph*{Explicit Reformulation to Standard IPGs}
The objective functions of the locally altruistic and selfish games can be reformulated into the standard IPG form by substituting $\y$-variables with $\x$-variables. These reformulations are detailed in Appendix  \ref{appendix:reform_standard_IPG}.

\subsection{Types and Network Structure in EBMC}
\label{sec:EBMC_types}
\vspaceminus{3pt}

We enrich the EBMC formulation with a finite set of \textit{types} $\mathcal{L}$, which may represent invasive species, commodity classes, or risk categories. For each vertex $j \in \I$ and type $l \in \mathcal{L}$, let $f_{j l} \in \{0,1\}$ indicate whether $j$ has type $l$, and let $n_{jk} \ge 0$ denote the interaction intensity from $j$ to $k$ (e.g., traffic or flow). An edge $(j,k)$ is type-$l$ risky if $f_{j l}=1$ and $f_{k l}=0$. We aggregate risk into the edge weight
\(
w_{jk} \;:=\; \sum_{l \in \mathcal{L}} f_{j l}\,\bigl(1 - f_{k l}\bigr)\, n_{jk},
\)
so only interactions that may transmit at least one type from $j$ to $k$ contribute to $w_{jk}$.

This structure induces a useful graph-theoretic dichotomy. When $|\mathcal{L}|=1$, vertices split into those with and without the unique type, and risky edges run only from the former to the latter, yielding a bipartite network. When $|\mathcal{L}|\ge 2$, vertices can differ across types, and both $(j,k)$ and $(k,j)$ may be risky (for different types), so the network is generally non-bipartite.

\subsection{A Resource Allocation Game among Minnesota Counties}
\label{sec:EBMC_AIS}

The EBMC games in Section~\ref{sec:EBMC_games} are motivated by aquatic invasive species (AIS) prevention in Minnesota (MN). Currently, 84 county planners manage AIS inspection programs independently, each with its own AIS prevention budget supported by direct state aid \citep{AIS_aid}. The Minnesota Department of Natural Resources (MN DNR) facilitates information sharing (e.g., workshops and tools such as AIS Explorer \citep{AIS_Explorer}) but does not enforce a formal inter-county coordination mechanism for placing inspection stations. Counties report inspection activity (e.g., hours) to MN DNR, which can be used to infer deployed stations, and the increasing use of AIS Explorer facilitates transparency about allocations. 

In this paper, we therefore model the \textit{current} practice as a non-cooperative game with complete information, where each county chooses how to allocate its own inspection stations and positive externalities across counties arise endogenously from these choices. The statewide goal of the MN DNR is captured by the social-welfare objective (total inspected risky boat movements). Designing regulatory interventions or incentive schemes that better align county-level decisions with statewide welfare is an important direction for future work but lies beyond the scope of this study. Instead, we focus on two utility specifications: a \textit{selfish} utility, where counties care only about risky boats entering lakes in their own $\I_i$, and a \textit{locally altruistic} utility, where they also value inspection of boats leaving their lakes (Section~\ref{sec:EBMC_games}). Our central questions are: \textit{given these utility specifications and network types, when do PNE solutions exist in the resulting EBMC games, how do the resulting PNE solutions compare--both theoretically and computationally--for both the selfish and locally altruistic formulations, and what social welfare levels do these PNE solutions achieve?}

The vertices $j \in \I$ represent lakes, partitioned by county so that $\I_i$ denotes the lakes controlled by county $i \in N$. For each ordered pair of lakes $(j,k)$, we let $n_{jk}$ denote the estimated number of boat trips from $j$ to $k$ over a given period, obtained from survey-based lake-to-lake movement data \citep{kao2021network}. Since Minnesota lakes host different AIS types, the edge weight $w_{jk}$ aggregates type-specific risk following the setup in Section~\ref{sec:EBMC_types}. Following the empirical AIS literature \citep{Haight2021,haight2023bi}, we focus on lake-to-lake risk driven by the \emph{most recent} trip on a given route; in particular, we do not model spread along multi-edge boat routes, both because current data and practice are organized at the origin--destination level and because infestation risk is driven primarily by short-term carryover from the immediately visited infested lake. Each county $i$ has a budget $\B_i$ representing the maximum number of inspection stations it can deploy, under the assumption that one station covers one lake.

\section{Pure Nash Equilibrium for EBMC Games} \label{sec:existence}
\vspaceminus{3pt}
As discussed in Section~\ref{sec:preliminaries}, the existence of a PNE is not guaranteed for general IPGs. In particular, when the interaction matrices $Q^i_p$ in \eqref{eq:IPG_obj} are dense, a player's utility is heavily influenced by externalities (i.e., the decisions of other players), making it unlikely that there are simple, general sufficient conditions for PNE existence. However, if an IPG is a potential game \citep{monderer1996potential}, then a PNE is guaranteed to exist. Building on this idea, we first show that locally altruistic EBMC games admit a PNE. We then establish sufficient conditions for the existence of a PNE in selfish EBMC games, with a particular focus on large-scale IPG settings where externalities tend to dominate internal utility contributions.

We define the objective function of the SW model \eqref{eq:SW_1} as the global function
\(
\tilde{\phi}(\y)
= \sum_{(j,k) \in \mathcal{A}} w_{jk} y_{jk}
= \sum_{(j,k) \in \mathcal{A}} w_{jk} \max (x_j,x_k)
= \sum_{(j,k) \in \mathcal{A}} w_{jk} (x_j+x_k-x_j x_k)
= \phi(\x).
\)
The next proposition shows that $\phi(\x)$ coincides with the sum of selfish utilities $u_i^{\text{Self}}(\x)$ as defined in \eqref{eq:SSP_0} where $\tilde u_i^{\text{Self}}(\y)$ is rewritten as $u_i^{\text{Self}}(\x)$.

\begin{proposition}[Social welfare as sum of selfish utilities]\label{lemma:selfish_sum}
Let $\mathcal{D} = (\I, \mathcal{A})$ be a directed graph and let the subsets of vertices $\I_i$, $i \in N$, form a partition of $\I$, i.e., $\I:=\bigsqcup_{i \in N} \I_i$. Then
\(
\mathcal A \;=\; \bigsqcup_{i \in N} \mathcal A^-[\mathcal{I}_i],
\)
and, for any feasible $(\x,\y)$ with $y_{jk} = \max(x_j,x_k)$, the social welfare objective satisfies
\(
\phi(\x)
\;=\;
\sum_{i \in N} u_i^{\text{Self}}(\x).
\)
\end{proposition}

The proof is given in Appendix~\ref{appendix:proofs}. Note that $\phi(\x)$ is not equal to the sum of players' utilities in the locally altruistic game. Thus, under the standard definition of social welfare (SW) as the sum of utilities, $\phi(\x)$ coincides with SW only for the selfish game. Nevertheless, since $\phi(\x)$ maximizes edge coverage over the entire network, we refer to it as the SW function throughout, including in the locally altruistic setting.

\vspaceminus{6pt}
\subsection{Existence of a PNE in the Locally Altruistic Games} \label{sec:alt_proof}
\vspaceminus{3pt}
We now show that locally altruistic EBMC games are exact potential games. A game is an \textit{exact potential game} if there exists a function $\psi: \X\rightarrow \mathbb{R}$ (an \textit{exact potential}) such that for all $i \in N$, any strategies $\bar\x^{\noti} \in \mathcal{X}_{\noti}$ and $\bar\x^{i}, \hat\x^{i} \in \mathcal{X}_{i}$ satisfy
\(
\psi(\hat\x^{i}, \bar\x^{\noti}) - \psi(\bar\x^{i}, \bar\x^{\noti})
=
u_i(\hat\x^{i}, \bar\x^{\noti}) - u_i(\bar\x^{i}, \bar\x^{\noti}).
\)
\citet{monderer1996potential} show that every potential game possesses at least one PNE: any strategy profile that maximizes the potential function. This existence result assumes that each player's strategy set $\mathcal{X}_i$ is nonempty and compact. In our setting, each $\mathcal{X}_i$ is a finite set of binary-feasible solutions (hence compact), and it is nonempty because the all-zero strategy is feasible. We now establish the following results; the proofs are given in Appendix~\ref{appendix:proofs}.

\begin{theorem} \label{thm:exact_potential_game}
The social welfare function $\phi$ is an exact potential function for locally altruistic games.
\end{theorem}
\begin{corollary} \label{cor:SW_is_PNE}
The strategy profile $\barxosw$ is a PNE for locally altruistic EBMC games.
\end{corollary}

This corollary  allows central planners to recommend an optimal strategy profile $\barxosw$ if all players are locally altruistic. Additionally, assuming that players have not decided on a utility function, central planners can encourage players to adopt a more altruistic approach by highlighting mutual benefits. Decision makers at any level, central planner or player level, can directly solve the SW model \eqref{eq:SW} and implement the solution.

\subsection{Existence of a PNE in the Selfish Games}
\label{sec:selfish_proof}
Next, we establish sufficient conditions for the existence of a PNE in selfish EBMC games, particularly in regimes where externalities from other players' decisions $\bar{\x}^{-i}$ dominate the utility generated by a player's own choices $\x^{i}$. We denote the selfish game with a single type by $\mathcal{G}^{\text{Self}}_{|\mathcal{L}|=1}$ and with multiple types by $\mathcal{G}^{\text{Self}}_{|\mathcal{L}|\ge 2}$. 

Let $\I = \I^{+} \cup \I^{0}$, where
\(
\I^{+} := \{ j\in\I : \exists l \in\mathcal{L}\ \text{s.t.}\ f_{jl}=1\}
\, \, \text{and} \, \, 
\I^{0} := \{ j\in\I : f_{jl}=0\ \text{for all}\ l\in\mathcal{L}\}.
\)
Thus, $\I^{+}$ is the set of vertices that have at least one type, while $\I^{0}$ is the set of vertices with no types. The following result formalizes these conditions; its proof is provided in Appendix~\ref{appendix:proofs}.
\begin{theorem}\label{thm:PNE-existence}
Let $\I^{\text{ng}}$ be the set of vertices selected in the non-game strategy profile $\bar\x_{\text{ng}}$. If $\I^{\text{ng}} \subseteq \I^{0}$, then $\bar\x_{\text{ng}}$ is a PNE for both $\mathcal{G}^{\text{Self}}_{|\mathcal{L}|=1}$ and $\mathcal{G}^{\text{Self}}_{|\mathcal{L}|\ge 2}$. Additionally, if $\I^{\text{ng}} \subseteq \I^{+}$, $\bar\x_{\text{ng}}$ is a PNE for $\mathcal{G}^{\text{Self}}_{|\mathcal{L}|=1}$. \Halmos
\end{theorem}

We now examine conditions aligned with Theorem~\ref{thm:PNE-existence}. For a vertex $j \in \mathcal{I}_i$, let $w(j)$ denote the total weight of inbound arcs incident to $j$ within the induced inbound arc set $\mathcal{A}^-[\mathcal{I}_i]$. In the random graphs we generated, the condition $\mathcal{I}_i^{\text{ng}} \subseteq \mathcal{I}_i^{0}$ for $\mathcal{G}^{\text{Self}}_{|\mathcal{L}|=1}$ frequently holds. This phenomenon arises from the structure of $\mathcal{A}^-[\mathcal{I}_i]$, where $w(k)$ for vertices $k \in \mathcal{I}_i^{0}$ often exceeds $w(j)$ for vertices $j \in \mathcal{I}_i^{+}$. As the number of players increases, the aggregate weight from incoming arcs typically exceeds that from internal arcs, leading to a non-game strategy profile $\bar{\x}_{\text{ng}}^i$ that favors vertices in $\mathcal{I}_i^{0}$.

Even when $\mathcal{I}_i^{\text{ng}} \subseteq \mathcal{I}_i^{0}$ is not strictly satisfied, we observe computationally (in randomly generated instances) that $\mathcal{G}^{\text{Self}}_{|\mathcal{L}|=1}$ almost always admits a PNE. Intuitively, large (though not strictly dominant) externalities can guide BRD to a PNE in only a few iterations. Typically, $\mathcal{I}_i^{\text{ng}}$ consists primarily of vertices in $\mathcal{I}_i^{0}$ due to the high volume of risky incoming arcs. If BRD starts from a conservative profile such as $\bar{\x}_{\text{ng}}$, players benefit iteratively: selecting vertices in $\mathcal{I}_i^{+}$ induces other players to select previously uncovered vertices in $\mathcal{I}_i^{0}$. Importantly, selecting vertices in $\mathcal{I}_i^{0}$ benefits only player $i$, whereas selecting vertices in $\mathcal{I}_i^{+}$ benefits others. Since $\mathcal{I}_i^{+}$ is finite, this improvement process eventually terminates, yielding a PNE. In Section~\ref{sec:6.2}, we confirm that all instances of the selfish game $\mathcal{G}^{\text{Self}}_{|\mathcal{L}|=1}$ admit a PNE.

However, this dynamic pattern typically holds under symmetric or balanced edge benefits. We obtain the counterexamples in Appendix~\ref{appendix:counterexamples} either via carefully calibrated edge weights and graph structures (for $|\mathcal{L}|=1$) or via multiple AIS types with conflicting priorities (for $|\mathcal{L}| \ge 2$).

\noindent\paragraph*{Summary and pointers.}
Section~\ref{sec:existence} establishes PNE existence for locally altruistic EBMC games via an exact potential and provides sufficient conditions (and counterexamples) for the selfish games.
For a brief discussion of the complexity of deciding PNE existence, see Appendix~\ref{appendix:reform_standard_IPG}.

\section{Computational Results} \label{sec:computational}

This section evaluates the effectiveness of RRR--BRD (Algorithm~\ref{alg:RRR_BRD_IPG}), BZR (Algorithm~\ref{alg:BZR}), and the ZR algorithm of \citet{dragotto2023zero} for computing PNEs and identifying high-welfare (best-found) PNEs in EBMC games and KPGs. To implement ZR, we use Algorithm~\ref{alg:BZR} with Lines~\ref{alg:BZR_2}--\ref{alg:additional_3} disabled. The main manuscript reports (i) summary results for RRR--BRD under multiple initial profiles and (ii) summary results and selected comparisons of BZR versus ZR for EBMC and KPG instances; complete results are provided in Appendix~\ref{appendix:CR-BRD}--\ref{appendix:full_BZR}. For instances where no PNE is found across all runs, we report approximate-PNE certificates in  Appendix~\ref{appendix:approximate_PNE}. Section~\ref{sec:6.1} describes the computational settings and performance metrics. Section~\ref{sec:6.2} reports results for EBMC games on randomly generated instances with varying numbers of players $(n=2,3,5,8,10,15,20,25,30)$. Section~\ref{sec:6.3} presents results for the Minnesota dataset with 84 county-level players and discusses managerial insights. Finally, Section~\ref{sec:7.4} reports results for KPGs using the instance-generation scheme of \citet{dragotto2023zero}, scaling from small $(n=2,3)$ to larger games $(n=5,8,10,15,20,25,30)$.

\subsection{Computational Setup and Metrics} \label{sec:6.1}
\vspaceminus{3pt}
All experiments were run on a machine with an Intel(R) Core(TM) i9-13900F CPU @ 3.10GHz (24 cores) and 64GB RAM. All algorithms (RRR--BRD, BZR, ZR) were implemented in Python and solved using Gurobi Optimizer (v11.0.3) with 16 threads. Equilibrium inequalities were enforced via Gurobi callbacks and lazy constraints. Code, instances, and results are available on GitHub.\footnote{\url{https://github.com/HyunwooLee0429/Best-response-dynamics-IPG}}

Model construction time for best-response problems is non-negligible, especially in large-scale IPGs where best responses must be solved repeatedly across players. To reduce overhead, we construct each player's best-response model once and then update only the components that change across solves (e.g., fixing $\x^{\noti}$), consistently across all algorithms. When integrating RRR--BRD trajectories into callbacks (e.g., within BZR), we also control the number of additional best-response inequalities generated from a trajectory; in most instances, we add at most $n$ such inequalities. For the Minnesota instance, we disable these additional cuts, since adding many inequalities can substantially slow the overall solving process.

Table~\ref{tab:abbreviations} summarizes the acronyms and shorthands used in the paper and in this section. While OSW and POS are defined in Section~\ref{sec:preliminaries}, we report time-limited approximations--denoted $\widetilde{\text{OSW}}$ and $\widetilde{\text{POS}}$--due to the scale of the instances. For both BZR and ZR, the time limit was 1800 seconds. The SW model~\eqref{eq:SW} was also solved with a 1800-second limit to compute $\phi(\barxswtime{1800})$. For RRR-BRD, we use two initial strategy profiles: $\barxzero$ and $\barxswtime{300}$.For standalone RRR-BRD, we set the maximum numbers of restarts and rounds to $L=10$ and $R=20$, respectively. When used as a subroutine within BZR, we use $L=3$ and $R=20$. Finally, all objective values ($\phi$) are reported in thousands and rounded to the nearest tenth in all tables.

\vspaceminus{15pt}
\begin{table}[htbp]
\centering
\caption{Acronyms and Shorthands}
\label{tab:abbreviations}
\small
\scalebox{0.9}{
\begin{tabular}{p{0.28\textwidth} | p{0.68\textwidth}}
\toprule
\textbf{Symbol / Abbreviation} & \textbf{Full Term / Meaning} \\
\midrule
$\barxzero$ & Initial strategy profile: all-zero vector \\
$\barxswt$   & Initial strategy profile obtained by solving the SW model with time limit (secs)  \\
$\bar{\x}_{\text{ng}}$ & Non-game strategy profile \\
$\hat\x_{\text{pne}}^1$ / $\hat\x_{\text{pne}}^*$ & First identified PNE / Best identified PNE (maximizing $\phi$) within time limit \\
$\phi(\cdot)$ & Social welfare function; values are reported in thousands and rounded to the nearest tenth. \\
$\widetilde{\text{OSW}}$ & Approximate optimal social welfare ($\phi(\barxswt)$)  \\
 $\widetilde{\text{POS}}$ & Approximated price of stability given by $\phi(\barxswt) / \phi(\hat\x_{\text{pne}}^*)$ \\
$\#_{\text{cuts(ZR)}}$ / $\#_{\text{cuts(BZR)}}$ & Number of equilibrium inequalities added by ZR / BZR \\
$\#_{\text{pne(ZR)}}$ / $\#_{\text{pne(BZR)}}$ & Number of distinct PNEs found by ZR / BZR \\
T(ZR$^{\text{1st}}$) / T(BZR$^{\text{1st}}$) & Time to find the first PNE using ZR / BZR \\
T(ZR) / T(BZR) & Total runtime of the ZR / BZR algorithm \\
\bottomrule
\end{tabular}
}
\end{table}

We report the number of players ($n$), the budget ratio (BG), and additional performance metrics.

\begin{itemize}
    \item \textbf{RRR-BRD metrics:} For each instance and each initial profile ($\barxzero$ and $\barxswtime{300}$), we report the best-found PNE (in terms of $\phi$), runtime, number of rounds, $\widetilde{\text{OSW}}$, and $\widetilde{\text{POS}}$ (computed using the higher-welfare PNE among the two runs). We also include reference values from BZR, such as the best PNE found and the best dual bound. The time to obtain $\barxswtime{300}$ is included in the reported runtime of RRR-BRD.
        
    \item \textbf{ZR and BZR metrics:} For each instance, we report the number of distinct PNEs found ($\#_{\text{pne(ZR)}}$, $\#_{\text{pne(BZR)}}$), the best-found PNE, the dual bound, time to the first PNE (e.g., T(BZR$^{\text{1st}}$)), total runtime (e.g., T(BZR)), $\widetilde{\text{OSW}}$, $\widetilde{\text{POS}}$ (based on the best-found PNE), and the number of equilibrium inequalities added (e.g., $\#_{\text{cuts(BZR)}}$). As noted in Remark~\ref{rmk:X_relax}, the gap between the best-found PNE and the dual bound can remain large at termination, reflecting the large-scale nature of IPG instances. The best-found PNE is certified as welfare-optimal when $\phi(\hat\x_{\text{pne}}^*)$ coincides with the dual bound (within a tolerance).
\end{itemize}

\subsection{EBMC Games with Random Dataset} \label{sec:6.2}
We generate random EBMC instances by varying the number of counties, budget levels, infestation patterns, and network density. The number of counties is chosen from
$\{2,3,5,8,10,15,20,25,30\}$, and each county controls $50$ lakes. For the single-AIS case ($|\mathcal{L}|=1$), each lake is independently infested with a probability drawn from $\{0.2,0.4,0.6,0.8,1.0\}$. For the multiple-AIS case ($|\mathcal{L}|\ge 2$), infestation is generated independently by AIS type, with type-specific probabilities drawn from nested candidate sets:
$\{0.2,0.4,0.6,0.8,1.0\}$ (type 1),
$\{0.2,0.4,0.6,0.8\}$ (type 2),
$\{0.2,0.4,0.6\}$ (type 3), and
$\{0.2,0.4\}$ (type 4);
this construction biases instances so that most lakes remain uninfested for at least one AIS type.
We consider three budget ratios, $\{0.3,0.5,0.8\}$, and set each county's budget as the number of infested lakes in that county multiplied by the chosen ratio. Arc weights are drawn i.i.d.\ uniformly from $[10,20]$. To avoid unrealistically dense (nearly complete) networks, we start from a complete directed graph and randomly delete arcs: $20\%$ when $|\mathcal{L}|=1$ and $50\%$ when $|\mathcal{L}|\ge 2$, with the higher deletion rate also helping control instance size. Overall, this procedure yields $27$ instances per case (single AIS and multiple AIS), corresponding to the $9$ county counts combined with the $3$ budget ratios.

Table~\ref{tab:EBMC_games_CR_BRD_sumary} reports standalone RRR--BRD performance over these 27 instances using two initial profiles, $\barxzero$ and $\barxswtime{300}$; complete results are provided in Appendix~\ref{appendix:CR-BRD}. Relative to the single-AIS game $\mathcal{G}^{\text{Self}}_{|\mathcal{L}|=1}$, the multiple-AIS game $\mathcal{G}^{\text{Self}}_{|\mathcal{L}|\ge 2}$ requires substantially more rounds on average, suggesting that strategic dynamics become markedly more complex when multiple AIS types are present.

\vspaceminus{12pt}
\begin{table}[htbp]
\caption{Summarized RRR-BRD Results with Different Initial Strategy Profiles for $\mathcal{G}^{\text{Self}}_{|\mathcal{L}|=1}$ and $\mathcal{G}^{\text{Self}}_{|\mathcal{L}|\ge 2}$.}
\centering
\small
\scalebox{0.8}{
\begin{tabular}{c c c c c c c c c c c c}
\toprule
type $\mathcal{L}$ & \multicolumn{2}{c}{Avg$^{\dagger}$ PNE ($\phi$)} & \multicolumn{2}{c}{Avg Price of Stability}  & \multicolumn{2}{c}{Avg References} & \multicolumn{2}{c}{Avg Time (secs)} & \multicolumn{2}{c}{Avg Round}   \\
\cmidrule(lr){2-3} \cmidrule(lr){4-5} \cmidrule(lr){6-7} \cmidrule(lr){8-9} \cmidrule(lr){10-11} 
& init:$\barxzero$ &   init:$\barxswtime{300}$ & $\widetilde{\text{OSW}}$ & $\widetilde{\text{POS}}$ & BZR($\phi$) & BZR(Bound) & init:$\barxzero$ &   init:$\barxswtime{300}$ & init:$\barxzero$ &   init:$\barxswtime{300}$  \\ \midrule
$|\mathcal{L}|=1$ & 786.8 & 1010.5 & 1055.5 & 1.051 & 1023.6 & 1089.5 & 7.5 & 212.1 & 2.0 & 1.6  \\
$|\mathcal{L}|\ge 2$  & 3658.0 & 3658.0 & 4010.0 & 1.097 & 3658.0 &  5409.3 & 25.6 & 289.5 & 4.0 & 4.7 \\ \bottomrule
\end{tabular}
}

\vspace{3pt}
\footnotesize{\textit{Note.} 
$\dagger$: Averages of $\phi$, OSW, and POS are calculated from instances where a PNE is found by both initial strategy profiles.
% Objective values ($\phi$) are reported in thousands and rounded to the nearest tenth. \\[-3pt]
}
\label{tab:EBMC_games_CR_BRD_sumary}
\vspaceminus{3pt}
\end{table}

Tables~\ref{tab:EBMC_comp_sumary} summarizes the performance of BZR and ZR on $\mathcal{G}^{\text{Self}}_{|\mathcal{L}|=1}$ and $\mathcal{G}^{\text{Self}}_{|\mathcal{L}|\ge 2}$. We report (out of 27 instances) how often a PNE was found, multiple PNEs were identified, the best PNE was obtained, and a tighter dual bound was achieved. ``Ties'' indicate the number of instances in which both algorithms either identify the same best PNE or attain the same dual bound. The full results for $\mathcal{G}^{\text{Self}}_{|\mathcal{L}|=1}$ appear in Table~\ref{tab:EBMCP_games_1}, while the full results for $\mathcal{G}^{\text{Self}}_{|\mathcal{L}|\ge 2}$ are reported in Appendix~\ref{appendix:full_BZR}. In both settings, BZR consistently outperforms ZR in (i) time to the first PNE, (ii) the number of distinct PNEs found, and (iii) the best PNE value identified.

\vspaceminus{12pt}
\begin{table}[htbp]
\caption{Summarized Comparison of BZR and ZR for $\mathcal{G}^{\text{Self}}_{|\mathcal{L}|=1}$ and $\mathcal{G}^{\text{Self}}_{|\mathcal{L}|\ge 2}$.}
\centering
\small
\scalebox{0.9}{
\begin{tabular}{c c c c c c c c c c c c c c c}
\toprule
type $\mathcal{L}$ & \multicolumn{2}{c}{\# PNEs Found}  & \multicolumn{2}{c}{Finding a PNE} & \multicolumn{3}{c}{Best PNE}  & \multicolumn{3}{c}{Dual Bounds} & Avg POS & \multicolumn{2}{c}{\# Avg EI Cuts}  \\
\cmidrule(lr){2-3} \cmidrule(lr){4-5} \cmidrule(lr){6-8} \cmidrule(lr){9-11} \cmidrule(lr){12-12} \cmidrule(lr){13-14}
 & $\#_\text{pne(ZR)}$ & $\#_\text{pne(BZR)}$ & ZR  & BZR &  Ties &  ZR & BZR & Ties &  ZR & BZR & $\widetilde{\text{POS}}$ & $\#_\text{cuts(ZR)}$ & $\#_\text{cuts(BZR)}$ \\ \midrule
$|\mathcal{L}|=1$ & 32 & 150 & 11 & 27 & 10 & 0 & 17 & 9 & 4 & 14 & 1.033 & 59.3 & 94.4  \\
$|\mathcal{L}|\ge 2$ & 3 & 29 & 3 & 23 & 3 & 0 & 20 & 5 & 18 & 4 & 1.097 & 66.0 & 116.3  \\
\bottomrule
\end{tabular}
}
\vspace{3pt}
\label{tab:EBMC_comp_sumary}
\end{table}

For the single-AIS case ($|\mathcal{L}|=1$), BZR typically finds a PNE within a few seconds when $n\le 10$, and within roughly 100 seconds for $15\le n\le 30$, indicating strong scalability. In contrast, ZR returns a PNE in only 11 of the 27 instances within the time limit; when both algorithms find a PNE, BZR more frequently identifies the best PNE. BZR also tends to provide tighter dual bounds in this setting.

For the multiple-AIS case ($|\mathcal{L}|\ge 2$), the performance gap widens: BZR finds a PNE in 23 of the 27 instances, whereas ZR succeeds in only 3 instances within the time limit. Although BZR continues to identify high-quality PNEs, its dual bounds are often not substantially tightened, even after adding additional equilibrium inequalities. This behavior is expected for three reasons: (i) the time spent in RRR--BRD can be significant in the hardest instances, (ii) dual progress is driven largely by the LP relaxation and general-purpose cutting planes, and (iii) the added equilibrium inequalities primarily enforce equilibrium feasibility rather than strongly strengthening the continuous relaxation. Since the primary goal of ZR/BZR is to compute one or more PNEs (and, when possible, a best PNE), we treat dual bounds as auxiliary metrics.

\vspaceminus{12pt}
\begin{table}[htbp]
\caption{Comparison of BZR and ZR for Selfish EBMC Games $\mathcal{G}^{\text{Self}}_{|\mathcal{L}|=1}$.}
\centering
\small
\scalebox{0.73}{
\begin{tabular}{rrrrrrrrrrrrrrrrrr}
\toprule
&& \multicolumn{2}{c}{\# PNEs Found} & \multicolumn{2}{c}{Best PNE ($\phi$)} & \multicolumn{2}{c}{Dual Bounds} & \multicolumn{2}{c}{Price of Stability} & \multicolumn{4}{c}{Time (secs)} & \multicolumn{2}{c}{\# EI Cuts}  \\
\cmidrule(lr){3-4} \cmidrule(lr){5-6} \cmidrule(lr){7-8} \cmidrule(lr){9-10} \cmidrule(lr){11-14} \cmidrule(lr){15-16}
$n$ & BG & $\#_\text{pne(ZR)}$ & $\#_\text{pne(BZR)}$ & ZR &  BZR  & ZR & BZR & $\widetilde{\text{OSW}}$ & $\widetilde{\text{POS}}$  & T(ZR$^{\text{1st}}$) & T(ZR) & T(BZR$^{\text{1st}}$) & T(BZR) &  $\#_\text{cuts(ZR)}$ & $\#_\text{cuts(BZR)}$  \\ \midrule
2  & 0.3 & 1          & 1           & 19.2  & 19.2            & 19.2            & 19.2            & 19.2   & 1.000 & 0.2    & 0.2    & 0.2  & 0.4    & 8   & 13  \\
   & 0.5 & 6          & \textbf{7}  & 22.3  & 22.3            & 22.3            & 22.3            & 22.3   & 1.000 & 0.9    & 1.7    & 0.2  & 2.2    & 11  & 25  \\
   & 0.8 & 1          & \textbf{6}  & 23.3  & 23.3            & 23.3            & 23.3            & 24.9   & 1.066 & 1.7    & 1.7    & 0.3  & 2.0    & 12  & 26  \\
3  & 0.3 & 1          & \textbf{6}  & 27.6  & 27.6            & 27.6            & 27.6            & 27.6   & 1.000 & 2.4    & 2.4    & 0.6  & 3.4    & 9   & 19  \\
   & 0.5 & \textbf{5} & 4           & 42.1  & 42.1            & 42.1            & 42.1            & 42.1   & 1.000 & 8.2    & 11.1   & 0.5  & 12.1   & 15  & 27  \\
   & 0.8 & 1          & 1           & 53.1  & 53.1            & 53.1            & 53.1            & 53.1   & 1.000 & 0.6    & 0.6    & 0.5  & 1.0    & 12  & 19  \\
5  & 0.3 & 1          & \textbf{8}  & 90.4  & 90.4            & 90.4            & 90.4            & 90.4   & 1.000 & 12.9   & 13.1   & 1.3  & 15.0   & 17  & 38  \\
   & 0.5 & 1          & \textbf{4}  & 100.2 & \textbf{100.6}  & 101.3           & \textbf{101.2}  & 100.6  & 1.000 & 1474.5 & 1800.5 & 1.9  & 1800.3 & 94  & 45  \\
   & 0.8 & 10         & \textbf{25} & 156.7 & 156.7           & 157.9           & \textbf{157.4}  & 156.7  & 1.000 & 1134.4 & 1800.9 & 1.5  & 1800.3 & 182 & 58  \\
8  & 0.3 & 3          & \textbf{5}  & 134.1 & 134.1           & 134.2           & 134.2           & 149.9  & 1.118 & 40.4   & 74.5   & 6.0  & 113.1  & 41  & 68  \\
   & 0.5 & 0          & \textbf{3}  & -     & \textbf{252.2}  & \textbf{260.1}  & 260.2           & 252.2  & 1.000 & -      & 1800.8 & 4.9  & 1800.2 & 28  & 51  \\
   & 0.8 & 0          & \textbf{10} & -     & \textbf{349.2}  & 379.6           & \textbf{353.4}  & 381.3  & 1.092 & -      & 1800.3 & 5.0  & 1800.6 & 28  & 83  \\
10 & 0.3 & 2          & \textbf{5}  & 317.2 & 317.2           & 317.3           & 317.3           & 317.2  & 1.000 & 535.8  & 1359.9 & 7.9  & 971.2  & 54  & 82  \\
   & 0.5 & 0          & \textbf{24} & -     & \textbf{361.7}  & 372.5           & \textbf{372.4}  & 361.7  & 1.000 & -      & 1801.5 & 5.2  & 1801.0 & 28  & 105 \\
   & 0.8 & 0          & \textbf{3}  & -     & \textbf{538.1}  & 558.8           & \textbf{549.3}  & 539.8  & 1.003 & -      & 1800.2 & 7.6  & 1800.2 & 31  & 59  \\
15 & 0.3 & 0          & \textbf{4}  & -     & \textbf{725.4}  & \textbf{743.5}  & 745.9           & 725.4  & 1.000 & -      & 1800.1 & 16.3 & 1800.0 & 51  & 94  \\
   & 0.5 & 0          & \textbf{6}  & -     & \textbf{708.4}  & \textbf{729.6}  & 730.1           & 857.5  & 1.210 & -      & 1800.1 & 22.3 & 1800.1 & 75  & 143 \\
   & 0.8 & 0          & \textbf{2}  & -     & \textbf{1216.5} & 1328.9          & \textbf{1249.2} & 1219.7 & 1.003 & -      & 1800.1 & 18.0 & 1800.1 & 49  & 91  \\
20 & 0.3 & 0          & \textbf{3}  & -     & \textbf{904.5}  & 937.7           & \textbf{922.5}  & 904.5  & 1.000 & -      & 1800.3 & 42.6 & 1800.1 & 79  & 138 \\
   & 0.5 & 0          & \textbf{3}  & -     & \textbf{1282.1} & \textbf{1469.8} & 1472.8          & 1529.1 & 1.193 & -      & 1800.3 & 38.2 & 1800.1 & 88  & 147 \\
   & 0.8 & 0          & \textbf{2}  & -     & \textbf{2024.4} & 2441.1          & \textbf{2207.2} & 2415.5 & 1.193 & -      & 1800.3 & 38.5 & 1800.2 & 78  & 136 \\
25 & 0.3 & 0          & \textbf{4}  & -     & \textbf{1765.0} & 1823.2          & \textbf{1819.5} & 1765.0 & 1.000 & -      & 1800.3 & 61.6 & 1800.1 & 93  & 166 \\
   & 0.5 & 0          & \textbf{4}  & -     & \textbf{2597.4} & 2782.7          & \textbf{2767.9} & 2597.7 & 1.000 & -      & 1800.3 & 61.1 & 1800.1 & 100 & 173 \\
   & 0.8 & 0          & \textbf{2}  & -     & \textbf{3307.4} & 3883.9          & \textbf{3541.1} & 3312.8 & 1.002 & -      & 1800.3 & 51.1 & 1800.3 & 86  & 156 \\
30 & 0.3 & 0          & \textbf{3}  & -     & \textbf{2448.1} & 2572.0          & \textbf{2556.0} & 2448.3 & 1.000 & -      & 1800.4 & 84.2 & 1800.2 & 98  & 180 \\
   & 0.5 & 0          & \textbf{2}  & -     & \textbf{3260.8} & 3509.6          & \textbf{3482.4} & 3263.8 & 1.001 & -      & 1800.4 & 90.9 & 1808.0 & 130 & 218 \\
   & 0.8 & 0          & \textbf{3}  & -     & \textbf{4910.4} & 5574.9          & \textbf{5399.1} & 4919.7 & 1.002 & -      & 1800.4 & 82.1 & 1800.3 & 103 & 189
   \\ \bottomrule
\end{tabular}
}

\vspace{3pt}
\footnotesize{\textit{Note.} 
Boldfaced values indicate better solutions found within the time limit, tighter dual bounds, or more PNEs.}
\label{tab:EBMCP_games_1}
\end{table}

\subsection{EBMC Games with Minnesota Dataset} \label{sec:6.3}
\vspaceminus{3pt}

Tables~\ref{tab:EBMC_minnesota_CR_BRD} and~\ref{tab:EBMCP_games_minnesota} report results for the selfish EBMC game \(\mathcal{G}^{\text{Self}}_{|\mathcal{L}|\ge 2}\) on the Minnesota dataset: Table~\ref{tab:EBMC_minnesota_CR_BRD} summarizes standalone RRR--BRD performance, and Table~\ref{tab:EBMCP_games_minnesota} compares BZR and ZR. The dataset, provided by the Minnesota Aquatic Invasive Species Research Center (MAISRC), contains lake-to-lake boat movement estimates among 9{,}182 lakes and uses 2018 infestation records, following the construction in \citet{kao2021network}. We exclude counties with no lakes or with no incoming or outgoing risky movements.

To reflect the scale of county-level inspection programs, we model statewide support corresponding to 436 inspection stations. Minnesota allocates roughly \$10 million per year to AIS inspection programs and employs approximately 800--1000 inspectors \citep{MDR2017}; assuming two inspectors per lake, this staffing level corresponds to operating about 400--500 inspection stations statewide. We allocate county budgets proportionally to county-level funding levels \citep{MDR2017}. We model four AIS types--Zebra Mussels, Starry Stonewort, Eurasian Watermilfoil, and Spiny Water Fleas--yielding \(\mathcal{G}^{\text{Self}}_{|\mathcal{L}|\ge 2}\).

Because this instance is extremely large (84 players, roughly 100 lakes per county on average, and millions of arcs), we use a 3600-second time limit for ZR, BZR, and the SW model, and a 600-second limit to compute the warm-start profile \(\barxswtime{600}\). Standalone RRR--BRD requires more than 400 seconds from either \(\barxzero\) or \(\barxswtime{600}\) (Table~\ref{tab:EBMC_minnesota_CR_BRD}), indicating a substantially more difficult best-response landscape than in our synthetic instances. Table~\ref{tab:EBMCP_games_minnesota} shows that BZR identifies multiple PNEs within the time limit, whereas ZR finds none.

\vspaceminus{12pt}
\begin{table}[htbp]
\caption{RRR-BRD Results for Selfish EBMC Games $\mathcal{G}^{\text{Self}}_{|\mathcal{L}|\ge 2}$ in Minnesota Dataset.}
\centering
\small
\scalebox{0.9}{
\begin{tabular}{r r r r r r r r r r r r r}
\toprule
& \multicolumn{2}{c}{CR-BRD ($\phi$)} & \multicolumn{2}{c}{Price of Stability} & \multicolumn{2}{c}{References} & \multicolumn{2}{c}{Time (secs)} & \multicolumn{2}{c}{Round} \\
\cmidrule(lr){2-3} \cmidrule(lr){4-5} \cmidrule(lr){6-7} \cmidrule(lr){8-9} \cmidrule(lr){10-11} 
$n$ & init:$\barxzero$ &   init:$\barxswtime{600}$ & $\widetilde{\text{OSW}}$ & $\widetilde{\text{POS}}$ & BZR($\phi$) & BZR(Bound) & init:$\barxzero$ &   init:$\barxswtime{600}$ & init:$\barxzero$ &   init:$\barxswtime{600}$  \\ \midrule
84  & 758.8   & 761.4  & 887.7   & 1.166 & 761.4 & 861.7 & 425.4  & 586.5   & 5 & 7  \\ \bottomrule
\end{tabular}
}

% \vspace{3pt}
% \footnotesize{\textit{Note.} Objective values ($\phi$) are presented in thousands and rounded to the nearest tenth.}
\label{tab:EBMC_minnesota_CR_BRD}
\end{table}
\vspaceminus{24pt}
\begin{table}[h!]
\caption{Comparison of BZR and ZR for Selfish EBMC Games $\mathcal{G}^{\text{Self}}_{|\mathcal{L}|\ge 2}$ in Minnesota Dataset.}
\centering
\small
\scalebox{0.80}{
\begin{tabular}{rrrrrrrrrrrrrrr}
\toprule
& \multicolumn{2}{c}{\# PNEs Found} & \multicolumn{2}{c}{Best PNE ($\phi$)} & \multicolumn{2}{c}{Dual Bounds} & \multicolumn{2}{c}{Price of Stability} & \multicolumn{4}{c}{Time (secs)} & \multicolumn{2}{c}{\# EI Cuts}  \\
\cmidrule(lr){2-3} \cmidrule(lr){4-5} \cmidrule(lr){6-7} \cmidrule(lr){8-9} \cmidrule(lr){10-13} \cmidrule(lr){14-15}
$n$ & $\#_\text{pne(ZR)}$ & $\#_\text{pne(BZR)}$ & ZR &  BZR  & ZR & BZR & $\widetilde{\text{OSW}}$ & $\widetilde{\text{POS}}$  & T(ZR$^{\text{1st}}$) & T(ZR) & T(BZR$^{\text{1st}}$) & T(BZR) &  $\#_\text{cuts(ZR)}$ & $\#_\text{cuts(BZR)}$   \\  \midrule
84  & 0  & \textbf{5} & -  & \textbf{761.4}  & 861.7 & 861.7            & 887.7             & 1.166           & -    & 3600.8 & 790.7    & 3600.8    & 363   & 363   \\ \bottomrule
\end{tabular}
}

\label{tab:EBMCP_games_minnesota}

\end{table}
\vspaceminus{12pt}
We emphasize that PNE existence in this instance is not guaranteed a priori by our sufficient condition (Theorem~\ref{thm:PNE-existence}). The Minnesota results therefore provide a computational witness of existence for the selfish game, complementing the locally altruistic case where a PNE is immediate via Corollary~\ref{cor:SW_is_PNE}. Table~\ref{tab:county-level} reports county-level selfish utilities on the Minnesota instance under several strategy profiles: the non-game profile \(\bar{\x}_{\text{ng}}\), the first PNE \(\hat{\x}_{\text{pne}}^{1}\), the best PNE \(\hat{\x}_{\text{pne}}^{*}\) found by BZR, and the welfare benchmark \(\barxswtime{3600}\). For each county \(i\), we compare utilities column-wise via
\(
u_i^{\text{Self}}(\bar{\x}_{\text{ng}}) \le u_i^{\text{Self}}(\hat{\x}_{\text{pne}}^{1}),
u_i^{\text{Self}}(\hat{\x}_{\text{pne}}^{1}) \le u_i^{\text{Self}}(\hat{\x}_{\text{pne}}^{*}),
u_i^{\text{Self}}(\hat{\x}_{\text{pne}}^{*}) \le u_i^{\text{Self}}(\barxswtime{3600}),
\)
and highlight any violations in bold. The complete 84-county table is provided in Appendix~\ref{appendix:county-level}.

Overall, the results suggest that (i) many counties benefit from strategic interaction relative to the non-game baseline, due to unintended positive spillovers; (ii) among equilibria, the best PNE improves utilities for most counties; and (iii) the welfare-oriented benchmark \(\barxswtime{3600}\) (and, in the locally altruistic game, the OSW solution) often yields the highest utilities, although exceptions exist (e.g., Sibley and Winona).

These comparisons illustrate why computed equilibria are policy-relevant benchmarks. A PNE is stable in the sense that no county has an incentive to deviate unilaterally, and the socially best PNE for the selfish game offers a tractable compromise between stability and statewide performance. Moreover, the locally altruistic specification highlights a concrete policy target: if counties can adopt (or are incentivized to adopt) utilities that internalize spillovers, then the welfare-optimal solution is self-enforcing--indeed, Corollary~\ref{cor:SW_is_PNE} shows that the OSW solution is a PNE under locally altruistic utilities. Nevertheless, the best selfish-game PNE can still deviate substantially from the welfare benchmark, as reflected by a POS of \(1.166\). This motivates future work on (i) incentive design that induces counties to internalize externalities and (ii) equilibrium-informed coordination that leverages equilibrium information from selfish games to improve social welfare.

\vspaceminus{12pt}
\begin{table}[h!]
\caption{County-level Utility Values Using Different Strategy Profiles.}
\label{tab:county-level}
\centering
\scalebox{0.80}{
\begin{tabular}{lrrrr@{\hspace{0.5cm}}|lrrrr@{\hspace{0.5cm}}}
\toprule
County & $u_c^{\text{Self}}(\bar{\x}_{\text{ng}})$ & $u_c^{\text{Self}}(\hat{\x}_{\text{pne}}^1)$ & $u_c^{\text{Self}}(\hat{\x}_{\text{pne}}^*)$ & $u_c^{\text{Self}}(\barxswtime{3600})$ & County & $u_c^{\text{Self}}(\bar{\x}_{\text{ng}})$ & $u_c^{\text{Self}}(\hat{\x}_{\text{pne}}^1)$ & $u_c^{\text{Self}}(\hat{\x}_{\text{pne}}^*)$ & $u_c^{\text{Self}}(\barxswtime{3600})$ \\ \midrule
Anoka    & 11292 & 12466 & 12495 & 15723 & Mcleod          & 7194          & 7479          & 7503          & 8933          \\
Becker   & 24219 & 26489 & 26497 & 31712 & Meeker          & 8866          & 9423          & 9428          & 9964          \\
Hennepin & 21136 & 22442 & 22474 & 27024 & Sherburne       & 5077          & 5557          & 5564          & 6698          \\
Hubbard  & 22780 & 24426 & 24444 & 29543 & \textbf{Sibley} & \textbf{2390} & \textbf{2388} & \textbf{2388} & \textbf{2429} \\
Isanti   & 3635  & 3876  & 3884  & 4227  & Stearns         & 20200         & 21544         & 21554         & 26076         \\
Lyon     & 2752  & 2925  & 2939  & 3314  & \textbf{Winona} & \textbf{1561} & \textbf{1579} & \textbf{1579} & \textbf{1444} \\
Mahnomen & 4177  & 4424  & 4428  & 6012  & Wright          & 22047         & 23782         & 23805         & 29464               \\
\bottomrule
\end{tabular}
}

\vspace{3pt}
\footnotesize{\textit{Note.} Counties, where the inequalities regarding utility value comparisons do not hold, are highlighted in bold.}
\end{table}

\vspaceminus{24pt}
\subsection{Large-scale Knapsack Problem Games Datasets and Results}
\label{sec:7.4}
\vspaceminus{3pt}
To further assess the effectiveness of RRR--BRD and BZR, we report extensive computational results for large-scale knapsack problem games (KPGs), a standard benchmark in the IPG literature \citep{carvalho2021cut, carvalho2022computing, dragotto2023zero, cronert2024equilibrium}. In a KPG, each player \(i\) solves a binary knapsack problem of the form
$
\max _{\x^{i}}\left\{\sum_{j=1}^{m} p_{j}^{i} x_{j}^{i}+\sum_{k=1, k \neq i}^{n} \sum_{j=1}^{m} f_{k, j}^{i} x_{j}^{i} x_{j}^{k}: \sum_{j=1}^{m} w_{j}^{i} x_{j}^{i} \leq \B^{i}, \x^{i} \in\{0,1\}^{m}\right\}$,
where \(x_j^i=1\) if player \(i\) selects item \(j\) (and \(0\) otherwise). The parameters \(p_j^i\), \(w_j^i\), and \(\B^i\) denote item profit, item weight, and player budget, respectively. Strategic interactions are captured by the bilinear coefficients \(f_{k,j}^i\), which contribute to player \(i\)'s payoff when both players \(i\) and \(k\) select item \(j\). To linearize $x_j^i x_j^k$, we introduce auxiliary variables $z_j^{i,k}$ imposing the following constraints:
\(z_j^{i,k} \geq x_j^i + x_j^k - 1, \,
z_j^{i,k} \leq x_j^i, \,
z_j^{i,k} \leq x_j^k, \, \forall j \in \{1,\dots,m\},\, \forall i \in N,\, \forall k \in N \setminus \{i\}.\)

Following \citet{dragotto2023zero}, we also incorporate two classes of problem-specific equilibrium inequalities--strategic dominance and strategic payoff inequalities--along with the standard equilibrium inequalities. For larger instances (\(n\ge 15\)), identifying dominance relations and minimal interaction sets for strategic payoff inequalities becomes computationally expensive, so we omit these inequalities.

We adopt the data generation scheme of \citet{dragotto2023zero} for interaction types (A) and (B), and use a modified version of type (C) to better match our experimental goal. In type (A), each \(f_{k,j}^i\) is set to a constant drawn uniformly from \([1,100]\); in type (B), each \(f_{k,j}^i\) is drawn independently from \([1,100]\). 
In the original type (C), \(f_{k,j}^i \sim \mathrm{Unif}[-100,100]\), which can induce very strong negative (adversarial) interactions; as also observed in \citet{dragotto2023zero}, such instances often fail to admit a PNE even for small sizes (\(n=2,3\)), and this issue becomes more pronounced as \(n\) grows. Because our convergence guarantees for RRR--BRD (and hence BZR's use of it as a subroutine) are conditional--\emph{on PNE existence}--we prioritize stress-testing instances that remain adversarial yet are not overwhelmingly likely to be PNE-free. Concretely, we modify type (C) by sampling \(f_{k,j}^i\) from \([-20,0]\), which preserves negative interactions but reduces their magnitude to an ``admissible'' level that more frequently yields equilibria in practice. 
Budgets are set as a fraction of total item weight:
\(
\B^i \in \left\{0.2\sum_{j=1}^m w_j^i,\ 0.5\sum_{j=1}^m w_j^i,\ 0.8\sum_{j=1}^m w_j^i\right\}.
\)
We start from the largest KPG setting in \citet{dragotto2023zero} (\(n\in\{2,3\}\), \(m=100\)) and scale to \(n\in\{2,3,5,8,10,15,20,25,30\}\) with \(m=100\), yielding 27 generated instances per interaction type.

Table~\ref{tab:KPG_CR_BRD_sumary} summarizes RRR--BRD performance under two initial profiles, \(\barxzero\) and \(\barxswtime{300}\); complete results appear in Appendix~\ref{appendix:CR-BRD}. Across interaction types, we observe that type (C) instances exhibit larger \(\widetilde{\text{POS}}\), and require substantially more rounds before reaching a PNE. These trends indicate that negative interactions produce markedly more complex best-response dynamics than types (A) and (B).

\vspaceminus{12pt}
\begin{table}[htbp]
\caption{Summarized RRR-BRD Results with Different Initial Strategy Profiles for Type A, B, and C KPG.}
\centering
\small
\scalebox{0.8}{
\begin{tabular}{c c c c c c c c c c c c}
\toprule
type & \multicolumn{2}{c}{Avg$^\dagger$ PNE ($\phi$)} & \multicolumn{2}{c}{Avg Price of Stability} & \multicolumn{2}{c}{Avg References} & \multicolumn{2}{c}{Avg Time (secs)} & \multicolumn{2}{c}{Avg Round}  \\
\cmidrule(lr){2-3} \cmidrule(lr){4-5} \cmidrule(lr){6-7} \cmidrule(lr){8-9} \cmidrule(lr){10-11} 
& init:$\barxzero$ &   init:$\barxswtime{300}$ & $\widetilde{\text{OSW}}$ &   $\widetilde{\text{POS}}$ & BZR($\phi$) & BZR(Bound) & init:$\barxzero$ &   init:$\barxswtime{300}$ & init:$\barxzero$ &   init:$\barxswtime{300}$  \\ \midrule
Type A & 700.7 & 711.0 & 714.1 & 1.012 & 711.2 & 715.6 & 2.6 & 192.9 & 7.7 & 3.9  \\
Type B  & 706.1 & 714.7 & 717.1 & 1.010 & 715.5 & 719.2 & 2.9 & 198.2 & 8.6 & 4.6  \\
Type C  & 14.0 & 14.1 & 16.9 & 1.171 & 14.2 & 16.3 & 1.8 & 73.2 & 29.5 & 16.5  \\
\bottomrule
\end{tabular}
}

\vspace{3pt}
\footnotesize{
\textit{Note.} $\dagger$: The averages of $\phi$, OSW, and POS are computed only from instances where a PNE is found under both initial solutions.}
\label{tab:KPG_CR_BRD_sumary}
\end{table}

Table~\ref{tab:KPG_comp_sumary} provides a summarized comparison of BZR and ZR across the three KPG datasets, reporting how often a PNE was found, multiple PNEs were identified, the best PNE was obtained, and a tighter dual bound was achieved. Full results for type (B) are given in Table~\ref{tab:KPG_B}, while those for types (A) and (C) appear in Appendix~\ref{appendix:full_BZR}. Overall, BZR consistently outperforms ZR across all interaction types in (i) time to the first PNE, (ii) the number of distinct PNEs found, and (iii) the best PNE value identified. For dual bounds, the advantage is clearer in types (A) and (B), whereas in type (C) the two methods exhibit complementary strengths: as shown in Table~\ref{tab:KPG_C} (Appendix~\ref{appendix:full_BZR}), BZR generally produces tighter dual bounds for \(n\le 15\), while ZR tends to perform better for \(n\ge 20\).

\vspaceminus{12pt}
\begin{table}[htbp]
\caption{Summarized Comparison of BZR and ZR for Type A, B, and C KPG.}
\centering
\small
\scalebox{0.9}{
\begin{tabular}{c c c c c c c c c c c c c c c}
\toprule
type & \multicolumn{2}{c}{\# PNEs Found}  & \multicolumn{2}{c}{Finding a PNE} & \multicolumn{3}{c}{Best PNE}  & \multicolumn{3}{c}{Dual Bounds} & Avg POS & \multicolumn{2}{c}{Avg \# EI Cuts}  \\
\cmidrule(lr){2-3} \cmidrule(lr){4-5} \cmidrule(lr){6-8} \cmidrule(lr){9-11} \cmidrule(lr){12-12} \cmidrule(lr){13-14}
 & $\#_\text{pne(ZR)}$ & $\#_\text{pne(BZR)}$ & ZR  & BZR &  Ties &  ZR & BZR & Ties &  ZR & BZR & $\widetilde{\text{POS}}$ & $\#_\text{cuts(ZR)}$ & $\#_\text{cuts(BZR)}$ \\ \midrule
Type A & 14 & 719 & 7 & 27 & 7 & 0 & 20 & 10 & 0 & 17& 1.010 & 1206.7 & 1012.0  \\
Type B & 12 & 1766 & 5 & 27 & 5 & 0 & 22 & 7 & 1 & 19 & 1.006 & 945.9 & 966.6  \\
Type C & 12 & 675 & 7 & 24 & 7 & 0 & 17 & 8 & 7 & 12 & 1.175 & 2180.9 & 2516.4 \\
\bottomrule
\end{tabular}
}
\vspace{3pt}
\label{tab:KPG_comp_sumary}
\end{table}

\vspace{9pt}

\begin{table}[htbp]
\caption{Comparison of BZR and ZR for Type B KPG.}
\centering
\small
\scalebox{0.73}{
\begin{tabular}{rrrrrrrrrrrrrrrrrr}
\toprule
&& \multicolumn{2}{c}{\# PNEs Found} & \multicolumn{2}{c}{Best PNE ($\phi$)} & \multicolumn{2}{c}{Dual Bounds} & \multicolumn{2}{c}{Price of Stability} & \multicolumn{4}{c}{Time (secs)} & \multicolumn{2}{c}{\# EI Cuts}  \\
\cmidrule(lr){3-4} \cmidrule(lr){5-6} \cmidrule(lr){7-8} \cmidrule(lr){9-10} \cmidrule(lr){11-14} \cmidrule(lr){15-16}
$n$ & BG & $\#_\text{pne(ZR)}$ & $\#_\text{pne(BZR)}$ & ZR &  BZR  & ZR & BZR & $\widetilde{\text{OSW}}$ & $\widetilde{\text{POS}}$  & T(ZR$^{\text{1st}}$) & T(ZR) & T(BZR$^{\text{1st}}$) & T(BZR) &  $\#_\text{cuts(ZR)}$ & $\#_\text{cuts(BZR)}$  \\ \midrule
2  & 0.2 & 3 & \textbf{9}   & 8.2  & 8.2             & 8.2             & 8.2             & 8.3    & 1.018 & 9.7   & 13.4   & 0.7  & 10.0   & 1944 & 1913 \\
   & 0.5 & 4 & \textbf{125} & 14.2 & 14.2            & 14.2            & 14.2            & 14.4   & 1.010 & 72.9  & 130.9  & 0.3  & 72.9   & 2281 & 2154 \\
   & 0.8 & 1 & \textbf{9}   & 19.1 & 19.1            & 19.1            & 19.1            & 19.1   & 1.002 & 1.2   & 1.4    & 0.3  & 2.2    & 1224 & 1239 \\
3  & 0.2 & 0 & \textbf{195} & -    & \textbf{15.6}   & 15.7            & \textbf{15.6}   & 15.9   & 1.023 & -     & 1800.3 & 1.1  & 1800.3 & 2679 & 2161 \\
   & 0.5 & 0 & \textbf{222} & -    & \textbf{28.9}   & 29.2            & \textbf{29.1}   & 29.4   & 1.016 & -     & 1800.3 & 0.6  & 1800.2 & 2744 & 2100 \\
   & 0.8 & 3 & \textbf{291} & 40.0 & 40.0            & 40.1            & 40.1            & 40.3   & 1.007 & 590.5 & 1800.3 & 1.2  & 1801.3 & 2287 & 2361 \\
5  & 0.2 & 0 & \textbf{84}  & -    & \textbf{40.5}   & 40.8            & 40.8            & 40.9   & 1.012 & -     & 1800.2 & 2.0  & 1800.2 & 663  & 832  \\
   & 0.5 & 0 & \textbf{129} & -    & \textbf{78.0}   & 78.8            & 78.8            & 78.8   & 1.011 & -     & 1800.4 & 2.2  & 1802.9 & 1377 & 1325 \\
   & 0.8 & 0 & \textbf{107} & -    & \textbf{108.3}  & 108.7           & \textbf{108.6}  & 108.7  & 1.003 & -     & 1800.5 & 4.0  & 1800.4 & 1790 & 983  \\
8  & 0.2 & 0 & \textbf{75}  & -    & \textbf{91.6}   & 92.7            & 92.7            & 92.6   & 1.011 & -     & 1800.3 & 4.4  & 1800.5 & 755  & 1055 \\
   & 0.5 & 0 & \textbf{89}  & -    & \textbf{188.6}  & 190.0           & 190.0           & 189.9  & 1.007 & -     & 1801.1 & 10.0 & 1801.3 & 874  & 1282 \\
   & 0.8 & 0 & \textbf{12}  & -    & \textbf{272.9}  & 273.4           & \textbf{273.2}  & 273.1  & 1.001 & -     & 1801.2 & 6.9  & 1800.8 & 998  & 176  \\
10 & 0.2 & 0 & \textbf{43}  & -    & \textbf{135.9}  & 137.1           & \textbf{137.0}  & 136.9  & 1.007 & -     & 1801.1 & 2.3  & 1800.4 & 523  & 721  \\
   & 0.5 & 0 & \textbf{28}  & -    & \textbf{287.8}  & 289.5           & \textbf{289.4}  & 289.0  & 1.004 & -     & 1800.8 & 6.1  & 1801.2 & 306  & 508  \\
   & 0.8 & 0 & \textbf{27}  & -    & \textbf{426.9}  & 428.4           & \textbf{428.3}  & 427.8  & 1.002 & -     & 1800.5 & 7.4  & 1801.4 & 667  & 435  \\
15 & 0.2 & 0 & \textbf{41}  & -    & \textbf{287.3}  & 290.6           & 290.6           & 289.5  & 1.008 & -     & 1801.9 & 20.4 & 1801.0 & 358  & 960  \\
   & 0.5 & 0 & \textbf{19}  & -    & \textbf{642.1}  & 645.7           & \textbf{645.5}  & 644.0  & 1.003 & -     & 1800.9 & 20.4 & 1801.5 & 252  & 466  \\
   & 0.8 & 0 & \textbf{12}  & -    & \textbf{963.3}  & 967.1           & \textbf{966.6}  & 965.2  & 1.002 & -     & 1800.9 & 23.0 & 1801.4 & 258  & 312  \\
20 & 0.2 & 0 & \textbf{16}  & -    & \textbf{488.4}  & 493.2           & \textbf{493.1}  & 491.2  & 1.006 & -     & 1800.7 & 13.4 & 1800.6 & 334  & 496  \\
   & 0.5 & 0 & \textbf{11}  & -    & \textbf{1123.1} & 1129.2          & \textbf{1129.1} & 1125.5 & 1.002 & -     & 1801.2 & 27.6 & 1801.3 & 309  & 402  \\
   & 0.8 & 0 & \textbf{10}  & -    & \textbf{1671.7} & 1677.5          & \textbf{1677.3} & 1672.9 & 1.001 & -     & 1800.8 & 24.5 & 1800.9 & 264  & 367  \\
25 & 0.2 & 0 & \textbf{12}  & -    & \textbf{760.9}  & 767.6           & \textbf{767.5}  & 763.5  & 1.003 & -     & 1800.8 & 21.6 & 1801.0 & 315  & 444  \\
   & 0.5 & 0 & \textbf{11}  & -    & \textbf{1716.7} & 1728.7          & \textbf{1728.5} & 1723.7 & 1.004 & -     & 1800.9 & 19.2 & 1800.9 & 338  & 448  \\
   & 0.8 & 0 & \textbf{9}   & -    & \textbf{2610.3} & 2623.1          & \textbf{2622.8} & 2614.3 & 1.002 & -     & 1800.9 & 28.9 & 1800.7 & 371  & 420  \\
30 & 0.2 & 0 & \textbf{10}  & -    & \textbf{1083.3} & 1094.7          & \textbf{1094.6} & 1088.8 & 1.005 & -     & 1801.0 & 34.1 & 1800.7 & 399  & 540  \\
   & 0.5 & 0 & \textbf{11}  & -    & \textbf{2459.8} & \textbf{2473.3} & 2473.4          & 2466.1 & 1.003 & -     & 1800.7 & 98.9 & 1801.0 & 237  & 584  \\
   & 0.8 & 0 & \textbf{8}   & -    & \textbf{3748.6} & 3764.1          & \textbf{3764.0} & 3750.8 & 1.001 & -     & 1800.8 & 27.0 & 1800.6 & 425  & 430 

 \\ \bottomrule
\end{tabular}
}

\vspace{3pt}
\footnotesize{\textit{Note.} 
Boldfaced values indicate better solutions found within the time limit, tighter dual bounds, or more PNEs.}
\label{tab:KPG_B}
\end{table}

\vspaceminus{12pt}
\paragraph*{How Interaction Coefficients and the Number of Players Affect $q_\mu[R]$.}
When interaction coefficients are predominantly positive, improving BR moves more often enter equilibrium basins, leading to larger $\tilde q_\mu[R]$ and faster PNE discovery; as interactions become more complex---e.g., when coefficients are mixed or negative---the dynamics can become more complicated and $q_\mu[R]$ may decrease. This viewpoint aligns with our EBMC variants: in $\G^{\text{Self}}_{|\mathcal{L}|=1}$, other players' selections are typically helpful for player $i$ (Section~\ref{sec:selfish_proof}), whereas in $\G^{\text{Self}}_{|\mathcal{L}|\ge 2}$, other players' updates can affect player $i$ either positively or negatively, yielding more complicated BR trajectories. Similarly, KPG types (A) and (B) behave like predominantly positive-interaction instances, while our type (C) is closer to a mixed regime. Finally, increasing the number of players can amplify externalities; for example, the original type (C) construction in \citet{dragotto2023zero} can exhibit a more adversarial landscape as externalities increase (Section~\ref{sec:7.4}), which may drive $q_\mu[R]$ to be very small. This is consistent with our results: $\G^{\text{Self}}_{|\mathcal{L}|=1}$ and KPG types (A) and (B) find PNEs in all instances, whereas $\G^{\text{Self}}_{|\mathcal{L}|\ge 2}$ fails in four instances and our KPG type (C) fails in three instances.

\section{Conclusion} \label{sec:Conclusion}
\vspaceminus{3pt}
Computing PNEs in IPGs remains challenging, primarily because most existing exact approaches rely on joint formulations that couple players' strategy sets, and computational evidence is therefore largely restricted to instances with only a few players and modest per-player model sizes. This scalability gap is particularly salient in large-scale resource-allocation settings with many stakeholders, such as the EBMC games studied in this paper. To address this challenge, we introduce round random-restart best-response dynamics (RRR--BRD) for IPGs and provide a probabilistic convergence guarantee whenever a PNE exists and the restart law assigns positive probability to basins that reach a PNE.

We further embed RRR--BRD as a randomized local-search subroutine within the ZR framework, yielding substantial practical gains in finding PNEs and high-social-welfare PNEs, and often improving dual bounds. Overall, these improvements translate into an order-of-magnitude enhancement in the tractable scale of equilibrium computation for our tested IPG classes. We also discuss how RRR--BRD can serve as a local-search subroutine within other IPG algorithms whenever integer-feasible profiles are available. On the modeling side, we study two EBMC game variants--locally altruistic and selfish--by specifying how players value edge coverage, and we establish sufficient conditions for PNE existence in each setting. Using the MN dataset, we illustrate how equilibrium outcomes depend on these behavioral specifications and provide managerial insights for allocating limited inspection resources at both the statewide and county levels.

Finally, our results point to two important directions for future research. First, our probabilistic guarantees are expressed through the instance-dependent success probability under restarts, $q_{\mu}[R]$, which is difficult to compute in general; deriving tractable estimators/bounds for $q_{\mu}[R]$ under additional structural assumptions, and designing restart laws that improve it, are promising extensions. Second, while we focus on equilibrium computation and comparative analysis under different utility specifications, we do not design incentive mechanisms; developing implementable policy levers that steer equilibria toward socially desirable outcomes, and analyzing welfare--stability trade-offs and sensitivity, are natural next steps.

\bigskip
\noindent\textbf{Acknowledgments.} 
H. Lee and R. Hildebrand were partially funded by AFOSR grant FA9550-21-1-0107.  H. Lee and {\.I.} E. B{\"u}y{\"u}ktahtak{\i}n have also been funded by the Grado Department of ISE at VT. S. Cai was partially funded by MAISRC Subaward  H009064601, and all authors were partially funded by the MAISRC Subaward H0113450M1. The authors are grateful to MAISRC for making the Minnesota dataset publicly accessible, and Amy Kinsley (MAISRC), Nick Phelps (MAISRC), Alex Bajcz (MAISRC), Bob Haight (USFS), Tina Fitzgerald (DNR) and Adam Doll (DNR) for pointing us to the Minnesota dataset and for helpful discussion on the AIS inspection problem in MN. Any opinions, findings, conclusions, or recommendations expressed in this study are those of the authors and do not necessarily reflect the views of the Air Force Office of Scientific Research and MAISRC.

% The authors express their gratitude to MAISRC for funding and making the Minnesota dataset publicly accessible. The opinions, findings, conclusions, or recommendations expressed in this study are solely those of the authors and do not necessarily represent the views of AFOSR or MAISRC.

\bibliography{references}

% bibliography style set in preamble

\clearpage
\renewcommand{\thetable}{A\arabic{table}}
\renewcommand{\thefigure}{A\arabic{figure}}
\setcounter{table}{0}
\setcounter{figure}{0}

\ecompanionsection{Acronyms and Notations}\label{appendix:acronyms}
Table~\ref{tab:acronyms} lists acronyms used throughout the paper. Tables~\ref{tab:notation_main}--\ref{tab:notation_app} summarize the notation.

\begin{table}[h!]
\centering
\small
\caption{List of Acronyms Used in the Manuscript and Appendix}
\label{tab:acronyms}
\begin{tabular}{@{}l p{0.78\linewidth}@{}}
\hline
\textbf{Acronym} & \textbf{Meaning} \\
\hline
IPG / IP & Integer programming game / integer program. \\
% IP & Integer program. \\
PNE & Pure Nash equilibrium. \\
BR & Best response (noun); best-response (adjective). \\
SCC & Strongly connected component in a directed graph. \\
DAG & Directed acyclic graph. \\
RR-BRD & Random-Restart best-response dynamics. \\
RRR-BRD & Round RR-BRD, i.e., a round-based implementation of RR-BRD. \\
EI & Equilibrium inequality. \\
ZR & Zero-regret (algorithm). \\
BZR & BRD-incorporated zero-regret (algorithm). \\
KPG & Knapsack problem game. \\
EBMC & Edge-weighted budgeted maximum coverage (game/model). \\
AIS & Aquatic invasive species. \\
OSW & Optimal social welfare (social optimum). \\
POS & Price of stability. \\
\hline
\end{tabular}
\end{table}

\begin{table}[h!]
\centering
\small
\caption{Summary of Notation Used in the Manuscript and Appendix (Core Notation)}
\label{tab:notation_main}
\begin{tabular}{@{}l p{0.78\linewidth}@{}}
\hline
\textbf{Symbol} & \textbf{Meaning} \\
\hline
\multicolumn{2}{@{}l}{\textbf{IPGs.}}\\
$\mathcal{G}$ & An integer programming game (IPG); the IPG tuple as defined in Section~2.1. \\
$N=\{1,\dots,n\}$ & Set of players; $n:=|N|$. \\
$m_i; m$ & Number of integer variables of player $i$; when $m_i$ is the same for all $i\in N$, we write $m$. \\
$\mathcal{X}_i$ & Polyhedrally bounded pure-integer set of player $i$ in the IPG (i.e., a finite strategy set). \\
$\X$ & Joint strategy set, $\X := \mathcal{X}_1 \times \cdots \times \mathcal{X}_n$. \\
$\x^i\in\mathcal{X}_i$; $\x=(\x^i)_{i\in N}$ & Player $i$'s strategy; joint strategy profile. \\
$u_i(\x^i,\x^{-i})$ & Utility (objective value) of player $i$ under profile $\x$. \\
$\phi(\x)$ & Social welfare function, $\phi(\x)=\sum_{i\in N} u_i(\x)$. \\
$\barxosw$ & OSW solution. \\
$\hat{\x}^*_{\text{pne}}$ & Best PNE (socially optimal PNE). \\
\hline
\multicolumn{2}{@{}l}{\textbf{Normal-Form Games, BR State Graph, and RRR-BRD.}}\\
$S_i$ & Set of pure strategies of player $i$ in the finite normal-form game. \\
$S=\prod_{i\in N} S_i$ & Set of pure strategy profiles (normal-form viewpoint). \\
$s=(s_i,s_{-i})$ & Strategy profile; $s_{-i}$ denotes the strategies of all players except $i$. \\
$\BR_i^+(s_{-i})$ & Set of (improving) best responses at $s_{-i}$. \\
SCC; sink SCC & Strongly connected component; a sink SCC has no outgoing edges in the condensation DAG. \\
Condensation DAG & Directed acyclic graph obtained by contracting SCCs of the BR state graph. \\
$\mu$ & Restart distribution over $S$ used by RRR-BRD. \\
$R$ & Round cap per attempt in RRR-BRD. \\
$q_\mu[R]$ & Probability that an attempt from $s\sim\mu$ reaches a PNE within $R$ rounds. \\
\hline
\multicolumn{2}{@{}l}{\textbf{Equilibrium Inequalities and Lifted Formulations.}}\\
$\mathcal{S}_{\mathrm{PNE}}(\mathcal{G})$ & Set of pure Nash equilibria of game $\mathcal{G}$. \\
$\mathcal{K}$ & Lifted/linearized formulation of the joint strategy profile. \\
$\mathcal{E}$ & Equilibrium closure. \\
$\Omega$ & Set over which equilibrium inequalities are generated/enforced. \\
\hline
\end{tabular}
\end{table}

\begin{table}[h!]
\centering
\small
\caption{Summary of Notation Used in the Manuscript and Appendix (Applications and Experiments)}
\label{tab:notation_app}
\begin{tabular}{@{}l p{0.78\linewidth}@{}}
\hline
\textbf{Symbol} & \textbf{Meaning} \\
\hline
\multicolumn{2}{@{}l}{\textbf{EBMC Games.}}\\
$i\in N$ & Player index in the EBMC model/games. \\
$\mathcal{I}_i$ & Set of vertices controlled by player $i$ (player-level decision set). \\
$x_j$ & Binary variable indicating whether vertex $j$ is selected. \\
$y_{jk}$ & Binary variable indicating whether arc $(j,k)$ is covered. \\
$\B_i$ & Budget of player $i$ (inspection budget). \\
$w_{jk}$ & Edge weight for edge $(j,k)$. \\
$\G^{\text{Self}}_{|\mathcal{L}|=1}$ & Selfish EBMC game with a single AIS type. \\
$\G^{\text{Self}}_{|\mathcal{L}| \ge 2}$ & Selfish EBMC game with multiple AIS types. \\
$\G^{\text{Alt}}$ & Locally altruistic EBMC game. \\
% $\G^{\text{Self}}$, $\G^{\text{Alt}}$ & Selfish EBMC game; locally altruistic EBMC game. \\
\hline
\multicolumn{2}{@{}l}{\textbf{KPGs.}}\\
$J$ & Set of items in the knapsack problem game. \\
$x^i_j\in\{0,1\}$ & Binary selection of item $j$ by player $i$. \\
$\B_i$ & Knapsack capacity of player $i$ (in KPG). \\
$p^i_j$ & Profit of item $j$ for player $i$. \\
$f^i_{k,j}$ & Bilinear coefficient associated with item $j$ for players $i$ and $k$. \\
\hline
\multicolumn{2}{@{}l}{\textbf{Computational Results.}}\\
$\tilde q_\mu[R]$ & $q_\mu[R]$ estimated via Monte Carlo simulation. \\
$\barxzero$ & Initial strategy profile (all-zero vector). \\
$\barxswt$ & Initial strategy profile obtained by solving the SW model with a time limit (seconds). \\
$\bar{\x}_{\text{ng}}$ & Non-game strategy profile. \\
$\hat\x_{\text{pne}}^1$ / $\hat\x_{\text{pne}}^*$ & First identified PNE / best identified PNE (maximizing $\phi$) within the time limit. \\
$\phi(\cdot)$ & Social welfare function; values are reported in thousands and rounded to the nearest tenth. \\
$\widetilde{\text{OSW}}$ & Approximate optimal social welfare, $\phi(\barxswt)$. \\
$\widetilde{\text{POS}}$ & Approximate price of stability, $\phi(\barxswt) / \phi(\hat\x_{\text{pne}}^*)$. \\
$\#_{\text{cuts(ZR)}}$ / $\#_{\text{cuts(BZR)}}$ & Number of equilibrium inequalities added by ZR / BZR. \\
$\#_{\text{pne(ZR)}}$ / $\#_{\text{pne(BZR)}}$ & Number of distinct PNEs found by ZR / BZR. \\
T(ZR$^{\text{1st}}$) / T(BZR$^{\text{1st}}$) & Time to find the first PNE using ZR / BZR. \\
T(ZR) / T(BZR) & Total runtime of the ZR / BZR algorithm. \\
\hline
\end{tabular}
\end{table}

\clearpage
\ecompanionsection{Proofs} \label{appendix:proofs}

\noindent \textbf{Proof of Theorem \ref{thm:rrr-one}}
\begin{proof}
(i) For attempt $r$, let
\(
I_r \;:=\; \mathbf{1}\{\tau_{\mathrm{PNE}}\le R \text{ in attempt } r\},
\)
i.e., $I_r=1$ iff the $r$-th attempt reaches (and certifies) a PNE within at most
$R$ rounds. Across attempts we independently resample $s\sim\mu$ and re-draw fresh
round permutations; hence $(I_r)_{r\ge1}$ are i.i.d.\ Bernoulli with parameter
$q_\mu[R]$. Let
\[
R^\ast:=\inf\{r\ge1: I_r=1\}
\]
be the number of attempts until success. Then
\[
\Pr(R^\ast>r)=(1-q_\mu[R])^r \xrightarrow[r\to\infty]{}\ 0,
\]
so $\Pr(R^\ast<\infty)=1$ and the algorithm finds a PNE almost surely.

\smallskip
(ii) Let $L_r$ be the number of BR solves in attempt $r$. For each failed attempt
$r<R^\ast$, exactly $R$ rounds are executed, so $L_r=nR$. On the successful attempt,
a PNE is first hit during one of the $R$ rounds, and the algorithm certifies it by
the end-of-round no-change test; thus $L_{R^\ast} \le nR$. Taking expectations and
using $\mathbb{E}[R^\ast]=1/q_\mu[R]$ (geometric),
$$
\mathbb{E}\!\Big[\sum_{r=1}^{R^\ast} L_r\Big]
\ \le\ nR\,\mathbb{E}[R^\ast]
\ =\ \frac{nR}{q_\mu[R]}.
$$
This establishes the bound in part (ii). 
\end{proof}

\medskip
\medskip
\noindent \textbf{Proof of Corollary \ref{cor:rrr-ipg}}
\begin{proof}
Because each $\mathcal{X}_i$ is a bounded integer polyhedron, it contains finitely many integer points, so
$\mathcal{X} = \prod_{i\in N} \mathcal{X}_i$ is finite and can be viewed as the pure strategy space $S$ of a
finite normal-form game. The RRR--BRD-IPG algorithm (Algorithm~\ref{alg:RRR_BRD_IPG}) is exactly RRR--BRD
(Algorithm~\ref{alg:rrrbrd}) applied to this finite game, with restart law $\mu$ induced by
\textsc{Random\_Generation}. If $\G$ admits a PNE and $q_\mu[R]>0$, then the hypotheses of
Theorem~\ref{thm:rrr-one} hold, so the almost-sure convergence and the bound on the expected number of BR
solves follow directly. 
\end{proof}

% \medskip
% \medskip
\clearpage
\noindent \textbf{Proof of Proposition \ref{lemma:selfish_sum}}
\begin{proof}
By Definition~\ref{def:Induced Subgraph}, $\mathcal A^-[\I_i] := \delta_-(\I_i) \cup \mathcal{A}_i$. Fix any edge $(j,k)\in \A$. Since $\I=\bigsqcup_{i\in N}\I_i$ is a partition, there is a \emph{unique} index $i\in N$ such that $k\in \I_i$. There are two cases:
(i) if $j\in \I_i$ as well, then $(j,k)\in \mathcal A_i \subseteq \mathcal A^-[\I_i]$;
(ii) if $j\notin \I_i$, then $(j,k)\in \delta_-(\I_i)\subseteq \mathcal A^-[\I_i]$.
Hence every edge $(j,k)\in \A$ belongs to $\bigcup_{i\in N}\mathcal A^-[\I_i]$.
Furthermore, since each edge $(j,k)$ is unique, it must exclusively belong to one set $\mathcal{A}^-[\I_i]$. Therefore, $\mathcal A = \bigsqcup_{i \in N} \mathcal A^-[\I_i]$.

Using this partition and the definition of $u_i^{\text{Self}}$, we obtain
$$
\phi(\x)
= \sum_{(j,k) \in \A} w_{jk} y_{jk}
= \sum_{i \in N} \sum_{(j,k) \in \A^-[\I_i]} w_{jk} y_{jk}
= \sum_{i \in N} \tilde{u}_i^{\text{Self}}(\y)
= \sum_{i \in N} u_i^{\text{Self}}(\x),
$$
which proves the claim.
\end{proof}

\medskip
\medskip
\noindent \textbf{Proof of Theorem \ref{thm:exact_potential_game}}
\begin{proof}
Fix a player $i \in N$. The arcs of the entire graph can be partitioned into those that are incident to $\I_i$ and those that are not, i.e.,
\(
\A = \A[\I_i]  \sqcup \A_{\noti},
\)
where $\A[\I_i] = \A_i \cup \delta_+(\I_i) \cup \delta_-(\I_i)$ and $\A_{\noti}$ collects all remaining arcs. For a given profile $(\x^i,\bar\x^{\noti})$, the potential can be written as
$$
\phi(\x) = \smashoperator[lr]{\sum_{j,k \in \I_i}} w_{jk} \cdot \max(x_j,x_k) + \smashoperator[lr]{\sum_{j \in \I_i, k \in \I_{\noti}}} w_{jk} \cdot \max(x_j,\bar{x}_k)  + \smashoperator[lr]{\sum_{j \in \I_{\noti},k \in \I_i}} w_{jk}\cdot \max(\bar{x}_j,x_k) + \smashoperator[lr]{\sum_{j,k \in \I_{\noti}}} w_{jk} \cdot \max(\bar{x}_j,\bar{x}_k)
$$
$$
\qquad = \sum_{(j,k) \in \A[\I_i]}  u_i^{\text{Alt}}(\x^{i}, \bar\x^{\noti})  + \smashoperator[lr]{\sum_{j,k \in \I_{\noti}}} w_{jk} \cdot \max(\bar{x}_j,\bar{x}_k).
$$
The second term depends only on $\bar\x^{\noti}$ and is therefore constant with respect to changes in $\x^i$. The first term is precisely $u_i^{\text{Alt}}(\x^i,\bar\x^{\noti})$ (where $\tilde u_i^{\text{Alt}}(\y)$ is rewritten as $u_i^{\text{Alt}}(\x)$). Hence, for any $\bar\x^i,\hat\x^i \in \X_i$,
$$
\phi(\hat\x^i,\bar\x^{\noti}) - \phi(\bar\x^i,\bar\x^{\noti})
=
u_i^{\text{Alt}}(\hat\x^i,\bar\x^{\noti})
-
u_i^{\text{Alt}}(\bar\x^i,\bar\x^{\noti}),
$$
which shows that $\phi$ is an exact potential function for the locally altruistic game. 
\end{proof}

\medskip
\medskip
\noindent \textbf{Proof of Corollary \ref{cor:SW_is_PNE}}
\begin{proof}
By Theorem~\ref{thm:exact_potential_game}, the locally altruistic EBMC games are exact potential games with potential function $\phi(\x)$, and $\barxosw$ is an optimal solution maximizing $\phi(\x)$ over $\x \in \mathcal{X}$. By Lemma~2.1 of \citet{monderer1996potential}, which states that any maximizer of a potential function is a PNE, $\barxosw$ is a PNE for locally altruistic EBMC games. 
\end{proof}

\clearpage
% \medskip
% \medskip
\noindent \textbf{Proof of Theorem~\ref{thm:PNE-existence}.}
\begin{proof}
\textbf{Case 1:} $\I^{\text{ng}}\subseteq \I^{0}$.
Fix $i\in N$. Since vertices in $\I^0$ have no types, other players’ selections in $\I^0$ create no relevant coverage benefit for player $i$ in the selfish utility. Hence, for all $\x^i\in\mathcal X_i$,
\[
u_i^{\text{self}}(\x^i,\bar\x_{\text{ng}}^{-i}) \;=\; u_i^{\text{self}}(\x^i,\mathbf{0}^{-i}).
\]
By definition of $\bar\x_{\text{ng}}^i$,
\[
\bar\x_{\text{ng}}^i \in \arg\max_{\x^i\in\mathcal X_i} u_i^{\text{self}}(\x^i,\mathbf{0}^{-i}),
\]
so $\bar\x_{\text{ng}}^i$ is also a best response to $\bar\x_{\text{ng}}^{-i}$. Thus $\bar\x_{\text{ng}}$ is a PNE (for both $|\mathcal L|=1$ and $|\mathcal L|\ge 2$).

\medskip
\medskip
\textbf{Case 2:} $\I^{\text{ng}}\subseteq \I^{+}$ and $|\mathcal L|=1$.
Fix $i\in N$. For any $\x^i\in\mathcal X_i$, overlap of covered edges implies
\[
u_i^{\text{self}}(\x^i,\bar\x_{\text{ng}}^{-i})
\;\le\;
u_i^{\text{self}}(\x^i,\mathbf{0}^{-i})
\;+\;
u_i^{\text{self}}(\mathbf{0}^{i},\bar\x_{\text{ng}}^{-i}).
\]
Since $\bar\x_{\text{ng}}^i$ maximizes $u_i^{\text{self}}(\cdot,\mathbf{0}^{-i})$,
\[
u_i^{\text{self}}(\x^i,\bar\x_{\text{ng}}^{-i})
\;\le\;
u_i^{\text{self}}(\bar\x_{\text{ng}}^i,\mathbf{0}^{-i})
\;+\;
u_i^{\text{self}}(\mathbf{0}^{i},\bar\x_{\text{ng}}^{-i}).
\]
Under $\mathcal G^{\text{Self}}_{|\mathcal L|=1}$, the instance is bipartite as in Section~\ref{sec:EBMC_types}, with each coverable edge having one endpoint in $\I^{+}$ and one in $\I^{0}$. With $\I^{\text{ng}}\subseteq\I^{+}$, there is no double counting of edges, so
\[
u_i^{\text{self}}(\bar\x_{\text{ng}}^i,\mathbf{0}^{-i})
+
u_i^{\text{self}}(\mathbf{0}^{i},\bar\x_{\text{ng}}^{-i})
=
u_i^{\text{self}}(\bar\x_{\text{ng}}^i,\bar\x_{\text{ng}}^{-i}).
\]
Therefore,
\[
u_i^{\text{self}}(\x^i,\bar\x_{\text{ng}}^{-i})
\;\le\;
u_i^{\text{self}}(\bar\x_{\text{ng}}^i,\bar\x_{\text{ng}}^{-i})
\qquad \forall \x^i\in\mathcal X_i,
\]
so $\bar\x_{\text{ng}}^i$ is a best response to $\bar\x_{\text{ng}}^{-i}$. Since $i$ is arbitrary, $\bar\x_{\text{ng}}$ is a PNE for $\mathcal G^{\text{Self}}_{|\mathcal L|=1}$. 
\end{proof}

\clearpage
\ecompanionsection{Random Feasible-Start Generators for KPG and EBMC}
\label{appendix:feas_generator}

We provide two randomized feasible-start generators for each IPG class studied in this paper:
(i) a \textit{maximal} generator, which tends to produce strategies that are close to capacity-filling and thus resemble typical best responses; and
(ii) a \textit{full-support} generator, which assigns strictly positive probability to every feasible pure strategy.
In our RRR--BRD implementation, we alternate between these two variants (e.g., using the maximal generator on odd-numbered restarts and the full-support generator on even-numbered restarts), although any mixture is admissible.

The design goals are:
(a) every feasible profile in $\mathcal{X}$ has nonzero probability under the full-support generator (for the theoretical guarantees in Section~\ref{subsec:rrrbrd}); and
(b) the generators remain lightweight (expected linear time) so that restarts are inexpensive.

\paragraph*{Notation.}
Let $N=\{1,\dots,n\}$ be the set of players/counties.
In KPG, let $J$ be the set of items, $w^i_{j}\ge 0$ the weight of item $j \in J$ for player $i\in N$, and $\B_i\ge 0$ the capacity.
A pure strategy for player $i$ is $\x^i\in\{0,1\}^{|J|}$ with
\(
\sum_{j\in J} w^i_{j}\, x^i_j \le \B_i.
\)
In EBMC, let $\I_i$ denote the lakes in county $i$ and $\B_i \in\mathbb{Z}_{\ge 0}$ its inspection budget; a pure strategy is $\x^i\in\{0,1\}^{|\I_i|}$ with
\(
\sum_{j \in \I_i} x_j \le \B_i.
\)

\subsection*{Generators for Knapsack Problem Games (KPG)}

\noindent\textbf{Maximal generator (random-order greedy).}
This generator matches the implementation used in our code: for each player, we take a randomized ordering of items and perform a single greedy pass, adding an item whenever it fits in the remaining capacity.

\begin{algorithm}[h!]
\caption{\textsc{MaxRandomGeneration-KPG} (random-order greedy)}
\label{alg:RG_KPG_max}
\begin{algorithmic}[1]
\small
\Require Items $J$, capacities $(\B_i)_{i\in N}$, weights $(w^i_{j})_{i\in N,\,j\in J}$
\Ensure Feasible profile $\bar{\x}\in \prod_{i\in N}\{\x^i\in\{0,1\}^{|J|}:\sum_j w^i_{j}x^i_j\le \B_i\}$
\For{\textbf{each} $i\in N$}
    \State $S \gets \emptyset$, \ $R \gets \B_i$
    \State $\pi \gets \Call{RandomPermutation}{J}$  \Comment{encodes shuffle}
    \For{\textbf{each} $j \in \pi$}
        \State $w \gets w^i_{j}$
        \If{$w \le R$}
            \State $S \gets S \cup \{j\}$; \ $R \gets R - w$
        \EndIf
    \EndFor
    \State Set $x^i_j \gets \mathbf{1}\{j\in S\}$ for all $j\in J$
\EndFor
\State \Return $\bar{\x} = (\x^i)_{i\in N}$
\end{algorithmic}
\end{algorithm}

\noindent
Algorithm~\ref{alg:RG_KPG_max} runs in $O(|N|\,|J|)$ time and always returns a feasible strategy for each player, since items are only accepted when the remaining capacity $R$ allows. Because all randomness sits in the permutation $\pi$, it yields a diverse collection of near-maximal solutions that typically resemble best responses.

\medskip
\noindent\textbf{Full-support generator (Bernoulli + rejection).}
The full-support generator uses a Bernoulli--rejection scheme. For each player $i$, we choose a Bernoulli parameter $p_i$ that reflects the ``density'' of the knapsack, typically on the order of the mean fill ratio $\B_i / W_i$, where $W_i := \sum_{j\in J} w^i_j$:

\begin{algorithm}[h!]
\caption{\textsc{RandomGeneration-KPG} (full-support, expected linear time)}
\label{alg:RG_KPG_full}
\begin{algorithmic}[1]
\small
\Require Items $J$, capacities $(\B_i)_{i\in N}$, weights $(w^i_{j})_{i\in N,\,j\in J}$
\Ensure Feasible profile $\bar{\x}\in \prod_{i\in N}\{\x^i\in\{0,1\}^{|J|}:\sum_j w^i_{j}x^i_j\le \B_i\}$
\For{\textbf{each} $i\in N$}
    \State $W_i \gets \sum_{j\in J} w^i_j$
    \State Choose $p_i$ in $(0,1)$, e.g.\ $p_i \gets \B_i / W_i$
    \Repeat
        \For{\textbf{each} $j \in J$}
            \State Draw $x^i_j \sim \mathrm{Ber}(p_i)$ independently
        \EndFor
    \Until{$\sum_{j\in J} w^i_j x^i_j \le \B_i$}  \Comment{reject if capacity violated}
\EndFor
\State \Return $\bar{\x} = (\x^i)_{i\in N}$
\end{algorithmic}
\end{algorithm}

\noindent
Every feasible knapsack pattern for player $i$ can be generated with positive probability in a single Bernoulli draw (because each coordinate is independent and $p_i \in (0,1)$), and is accepted whenever it satisfies the capacity constraint. Thus, after accounting for the rejection step, \textsc{RandomGeneration-KPG} induces a full-support distribution over the feasible set. For moderate $p_i$ and capacities, the expected number of rejection loops is small, so the expected runtime remains linear in $|J|$ per player.

In RRR--BRD, we alternate these two generators across restarts (e.g., \textsc{MaxRandomGeneration-KPG} for odd restarts and \textsc{RandomGeneration-KPG} for even restarts), combining capacity-filling starts with fully supported random perturbations.

\subsection*{Generators for EBMC}

For EBMC, budgets are cardinality bounds, so we can work directly with subsets of lakes.

\medskip
\noindent\textbf{Maximal generator (uniform cardinality sampling).}
Here we sample a subset of lakes of size $b_i := \min\{\B_i,|\I_i|\}$ uniformly without replacement.

\begin{algorithm}[h!]
\caption{\textsc{MaxRandomGeneration-EBMC} (uniform cardinality)}
\label{alg:RG_EBMC_max}
\begin{algorithmic}[1]
\small
\Require County lakes $(\I_i)_{i\in N}$, budgets $(\B_i)_{i\in N}$
\Ensure Feasible profile with $\sum_{j \in \I_i} x_j \le \B_i$ for all $i\in N$
\For{\textbf{each} $i\in N$}
    \State $b_i \gets \min\{\B_i,\,|\I_i|\}$
    \State $S \gets \Call{SampleWithoutReplacement}{\I_i,\,b_i}$ \Comment{uniform over $\binom{|\I_i|}{b_i}$}
    \State Set $x_j \gets \mathbf{1}\{j\in S\}$ for all $j\in \I_i$
\EndFor
\State \Return $\bar{\x} = (\x^i)_{i\in N}$
\end{algorithmic}
\end{algorithm}

\noindent
This generator runs in $O(|\I_i|)$ per county (via a shuffle-and-take routine) and samples uniformly among all maximal (cardinality-$b_i$) feasible strategies.

\medskip
\noindent\textbf{Full-support generator (Bernoulli + rejection).}
To obtain full support in EBMC, we again use a Bernoulli--rejection scheme. Here the natural density proxy is the ratio $\B_i / |\I_i|$:

\begin{algorithm}[h!]
\caption{\textsc{RandomGeneration-EBMC} (full-support)}
\label{alg:RG_EBMC_full}
\begin{algorithmic}[1]
\small
\Require County lakes $(\I_i)_{i\in N}$, budgets $(\B_i)_{i\in N}$
\Ensure Feasible profile with $\sum_{j \in \I_i} x_j \le \B_i$ for all $i\in N$
\For{\textbf{each} $i\in N$}
    \State Choose $p_i$ in $(0,1)$, e.g.\ $p_i \gets \B_i / |\I_i|$
    \Repeat
        \For{\textbf{each} $j \in \I_i$}
            \State Draw $x_j \sim \mathrm{Ber}(p_i)$ independently
        \EndFor
    \Until{$\sum_{j\in \I_i} x_j \le \B_i$}
\EndFor
\State \Return $\bar{\x} = (\x^i)_{i\in N}$
\end{algorithmic}
\end{algorithm}

\noindent
As in the KPG case, every feasible subset of $\I_i$ has positive probability under this procedure, and the expected number of rejection loops remains small for moderate $p_i$ and budgets.

\medskip
\medskip
\noindent\textbf{Remark.}
In summary, for both KPG and EBMC we alternate between maximal and
full-support generators across restarts. Formally, let $\mu^{\max}$ and
$\mu^{\mathrm{full}}$ denote the restart laws on $\mathcal{X}$ induced
when all restarts use the maximal generator and the full-support
generator, respectively. If a PNE exists and the full-support
generator is used (so the restart law is $\mu^{\mathrm{full}}$), then
$q_{\mu^{\mathrm{full}}}[R] > 0$ for all $R \ge 1$. In practice, however, the maximal generator often produces starting
profiles that are closer to best responses and can empirically yield
larger success probabilities ($q_{\mu^{\mathrm{max}}}[R] >$ 
$q_{\mu^{\mathrm{full}}}[R]$), even though these quantities are not
computable in closed form. Alternating the two variants balances
theoretical robustness (via the full-support generator) with the
potential efficiency of more structured, capacity-filling starts.

\clearpage
\ecompanionsection{Extremely Small $q_\mu[R]$} \label{appendix:q_mu_R_small}

\begin{figure}[h!]
\centering
\resizebox{0.5\textwidth}{!}{%
\begin{tikzpicture}[
  >=Latex,
  node distance=12mm and 18mm,
  label distance=2pt,
  every label/.style={font=\footnotesize},
  scc/.style={draw, rounded corners, minimum width=26mm, minimum height=10mm, align=center},
  sink/.style={draw, double, circle, minimum size=10mm, align=center, thick},
  arr/.style={->, line width=0.6pt}
]

% --- Many transient SCC singletons feeding into S ---
\node[scc, label=below:{many singleton SCCs}] (T) {};

% --- Non-PNE sink S (cycle) ---
\node[sink, right=30mm of T, label=below:{non\mbox{-}PNE sink $O$}] (S) {};

% --- Isolated PNE sink P (singleton) ---
\node[sink, below=15mm of S, label=above:{PNE sink $P$}] (P) {};

% --- Edge into S; P isolated ---
\draw[arr] (T) -- (S);

% --- Internal dots: show that T is a big collection; S is a cycle; P is singleton ---
% T: lots of dots (schematic)
\fill ($(T.center)+(-7pt,2pt)$) circle (1.2pt);
\fill ($(T.center)+(-1pt,4pt)$) circle (1.2pt);
\fill ($(T.center)+(6pt,2pt)$) circle (1.2pt);
\fill ($(T.center)+(-5pt,-3pt)$) circle (1.2pt);
\fill ($(T.center)+(3pt,-3pt)$) circle (1.2pt);

% S: cycle dots
\fill ($(S.center)+(-4pt,3pt)$) circle (1.2pt) coordinate (s1);
\fill ($(S.center)+(4pt,3pt)$)  circle (1.2pt) coordinate (s2);
\fill ($(S.center)+(0pt,-4pt)$) circle (1.2pt) coordinate (s3);
\draw[->, shorten >=1pt, shorten <=1pt] (s1) .. controls +(-6pt,6pt) and +(-6pt,6pt) .. (s2);
\draw[->, shorten >=1pt, shorten <=1pt] (s2) .. controls +(6pt,-8pt) and +(6pt,-8pt) .. (s3);
\draw[->, shorten >=1pt, shorten <=1pt] (s3) .. controls +(-8pt,-6pt) and +(-8pt,-6pt) .. (s1);

% P: singleton dot
\fill ($(P.center)+(0pt,0pt)$) circle (1.2pt);

\end{tikzpicture}%
}
\caption{A Condensation DAG Showing That $q_\mu[R]$ Can Be Extremely Small.}
\label{fig:connected-condensation-dag_small}
\end{figure}

Figure~\ref{fig:connected-condensation-dag_small} highlights that the instance-dependent success probability
$q_\mu[R]$ \eqref{eq:q_mu_R} in RRR--BRD can be extremely small even when a PNE exists. Assuming that the restart law
$\mu$ has full support on the restart set and that $R$ is sufficiently large, the diagram can be interpreted as
the condensation DAG of the best-response state graph. In the schematic, $O$ is an absorbing non-PNE sink
SCC, $P$ is an absorbing PNE sink SCC, and the remaining node(s) represent (possibly many) transient SCCs
that flow into $O$. In particular, $P$ may be a singleton and may be \emph{unreachable} from any non-PNE start
under best-response moves, so that a run succeeds only if the restart lands directly in $P$. Under full-support
restart laws (e.g., uniform sampling over feasible profiles), the success probability $q_\mu[R]$ can therefore be
made arbitrarily small by scaling the number of transient states (and hence their $\mu$-mass) relative to $P$.

\paragraph*{A knapsack-style IPG inducing vanishing $q_\mu[R]$.}
We now give a concrete IPG with a single knapsack constraint per player for which $q_\mu[R]$ decreases
exponentially in $M$, matching the structure in Figure~\ref{fig:connected-condensation-dag_small}.
Consider a two-player IPG ($N=\{1,2\}$) with $m:=M$ binary decisions indexed by $j\in\{1,\dots,M\}$.
Each player $i$ chooses a binary vector $\x^i\in\{0,1\}^m$ subject to the single knapsack constraint
\[
\sum_{j=1}^M x^i_j \le M,
\]
so every subset is feasible.

We use the standard IPG utility form
\[
u_i(\x^i,\x^{\noti}) = (\mathbf{d}^i)^\top \x^i + \sum_{p\in N\setminus\{i\}} (\x^p)^\top Q_p^i \x^i,
\]
with $Q^i\equiv 0$ (no within-player quadratic term).
Define the cyclic successor map on indices
$g:\{1,\dots,M\}\to\{1,\dots,M\}$ by $g(j)=j+1$ for $j=1,\dots,M-1$ and $g(M)=1$.

Set linear coefficients
\(
\mathbf{d}^1=\mathbf{d}^2=-c\,\mathbf{1} \ \text{for some }c>0,
\)
and define the interaction matrices by the following nonzero entries (all others are $0$):
$$
Q_2^1[j,j]=T\ (j=1,\dots,M),\qquad
Q_1^2[g(j),j]=T\ (j=1,\dots,M),
$$
for some $T>c$. Under these utilities, player~1’s unique best response to $\x^2$ is to choose exactly the same
subset $\x^1=\x^2$, and player~2’s unique best response to $\x^1$ is to choose the shifted subset
$x^2_{g(j)}=x^1_j$ for all $j$.

\paragraph*{Identification of the sinks $O$ and $P$ and the transient region.}
The profile $(\x^1,\x^2)=(\mathbf{0},\mathbf{0})$ is a PNE: each player obtains payoff $0$, while any deviation to a
nonzero subset yields negative payoff due to the linear cost. Thus the singleton $\{(\mathbf{0},\mathbf{0})\}$ is
the PNE sink $P$ in Figure~\ref{fig:connected-condensation-dag_small}.

In contrast, no profile with $\x^1\neq\mathbf{0}$ or $\x^2\neq\mathbf{0}$ can be a PNE. If $\x^2\neq\mathbf{0}$,
then player~1’s unique best response is $\x^1=\x^2\neq\mathbf{0}$. If $\x^1\neq\mathbf{0}$, then player~2’s unique
best response is the shifted subset $g(\x^1)\neq \x^1$, and hence $\x^2$ must be updated unless it already equals
that shift. Restricting to the ``active'' profiles with $\x^1\neq\mathbf{0}$ and $\x^2\neq\mathbf{0}$, best-response
updates cycle on the orbit induced by $g$, forming an absorbing strongly connected component $O$ containing no PNE.
All other non-PNE profiles are transient singleton SCCs that flow into $O$ under improving best responses. This
matches the schematic: a (possibly very large) transient region feeding into the non-PNE sink $O$, with a separate
singleton PNE sink $P$.

Moreover, $(\mathbf{0},\mathbf{0})$ is not reachable from any non-PNE start under improving best-response moves:
when the opponent plays a nonzero subset, deviating to $\mathbf{0}$ yields payoff $0$, whereas the prescribed best
response yields strictly positive payoff because $T>c$. Thus, all non-PNE starts eventually enter $O$ and remain
there.

\paragraph*{Implication for $q_\mu[R]$ under full-support restarts.}
Let the restart law $\mu$ have full support over the restart set, e.g., uniform over
$\{0,1\}^m\times\{0,1\}^m$. In this instance, $P=\{(\mathbf{0},\mathbf{0})\}$ is not reachable from any non-PNE start
under improving best responses, whereas all other starts enter the non-PNE sink $O$ and fail for any finite round
cap $R$. Consequently,
\[
q_\mu[R]=\mu(P)=\mu\big((\mathbf{0},\mathbf{0})\big)=\frac{1}{2^{2M}},
\]
which can be made arbitrarily small by increasing $M$. Equivalently, the expected number of restarts required to hit
a PNE is $1/q_\mu[R] = 2^{2M}$, so RRR--BRD (Algorithm~\ref{alg:RRR_BRD_IPG}) can be exponentially weak in
expectation on this instance.

Moreover, since the number of restarts until success is geometric, tail probabilities are non-negligible:
\[
\Pr(R^\ast>r)=(1-q_\mu[R])^r.
\]
With \(q_\mu[R]=2^{-2M}\), we have
\[
\Pr(R^\ast>2^{2M})=(1-2^{-2M})^{2^{2M}}\approx e^{-1},
\]
so even at the mean scale \(2^{2M}\) there remains a constant probability of requiring more restarts (and hence more
work).

\clearpage
\ecompanionsection{Full Simulation Results} \label{appendix:full_simulation}
Tables~\ref{tab:BRR_BRD_normal_form_full}--\ref{tab:BRR_BRD_KPGs_full} report full simulation results for the normal-form reductions and KPG instances. Most settings exhibit large $\tilde q_\mu[R]$ and small average rounds, while a minority of instances have much smaller $\tilde q_\mu[R]$, leading to larger ESB and longer runs; this instance dependence is most visible in the harder KPG configurations (notably type C).

\begin{table}[h!]
    \caption{Full Simulation Results for Normal-form reductions}
    \label{tab:BRR_BRD_normal_form_full}
    \centering
    \scalebox{0.85}{
    \begin{tabular}{cccrrrrrr}
    \toprule
    $n$ & $k$ & inst &  trials & successes & $\tilde{q_\mu}[R]$ & avg\_rounds\_all & ESB ($\frac{nR}{q_\mu[R]}$) & Time (secs) \\ \midrule
    2 & 3 & 1  & 64   & 64   & 1.00 & 2.84 & 20.0  & 0.01 \\
    2 & 3 & 2  & 64   & 64   & 1.00 & 2.83 & 20.0  & 0.00 \\
    2 & 3 & 3  & 64   & 64   & 1.00 & 2.52 & 20.0  & 0.00 \\
    2 & 3 & 4  & 64   & 64   & 1.00 & 2.58 & 20.0  & 0.00 \\
    2 & 3 & 5  & 64   & 64   & 1.00 & 2.66 & 20.0  & 0.00 \\
    2 & 3 & 6  & 64   & 64   & 1.00 & 2.44 & 20.0  & 0.00 \\
    2 & 3 & 7  & 64   & 64   & 1.00 & 2.73 & 20.0  & 0.00 \\
    2 & 3 & 8  & 64   & 64   & 1.00 & 2.63 & 20.0  & 0.00 \\
    2 & 3 & 9  & 64   & 64   & 1.00 & 2.39 & 20.0  & 0.00 \\
    2 & 3 & 10 & 64   & 64   & 1.00 & 2.34 & 20.0  & 0.00 \\
    2 & 4 & 1  & 256  & 256  & 1.00 & 2.34 & 20.0  & 0.02 \\
    2 & 4 & 2  & 256  & 64   & 0.25 & 8.04 & 80.0  & 0.11 \\
    2 & 4 & 3  & 256  & 256  & 1.00 & 3.18 & 20.0  & 0.05 \\
    2 & 4 & 4  & 256  & 256  & 1.00 & 2.62 & 20.0  & 0.03 \\
    2 & 4 & 5  & 256  & 256  & 1.00 & 2.84 & 20.0  & 0.05 \\
    2 & 4 & 6  & 256  & 256  & 1.00 & 2.88 & 20.0  & 0.04 \\
    2 & 4 & 7  & 256  & 256  & 1.00 & 2.20 & 20.0  & 0.02 \\
    2 & 4 & 8  & 256  & 256  & 1.00 & 3.63 & 20.0  & 0.05 \\
    2 & 4 & 9  & 256  & 256  & 1.00 & 2.13 & 20.0  & 0.02 \\
    2 & 4 & 10 & 256  & 256  & 1.00 & 2.41 & 20.0  & 0.04 \\
    3 & 3 & 1  & 512  & 410  & 0.80 & 6.32 & 37.5  & 0.13 \\
    3 & 3 & 2  & 512  & 287  & 0.56 & 6.58 & 53.5  & 0.12 \\
    3 & 3 & 3  & 512  & 176  & 0.34 & 7.56 & 87.3  & 0.14 \\
    3 & 3 & 4  & 512  & 332  & 0.65 & 5.86 & 46.3  & 0.13 \\
    3 & 3 & 5  & 512  & 45   & 0.09 & 9.37 & 341.3 & 0.16 \\
    3 & 3 & 6  & 512  & 512  & 1.00 & 2.89 & 30.0  & 0.06 \\
    3 & 3 & 7  & 512  & 69   & 0.13 & 8.98 & 222.6 & 0.20 \\
    3 & 3 & 8  & 512  & 334  & 0.65 & 5.79 & 46.0  & 0.12 \\
    3 & 3 & 9  & 512  & 310  & 0.61 & 6.66 & 49.5  & 0.14 \\
    3 & 3 & 10 & 512  & 512  & 1.00 & 3.80 & 30.0  & 0.09 \\
    3 & 4 & 1  & 4096 & 4096 & 1.00 & 3.42 & 30.0  & 1.09 \\
    3 & 4 & 2  & 4096 & 2542 & 0.62 & 6.86 & 48.3  & 2.20 \\
    3 & 4 & 3  & 4096 & 612  & 0.15 & 9.31 & 200.8 & 2.95 \\
    3 & 4 & 4  & 4096 & 3173 & 0.77 & 6.04 & 38.7  & 1.92 \\
    3 & 4 & 5  & 4096 & 4096 & 1.00 & 3.41 & 30.0  & 1.16 \\
    3 & 4 & 6  & 4096 & 2542 & 0.62 & 6.47 & 48.3  & 2.10 \\
    3 & 4 & 7  & 4096 & 3566 & 0.87 & 4.65 & 34.5  & 1.59 \\
    3 & 4 & 8  & 4096 & 943  & 0.23 & 8.36 & 130.3 & 2.62 \\
    3 & 4 & 9  & 4096 & 4096 & 1.00 & 3.93 & 30.0  & 1.30 \\
    3 & 4 & 10 & 4096 & 3702 & 0.90 & 4.58 & 33.2  & 1.61  \\
    \bottomrule
    \end{tabular}   
    }
\end{table}

\begin{table}[h!]
\caption{Full Simulation Results for KPGs}
\label{tab:BRR_BRD_KPGs_full}
    \centering
    \scalebox{0.85}{
    \begin{tabular}{ccrrrrrrrr}
    \toprule
    type  & $n$ & items   & budget & trials & successes  & $\tilde q_\mu[R]$ & avg\_rounds\_all & ESB ($\frac{nR}{q_\mu[R]}$) &  Time (secs) \\ \midrule
    A & 2 & 25  & 2 & 200 & 200 & 1.00 & 2.50  & 40.0    & 0.30 \\
    A & 2 & 25  & 5 & 200 & 200 & 1.00 & 2.72  & 40.0    & 0.34 \\
    A & 2 & 25  & 8 & 200 & 200 & 1.00 & 2.94  & 40.0    & 0.44 \\
    A & 2 & 50  & 2 & 200 & 200 & 1.00 & 3.18  & 40.0    & 0.58 \\
    A & 2 & 50  & 5 & 200 & 200 & 1.00 & 2.58  & 40.0    & 0.56 \\
    A & 2 & 50  & 8 & 200 & 200 & 1.00 & 2.71  & 40.0    & 0.66 \\
    A & 2 & 75  & 2 & 200 & 200 & 1.00 & 3.25  & 40.0    & 0.75 \\
    A & 2 & 75  & 5 & 200 & 200 & 1.00 & 3.57  & 40.0    & 1.14 \\
    A & 2 & 75  & 8 & 200 & 200 & 1.00 & 3.46  & 40.0    & 1.62 \\
    A & 2 & 100 & 2 & 200 & 200 & 1.00 & 3.26  & 40.0    & 1.03 \\
    A & 2 & 100 & 5 & 200 & 200 & 1.00 & 3.66  & 40.0    & 1.56 \\
    A & 2 & 100 & 8 & 200 & 200 & 1.00 & 3.17  & 40.0    & 4.50 \\
    A & 3 & 25  & 2 & 200 & 200 & 1.00 & 3.21  & 60.0    & 0.60 \\
    A & 3 & 25  & 5 & 200 & 200 & 1.00 & 3.40  & 60.0    & 0.79 \\
    A & 3 & 25  & 8 & 200 & 200 & 1.00 & 3.15  & 60.0    & 0.75 \\
    A & 3 & 50  & 2 & 200 & 200 & 1.00 & 3.59  & 60.0    & 0.99 \\
    A & 3 & 50  & 5 & 200 & 200 & 1.00 & 3.17  & 60.0    & 1.25 \\
    A & 3 & 50  & 8 & 200 & 200 & 1.00 & 3.78  & 60.0    & 1.56 \\
    A & 3 & 75  & 2 & 200 & 200 & 1.00 & 4.34  & 60.0    & 1.71 \\
    A & 3 & 75  & 5 & 200 & 200 & 1.00 & 3.42  & 60.0    & 1.81 \\
    A & 3 & 75  & 8 & 200 & 200 & 1.00 & 4.13  & 60.0    & 2.35 \\
    A & 3 & 100 & 5 & 200 & 200 & 1.00 & 4.66  & 60.0    & 7.34 \\
    A & 3 & 100 & 8 & 200 & 200 & 1.00 & 4.02  & 60.0    & 9.56 \\
    B & 2 & 25  & 2 & 200 & 200 & 1.00 & 3.08  & 40.0    & 0.34 \\
    B & 2 & 25  & 5 & 200 & 200 & 1.00 & 3.22  & 40.0    & 0.45 \\
    B & 2 & 25  & 8 & 200 & 200 & 1.00 & 2.85  & 40.0    & 0.42 \\
    B & 2 & 50  & 2 & 200 & 200 & 1.00 & 3.52  & 40.0    & 0.63 \\
    B & 2 & 50  & 5 & 200 & 200 & 1.00 & 3.71  & 40.0    & 0.87 \\
    B & 2 & 50  & 8 & 200 & 200 & 1.00 & 3.04  & 40.0    & 0.70 \\
    B & 2 & 75  & 2 & 200 & 200 & 1.00 & 3.28  & 40.0    & 0.88 \\
    B & 2 & 75  & 5 & 200 & 200 & 1.00 & 3.97  & 40.0    & 1.21 \\
    B & 2 & 75  & 8 & 200 & 200 & 1.00 & 3.69  & 40.0    & 2.95 \\
    B & 2 & 100 & 2 & 200 & 200 & 1.00 & 3.26  & 40.0    & 1.08 \\
    B & 2 & 100 & 5 & 200 & 200 & 1.00 & 3.58  & 40.0    & 5.61 \\
    B & 2 & 100 & 8 & 200 & 200 & 1.00 & 3.31  & 40.0    & 5.99 \\
    B & 3 & 25  & 2 & 200 & 200 & 1.00 & 3.45  & 60.0    & 0.82 \\
    B & 3 & 25  & 5 & 200 & 200 & 1.00 & 3.64  & 60.0    & 0.83 \\
    B & 3 & 25  & 8 & 200 & 200 & 1.00 & 3.46  & 60.0    & 0.74 \\
    B & 3 & 50  & 2 & 200 & 200 & 1.00 & 4.05  & 60.0    & 1.26 \\
    B & 3 & 50  & 5 & 200 & 200 & 1.00 & 4.43  & 60.0    & 1.57 \\
    B & 3 & 50  & 8 & 200 & 200 & 1.00 & 3.63  & 60.0    & 1.34 \\
    B & 3 & 75  & 5 & 200 & 200 & 1.00 & 4.55  & 60.0    & 2.19 \\
    B & 3 & 75  & 8 & 200 & 200 & 1.00 & 4.11  & 60.0    & 2.37 \\
    C & 2 & 25  & 8 & 200 & 200 & 1.00 & 2.62  & 40.0    & 0.27 \\
    C & 2 & 50  & 2 & 200 & 200 & 1.00 & 3.66  & 40.0    & 0.63 \\
    C & 2 & 50  & 8 & 200 & 200 & 1.00 & 3.04  & 40.0    & 0.38 \\
    C & 2 & 75  & 2 & 200 & 3   & 0.02 & 19.75 & 2666.7  & 5.36 \\
    C & 2 & 75  & 8 & 200 & 74  & 0.37 & 13.85 & 108.1   & 2.92 \\
    C & 2 & 100 & 2 & 200 & 156 & 0.78 & 7.53  & 51.3    & 2.59 \\
    C & 3 & 25  & 2 & 200 & 1   & 0.01 & 19.92 & 12000.0 & 3.42  \\
    \bottomrule
\end{tabular}
}
\end{table}

\clearpage
\ecompanionsection{Example for Best-Response Dynamics incorporated Zero-Regret Algorithm} \label{appendix:BZR_example}

We exemplify how RRR--BRD, BZR, and ZR operate on a small two-player instance. In this example the performance gap is not pronounced, but the mechanics are transparent.

$$
\textbf{Player 1:} \quad
\max_{\x^1 \in \{0,1\}^3} \quad 3x^1_1 + 2x^1_2 + 2x^1_3 - 4x^1_1 x^2_1  - 5 x^1_2 x^2_2 - 3 x^1_3 x^2_3   \quad \text{s.t.} \quad x^1_1 + x^1_2 + x^1_3 \leq 2.
$$
$$
\textbf{Player 2:} \quad
\max_{\x^2 \in \{0,1\}^3} \quad 2x^2_1 + x^2_2 + 0x^2_3 +5 x^2_1 x^1_1 + 2x^2_2 x^1_2 + x^2_3 x^1_3 \quad \text{s.t.} \quad x^2_1 + x^2_2 + x^2_3 \leq 2.
$$

\noindent
The game has a unique PNE
\(
\hat{\x}^*_{\text{pne}}= \{(0,0,1), (1,1,0)\}
\)
with social welfare \(\phi(\hat{\x}^*_{\text{pne}})=5\).
We use \(\barxzero=\{(0,0,0),(0,0,0)\}\) as a starting profile.

\smallskip
\noindent
\textbf{RRR--BRD:}
Assume the (round-wise) random playing sequences are
\([1,2] \rightarrow [2,1] \rightarrow [1,2]\).
Starting from \(\barxzero\), RRR--BRD follows the BR path
$$
\{(0,0,0),(0,0,0)\}
\xrightarrow{1}
\{(1,1,0),(0,0,0)\}
\xrightarrow{2}
\{(1,1,0),(1,1,0)\}
\xrightarrow{1}
\{(0,0,1),(1,1,0)\},
$$
and the final round \([1,2]\) certifies this profile via the end-of-round no-change test.

\smallskip
\noindent
\textbf{BZR:}
In our implementation, BZR first calls the separation oracle at an incumbent integer solution. Suppose the MIP solver produces
\(\{(0,0,0),(0,0,0)\}\).
This profile is not a PNE: player 1 has BR \((1,1,0)\) and player 2 has BR \((1,1,0)\), yielding two violated equilibrium inequalities (EIs), which are added:
$$
5-4x^2_1 -5x^2_2
\leq
3x^1_1 + 2x^1_2 + 2x^1_3 - 4z_1  - 5 z_2 - 3 z_3, \quad
3 + 5x^1_1 + 2x^1_2
\leq
2x^2_1 + x^2_2 + 0x^2_3 + 5z_1 + 2z_2 + z_3.
$$
Since \(\{(0,0,0),(0,0,0)\}\) is not a PNE, BZR then invokes RRR--BRD and finds the PNE
\(\{(0,0,1),(1,1,0)\}\)
under \([1,2]\rightarrow[2,1]\rightarrow[1,2]\).
Along this BRD trajectory, player 1's BRs are \((1,1,0)\) and \((0,0,1)\), and player 2's BR is \((1,1,0)\); thus player 1's BR \((0,0,1)\) provides an additional inequality:
$$
2-3x^2_3
\leq
3x^1_1 + 2x^1_2 + 2x^1_3 - 4z_1  - 5 z_2 - 3 z_3.
$$
After adding these inequalities, BZR resolves and terminates once the separation oracle confirms that the incumbent
\(\{(0,0,1),(1,1,0)\}\)
(maximizing social welfare) is indeed the best PNE.

\smallskip
\noindent
\textbf{ZR:}
ZR first checks \(\{(1,0,1), (1,1,0)\}\) (social welfare \(=9\)), generating the violated EI from player 1's BR \((0,0,1)\):
$$
2-3x^2_3
\leq
3x^1_1 + 2x^1_2 + 2x^1_3 - 4z_1  - 5 z_2 - 3 z_3.
$$
Next it evaluates \(\{(1,1,0), (1,0,1)\}\) (social welfare \(=8\)), generating two violated EIs from player 1's BR \((0,1,0)\) and player 2's BR \((1,1,0)\):
$$
2 - 5x^2_2
\leq
3x^1_1 + 2x^1_2 + 2x^1_3 - 4z_1 - 5z_2 - 3z_3,
$$
$$
3 + 5x^1_1 + 2x^1_2
\leq
2x^2_1 + x^2_2 + 0x^2_3 + 5z_1 + 2z_2 + z_3.
$$
Finally, \(\{(0,0,1), (1,1,0)\}\) is tested and confirmed as the best PNE.

\smallskip
\noindent
\textbf{Comparison:}
BZR requires one separation call at the initial incumbent, one RRR--BRD run, and three EIs in total (two from the initial separation and one from the BRD trajectory), resulting in two separation calls overall (including the final call that certifies the best PNE). ZR requires three separation calls and generates three EIs. This gap can widen substantially in large-scale IPGs where many separation rounds may be needed. We include this example for \textit{illustrative} purposes; in the large-scale IPG instances studied in this paper, the performance gap between BZR and ZR is often substantial.

\clearpage
% === MATERIAL 3: Exponential Gap Construction (concise) ===
\ecompanionsection{Exponential Separation Between ZR and BZR} \label{appendix:exponential-gap}

This section gives a simple worst-case construction showing an exponential
separation between ZR and BZR when \emph{maximizing social welfare over PNEs}
(i.e., the welfare-maximizing equilibrium problem). We use the $n$-fold product
of the classical Prisoner’s Dilemma and the standard welfare $\phi=u_1+u_2$ to
illustrate that there can be exponentially many high-welfare non-equilibria,
forcing welfare-guided equilibrium search to reject many candidates before
certifying the optimal PNE. (We use $O(\cdot)$ for upper bounds and
$\Theta(\cdot)$ when the bound is tight up to constant factors.)

\paragraph*{Construction.}
Two players choose $\x,\y\in\{0,1\}^n$. Each coordinate $j\in\{1,\ldots,n\}$ is an
independent Prisoner’s Dilemma, where $0=\text{Cooperate}$ and $1=\text{Defect}$,
with per-coordinate payoffs shown in Table~\ref{tab:pd_payoffs}.

\vspace{10pt}
\begin{table}[h!]
    \centering
    \caption{Per-Coordinate Payoffs for the Prisoner’s Dilemma Component Game in $\mathcal{P}_n$.}
    \label{tab:pd_payoffs}    \begin{tabular}{c|cc}
    & $y_j = 0$ (C) & $y_j = 1$ (D) \\
    \hline
    $x_j = 0$ (C) & $(3, 3)$ & $(0, 5)$ \\
    $x_j = 1$ (D) & $(5, 0)$ & $(1, 1)$ \\
    \end{tabular}

    \vspace{3pt}
    \footnotesize{\textit{Note.} ($0=$ cooperate, $1=$ defect)}
\end{table}

\paragraph*{Structure.}
Defection strictly dominates cooperation in each coordinate for both players.
Hence the unique PNE is $(\mathbf{1},\mathbf{1})$ with $\phi(\mathbf{1},\mathbf{1})=2n$,
while the social optimum is $(\mathbf{0},\mathbf{0})$ with $\phi(\mathbf{0},\mathbf{0})=6n$.
Moreover, every profile other than $(\mathbf{1},\mathbf{1})$ has strictly larger
welfare, so exactly $4^n-1$ profiles satisfy $\phi>2n$.

\paragraph*{Complexity gap.}
ZR must reject all $4^n-1$ non-PNE profiles with $\phi>2n$ before it can certify
that $(\mathbf{1},\mathbf{1})$ is the optimal PNE, yielding $\Theta(4^n)$
separation-oracle calls in the worst case. By contrast, BZR finds the PNE in
$O(1)$ best-response steps (player~1 best-responds to $\mathbf{1}$, then player~2
best-responds to $\mathbf{1}$). Verification is still exponential and in fact
$\Theta(2^n)$ in this construction: because defection is dominant, equilibrium
inequalities are essentially non-restrictive and do not cut off high-welfare
non-PNE profiles, so BZR must rule out $(\mathbf{1},\y)$ for each $\y\neq\mathbf{1}$.
There are $2^n-1$ such $\y$, and for each one player~2’s best response to
$\mathbf{1}$ is $\mathbf{1}\neq\y$, hence $(\mathbf{1},\y)$ is not a PNE. Thus
verification requires $\Theta(2^n)$ checks and the overall separation is
$\Theta(2^n)$.

\vspace{10pt}
\begin{table}[h!]
    \centering
    \caption{Worst-Case Complexity Comparison on $\mathcal{P}_n$: ZR versus BZR.}
    \label{tab:pd_zr_bzr_gap}
    \begin{tabular}{lccc}
    \toprule
    \textbf{Algorithm} & \textbf{Finding} & \textbf{Verification} & \textbf{Total} \\
    \midrule
    ZR  & --    & (included)   & $\Theta(4^n)$ \\
    BZR & $O(1)$ & $\Theta(2^n)$ & $\Theta(2^n)$ \\
    \midrule
    \multicolumn{3}{l}{\textbf{Gap:}} & $\Theta(2^n)$ \\
    \bottomrule
    \end{tabular}
\end{table}

\paragraph*{Takeaway.}
This dominant-strategy example shows that certifying the welfare-optimal PNE can
be exponentially harder than finding a PNE.

% === END MATERIAL 3 ===

\clearpage
\ecompanionsection{Illustration: A Two-Player Selfish EBMC Game} \label{appendix:two-player}
Figure~\ref{fig:selfish interactions_k=2} illustrates the interplay between two selfish players, each controlling three vertices and selecting one under a budget constraint of one. The bold circles and the dashed lines represent the selected vertices and the edges covered by the choices of players, respectively. For example: In Fig. (\ref{fig:subgraph_selfish_k=2_0}), edges from $B_1$ to $A_1$ are covered at vertex $A_1$, while those from $B_1$ to $B_2$ are not covered. Assume that the vertices differ in status between two task types. Following standard BRD, Player A first chooses vertex $A_1$. Player B benefits from this choice because the edge $(A_1, B_1)$ is now covered. Subsequently, Player B chooses vertex $B_1$, covering additional arcs. This in turn activates the edge $(B_1, A_1)$, allowing Player A to consider selecting vertex $A_3$ as a new best response. Player B maintains the same choice. At this point, the game reaches a PNE, as neither player has an incentive to deviate from their current strategy. This example illustrates how one player's best response may incidentally benefit the other, even in a selfish setting. However, these benefits are not intentional but rather emergent consequences of self-interested decisions. Importantly, mutual gains observed during the BRD do not guarantee monotonic increases in utility for all players. For example, Player B benefits from Player A's initial choice of $A_1$, but the subsequent choice of $A_3$ does not influence Player B's utility, so that it continues to prefer vertex $B_1$. From Player B's perspective, Player A's earlier strategy (choosing $A_1$ alone) yields a higher selfish objective value than the final strategy (choosing $A_3$).

\vspaceminus{15pt}
\begin{figure}[h]
\begin{subfigure}{.3\textwidth}
  \centering
        \resizebox{\textwidth}{!}{
        \begin{tikzpicture}[auto]
            \tikzstyle{node_A} = [draw, fill=blue!60, circle, minimum size=2.5em, text=white]
            \tikzstyle{node_B} = [draw, fill=yellow!60, circle, minimum size=2.5em, text=black]
            \tikzstyle{arrow} = [->,>=stealth, bend left, line width=1.1pt]
            \tikzstyle{arrow_dashed} = [dashed,->,>=stealth, bend left, line width=1.1pt]
            \tikzstyle{arrow_dashed_right} = [dashed,->,>=stealth, bend right, line width=1.1pt]

            \node[node_A, draw=black, line width = 0.75mm] (A1) at (2,6) {A$_1$};
            \node[node_A] (A2) at (1,4.5) {A$_2$};
            \node[node_A] (A3) at (2,3) {A$_3$};
            \node[node_B] (B1) at (5,6) {B$_1$};
            \node[node_B] (B2) at (5,3) {B$_2$};
            \node[node_B] (B3) at (6.5,4.5) {B$_3$};

            \draw[arrow_dashed] (A1) to node[pos=0.7] {1} (A2);
            \draw[arrow_dashed] (A1) to node[pos=0.7] {1} (A3);
            \draw[arrow_dashed] (A1) to node[pos=0.5] {3} (B1);
            \draw[arrow_dashed_right] (A1) to node[pos=0.6] {1} (B3);
            \draw[arrow_dashed] (B1) to node[pos=0.5] {3} (A1);
            \draw[arrow] (B1) to node[pos=0.7] {3} (B2);
            \draw[arrow] (B1) to node[pos=0.7] {3} (B3);
            \draw[arrow] (B2) to node[pos=0.7] {2} (A3);
        \end{tikzpicture}
                }
        \caption{Player A first chooses vertex A$_1$}
    \label{fig:subgraph_selfish_k=2_0}
\end{subfigure}
\quad
\begin{subfigure}{.3\textwidth}
  \centering
        \resizebox{\textwidth}{!}{
        \begin{tikzpicture}[auto]
            \tikzstyle{node_A} = [draw, fill=blue!60, circle, minimum size=2.5em, text=white]
            \tikzstyle{node_B} = [draw, fill=yellow!60, circle, minimum size=2.5em, text=black]
            \tikzstyle{arrow} = [->,>=stealth, bend left, line width=1.1pt]
            \tikzstyle{arrow_dashed} = [dashed,->,>=stealth, bend left, line width=1.1pt]
            \tikzstyle{arrow_dashed_right} = [dashed,->,>=stealth, bend right, line width=1.1pt]

            \node[node_A, draw=black, line width = 0.75mm] (A1) at (2,6) {A$_1$};
            \node[node_A] (A2) at (1,4.5) {A$_2$};
            \node[node_A] (A3) at (2,3) {A$_3$};
            \node[node_B, draw=black, line width = 0.75mm] (B1) at (5,6) {B$_1$};
            \node[node_B] (B2) at (5,3) {B$_2$};
            \node[node_B] (B3) at (6.5,4.5) {B$_3$};

            \draw[arrow_dashed] (A1) to node[pos=0.7] {1} (A2);
            \draw[arrow_dashed] (A1) to node[pos=0.7] {1} (A3);
            \draw[arrow_dashed] (A1) to node[pos=0.5] {3} (B1);
            \draw[arrow_dashed_right] (A1) to node[pos=0.6] {1} (B3);
            \draw[arrow_dashed] (B1) to node[pos=0.5] {3} (A1);
            \draw[arrow_dashed] (B1) to node[pos=0.7] {3} (B2);
            \draw[arrow_dashed] (B1) to node[pos=0.7] {3} (B3);
            \draw[arrow] (B2) to node[pos=0.7] {2} (A3);
        \end{tikzpicture}
        }    \caption{Player B first chooses vertex B$_1$}
    \label{fig:subgraph_selfish_k=2_1}
\end{subfigure}
\quad
\begin{subfigure}{.32\textwidth}
  \centering
        \resizebox{.95\textwidth}{!}{
        \begin{tikzpicture}[auto]
            \tikzstyle{node_A} = [draw, fill=blue!60, circle, minimum size=2.5em, text=white]
            \tikzstyle{node_B} = [draw, fill=yellow!60, circle, minimum size=2.5em, text=black]
            \tikzstyle{arrow} = [->,>=stealth, bend left, line width=1.1pt]
            \tikzstyle{arrow_right} = [->,>=stealth, bend right, line width=1.1pt]
            \tikzstyle{arrow_dashed} = [dashed,->,>=stealth, bend left, line width=1.1pt]
            \tikzstyle{arrow_dashed_right} = [dashed,->,>=stealth, bend right, line width=1.1pt]

            \node[node_A] (A1) at (2,6) {A$_1$};
            \node[node_A] (A2) at (1,4.5) {A$_2$};
            \node[node_A, draw=black, line width = 0.75mm] (A3) at (2,3) {A$_3$};
            \node[node_B, draw=black, line width = 0.75mm] (B1) at (5,6) {B$_1$};
            \node[node_B] (B2) at (5,3) {B$_2$};
            \node[node_B] (B3) at (6.5,4.5) {B$_3$};

            \draw[arrow] (A1) to node[pos=0.7] {1} (A2);
            \draw[arrow_dashed] (A1) to node[pos=0.7] {1} (A3);
            \draw[arrow_dashed] (A1) to node[pos=0.5] {3} (B1);
            \draw[arrow_right] (A1) to node[pos=0.6] {1} (B3);
            \draw[arrow_dashed] (B1) to node[pos=0.5] {3} (A1);
            \draw[arrow_dashed] (B1) to node[pos=0.7] {3} (B2);
            \draw[arrow_dashed] (B1) to node[pos=0.7] {3} (B3);
            \draw[arrow_dashed] (B2) to node[pos=0.7] {2} (A3);
        \end{tikzpicture}
        }
        \caption{Player A now can choose vertex A$_3$}
    \label{fig:subgraph_selfish_k=2_2}
\end{subfigure}
\caption{The Interactions in the Selfish Game with Two Players.}
\label{fig:selfish interactions_k=2}
\vspaceminus{6pt}
\end{figure}
\vspaceminus{15pt}

\clearpage
\ecompanionsection{EBMC as an IPG: Reformulation and Complexity Implications}
\label{appendix:reform_standard_IPG}

The objective functions of the locally altruistic and selfish EBMC games can be reformulated in the
standard (quadratic) IPG form by eliminating the edge variables $\y$ and expressing utilities purely in terms of
vertex variables $\x$. Throughout, vertex decisions are binary ($x_j\in\{0,1\}$), and edge coverage is
defined by $y_{jk}=\max\{x_j,x_k\}$. Hence, for every arc $(j,k)$ we have the equation
\(
y_{jk}=x_j+x_k-x_jx_k.
\)
Using this equation, for each player $i\in N$ we can write the EBMC utility as
$$
u_i(\x^i,\x^{\not i})
\;=\;
(\x^i)^\top Q^i \x^i
\;+\;
\sum_{p\in N\setminus\{i\}} (\x^{p})^\top Q_{p}^i \x^i
\;+\;
(\mathbf{d}^i)^\top \x^i
\;+\;
c_i(\x^{\not i}),
$$
where $Q^i$ captures within-player quadratic terms, $Q_p^i$ captures cross-player interactions (rows indexed by variables in $\x^p$ and columns by variables in $\x^i$), and
$\mathbf d^i$ captures linear terms in $\x^i$. The last summation collects all terms that depend only on opponents' decisions $\x^{\not i}$. Since $c_i(\x^{\not i})$ does not depend on $\x^i$, it is an additive constant from player $i$'s perspective and therefore does not affect best responses or equilibria.

\textbf{Coefficient construction.}
The reformulation follows by expressing the three core utility components
\(
\sum_{(j,k)\in \A_i} w_{jk} y_{jk},
\sum_{(j,k)\in \delta_+(\I_i)} w_{jk} y_{jk},
\sum_{(j,k)\in \delta_-(\I_i)} w_{jk} y_{jk}
\)
in terms of $\x$ via $y_{jk}=x_j+x_k-x_jx_k$. Below we describe how each arc set contributes to the coefficients
in $u_i(\x^i,\x^{\not i})$. (As usual, if one stores $Q^i$ as a symmetric matrix, then a term $-w_{jk}x_j x_k$
may be split across $(j,k)$ and $(k,j)$ to avoid double counting.)

\begin{itemize}
\item \textbf{Internal arcs.}
For an internal arc $(j,k)\in \A_i$ (so $j,k\in \I_i$), the contribution
\(
w_{jk}(x_j + x_k - x_j x_k)
\)
adds $w_{jk}$ to the $j$-th and $k$-th entries of $\mathbf d^i$, and adds $-w_{jk}$ to the $(j,k)$ entry of $Q^i$
(representing the bilinear term $-w_{jk}x_j x_k$).

\item \textbf{Outgoing arcs.}
For an outgoing arc $(j,k)\in \delta_+(\I_i)$ with $j\in \I_i$ and $k\in \I_p$ for some $p\neq i$, the contribution
\(
w_{jk}(x_j + x_k - x_j x_k)
\)
adds $w_{jk}$ to the $j$-th entry of $\mathbf d^i$ (the $x_j$ term),
adds the bilinear coefficient $-w_{jk}$ to the $(k,j)$ entry of $Q_p^i$ (row indexed by $x_k$, column indexed by $x_j$),
and adds the opponent-only linear term $w_{jk}x_k$ into $c_i(\x^{\not i})$.

\item \textbf{Incoming arcs.}
For an incoming arc $(j,k)\in \delta_-(\I_i)$ with $j\in \I_p$ for some $p\neq i$ and $k\in \I_i$, the contribution
\(
w_{jk}(x_j + x_k - x_j x_k)
\)
adds $w_{jk}$ to the $k$-th entry of $\mathbf d^i$ (the $x_k$ term),
adds the bilinear coefficient $-w_{jk}$ to the $(j,k)$ entry of $Q_p^i$ (row indexed by $x_j$, column indexed by $x_k$),
and adds the opponent-only linear term $w_{jk}x_j$ into $c_i(\x^{\not i})$.
\end{itemize}

Applying these transformations to the appropriate arc sets yields an explicit representation of EBMC games
in the standard (quadratic) IPG format. In particular, the interaction structure of EBMC, captured by the family
$\{Q_p^i\}_{p\neq i}$, mirrors that of quadratic IPGs (qIPGs).

Finally, although $u_i(\x^i,\x^{\not i})$ includes the term $c_i(\x^{\not i})$, this term depends only on $\x^{\not i}$
and therefore does not affect player $i$'s best response; it can be safely ignored when computing best responses or
separating equilibrium inequalities.

\medskip
\paragraph*{Complexity considerations.}
Both the selfish and locally altruistic EBMC games are instances of quadratic integer programming games (qIPGs):
each player solves a binary optimization problem over an individual feasible set $\X_i$, and utilities admit the
qIPG representation. Moreover, any term depending only on opponents' decisions can be absorbed into
$c_i(\x^{\not i})$ and therefore does not affect best responses or equilibria.
\citet{carvalho2018existence} show that deciding whether a general IPG admits a pure Nash equilibrium (PNE) is
$\Sigma_2^p$-complete; consequently, the PNE-existence problem restricted to EBMC games lies in $\Sigma_2^p$.
In the locally altruistic EBMC game, we establish that the game is an exact potential game and hence a PNE always exists
(each $\X_i$ is finite and nonempty). For the selfish EBMC game, PNE existence is not guaranteed; while the decision problem
lies in $\Sigma_2^p$, establishing $\Sigma_2^p$-hardness (and thus completeness) for the EBMC subclass would require a dedicated
reduction, which we leave for future work.

\clearpage
\ecompanionsection{Counterexamples for the Existence of PNE} \label{appendix:counterexamples}
This section presents counterexamples for the existence of a PNE in the EBMC games $\mathcal{G}^{\text{Self}}_{|\mathcal{L}|=1}$ and $\mathcal{G}^{\text{Self}}_{|\mathcal{L}| \ge 2}$. For $\mathcal{G}^{\text{Self}}_{|\mathcal{L}|=1}$, we identify a counterexample using modest integer weights between 1 and 10 with a total of 9 vertices and two players with budgets 2 and 1. Table \ref{tab:no_PNE_k=1} shows that there is a unilateral selfish deviation for any strategy profile. We demonstrate the existence of a selfish deviation for any strategy profile rather than showing cycles, as the latter case does not conclusively indicate the absence of a PNE.

\begin{figure}[h!]
  \begin{minipage}[b]{.5\textwidth}
  \centering
    \begin{tikzpicture}[node distance=2cm and 1cm, auto]
    \tikzstyle{node_red} = [draw, line width=1mm, fill=red!60, circle, minimum size=2.5em, text=white]
    \tikzstyle{node_green} = [draw, fill=green!60, circle, minimum size=2.5em, text=white]
    \tikzstyle{node_red_chosen} = [draw,line width=1mm, fill=red!60, circle, minimum size=2.5em, text=white]
    \tikzstyle{node_green_chosen} = [draw, fill=green!60, circle, minimum size=2.5em, text=white]
    \tikzstyle{node_A} = [draw, fill=blue!60,circle, minimum size=2.5em, text=white]
    \tikzstyle{node_B} = [draw, fill=yellow!60,circle, minimum size=2.5em, text=black]
    \tikzstyle{node_A_x} = [draw, fill=blue!60, line width=1mm,circle, minimum size=2.5em, text=white]
    \tikzstyle{node_B_x} = [draw, fill=yellow!60, line width=1mm,circle, minimum size=2.5em, text=black]
    \tikzstyle{arrow} = [->,>=stealth]
    
    \node[node_A] (A1) at (0,0) {A$_1$};
    \node[node_A] (A2) at (0,-1.5) {A$_2$};
    \node[node_A] (A3) at (0,-3) {A$_3$};
    \node[node_B] (B1) at (0,-4.5) {B$_1$};
    
    \node[node_A] (A4) at (4,0) {A$_4$};
    \node[node_A] (A5) at (4,-1.5) {A$_5$};
    \node[node_A] (A6) at (4,-3) {A$_6$};
    \node[node_B] (B2) at (4,-4.5) {B$_2$};
    \node[node_B] (B3) at (4,-6) {B$_3$};
    
    \draw[arrow] (A1) to node[pos=0.15] {5} (A4);
    \draw[arrow] (A1) to node[pos=0.15] {2} (A5);
    \draw[arrow] (A1) to node[pos=0.15] {2} (A6);
    
    \draw[arrow] (A2) to node[pos=0.1] {5} (A4);
    \draw[arrow] (A2) to node[pos=0.15] {2} (A5);
    \draw[arrow] (A2) to node[pos=0.1] {2} (A6);

    \draw[arrow] (A3) to node[pos=0.15] {3} (A5);
    \draw[arrow] (A3) to node[pos=0.15] {3} (A6);
    \draw[arrow] (A3) to node[pos=0.1] {2} (B2);
    
    \draw[arrow] (B1) to node[pos=0.05] {10} (A4);
    \draw[arrow] (B1) to node[pos=0.1] {1} (B2);
    \draw[arrow] (B1) to node[pos=0.1] {1} (B3);

\end{tikzpicture}
    \vspaceminus{1.0cm}
    \captionof{figure}{No PNE for $\mathcal{G}^{\text{Self}}_{|\mathcal{L}|=1}$.} \label{fig:no_PNE_k=1}
  \end{minipage}
  \begin{minipage}[b]{.5\textwidth}
    \captionof{table}{Strategy Profiles and Selfish Deviations.} \label{tab:no_PNE_k=1}
    \vspace{0.1cm}
    \scalebox{0.75}{
    \begin{tabular}{ccccc}
    \toprule
    Strategy profiles & Selfish unilateral deviation \\
    \midrule
    $x_{A_3} = 1, \x^B \neq (1,0,0) $  & $\hat{\x}^B = (1,0,0)$\\
    $x_{A_3} \neq 1, \x^B \neq (0,1,0) $  & $\hat{\x}^B = (0,1,0)$\\
    $x_{B_1} = 1, \x^A \neq (1,1,0,0,0,0)$  & $\hat{\x}^A = (1,1,0,0,0,0)$\\
    $x_{B_1} \neq 1, \x^A \neq (0,0,1,1,0,0)$  & $\hat{\x}^A = (0,0,1,1,0,0)$\\
    \bottomrule
    \end{tabular}}
    \vspace{0.2cm}

    \footnotesize{This table displays all potential strategy profiles for players A and B. A has a budget of 2 while B has a budget of 1.  Each row describes a set of strategy profiles. The first row contains the strategy profile $\x^A = (0,0,1,1,0,0)$ and $\x^B = (0,1,0)$.  In this case, player B would deviate to $\hat{\x}^B = (1,0,0)$. In another scenario, if $\x^A = (1,0,1,0,0,0)$ and $\x^B = (0,1,0)$, possible deviations are $\hat{\x}^B = (1,0,0)$ for player B and $\hat{\x}^A = (0,0,1,1,0,0)$ for player A.}
     \end{minipage}
\end{figure}
Compared to $\mathcal{G}^{\text{Self}}_{|\mathcal{L}|=1}$, counterexamples in $\mathcal{G}^{\text{Self}}_{|\mathcal{L}| \ge 2}$ are more readily identifiable. An example is presented in Figure \ref{fig:no_PNE_k=2}, where the vertices $A_1$ and $A_2$ are of type-1, and the vertices $B_1$ and $B_2$ are of type-2. Assuming that each player has only one budget, no PNE exists; for every possible strategy profile, there is a selfish unilateral deviation, as detailed in Table \ref{tab:no_PNE_k=2}.

\begin{figure}[h!]
  \begin{minipage}[b]{.5\textwidth}
  \centering
    \begin{tikzpicture}[node distance=2cm and 1cm, auto]
    \tikzstyle{node_red} = [draw, line width=1mm, fill=red!60, circle, minimum size=2.5em, text=white]
    \tikzstyle{node_green} = [draw, fill=green!60, circle, minimum size=2.5em, text=white]
        \tikzstyle{node_red_chosen} = [draw,line width=1mm, fill=red!60, circle, minimum size=2.5em, text=white]
    \tikzstyle{node_green_chosen} = [draw, fill=green!60, circle, minimum size=2.5em, text=white]
    \tikzstyle{node_A} = [draw, fill=blue!60,circle, minimum size=2.5em, text=white]
    \tikzstyle{node_B} = [draw, fill=yellow!60,circle, minimum size=2.5em, text=black]
       \tikzstyle{node_A_x} = [draw, fill=blue!60, line width=1mm,circle, minimum size=2.5em, text=white]
    \tikzstyle{node_B_x} = [draw, fill=yellow!60, line width=1mm,circle, minimum size=2.5em, text=black]
    % \tikzstyle{arrow} = [thick,->,>=stealth]
    \tikzstyle{arrow} = [->,>=stealth, bend left]
    
    \node[node_A] (A1) at (0,0) {A$_1$};
    \node[node_A] (A2) at (0,-3) {A$_2$};
    \node[node_B] (B1) at (3,0) {B$_1$};
    \node[node_B] (B2) at (3,-3) {B$_2$};
    
    %\node[draw,ellipse,fit=(A1) (A2),inner sep=15pt,label=above:County A] {};
    %\node[draw,ellipse,fit=(B1) (B2) (B3) (B4),inner sep=15pt,label=above:County B] {};
    
    \draw[arrow] (A1) -- node[above] {1} (B1);
    \draw[arrow] (A2) -- node[above] {2} (B2);
    \draw[arrow] (B1) to[bend right=25] node[above] {2} (A2);
    \draw[arrow] (B2) to[bend left=25] node[above] {1} (A1);
    % \draw[arrow, red] (B1) to[bend right=25] node[above] {2} (A2);
    % \draw[arrow, red] (B2) to[bend left=25] node[above] {1} (A1);
    %\draw[arrow] (B2) -- node[above] {1} (A1);
\end{tikzpicture}
    \captionof{figure}{No PNE for $\mathcal{G}^{\text{Self}}_{|\mathcal{L}| \ge 2}$.} \label{fig:no_PNE_k=2}
  \end{minipage}
  \begin{minipage}[b]{.5\textwidth}
    \captionof{table}{Strategy Profiles and Selfish Deviations.} \label{tab:no_PNE_k=2}
    \vspace{0.1cm}
    \scalebox{0.9}{
    \begin{tabular}{ccccc}
    \toprule
    Strategy profiles  & Selfish unilateral deviation \\
    \midrule
    $x_{A_1} = 1, \x^B \neq (0,1)$  & $\hat{\x}^B = (0,1)$\\
    $x_{A_1} \neq 1, \x^B \neq (1,0)$  & $\hat{\x}^B = (1,0)$\\
    $x_{B_1} = 1, \x^A \neq (1,0)$  & $\hat{\x}^A = (1,0)$\\
    $x_{B_1} \neq 1, \x^A \neq (0,1)$  & $\hat{\x}^A = (0,1)$\\
    \bottomrule
    \end{tabular}}
    \vspace{0.2cm}

    \footnotesize{This table displays all potential strategy profiles for players A and B. For instance, if $\x^A = (1,0)$ and $\x^B = (1,0)$, player B would deviate to $\hat{\x}^B = (0,1)$.}
  \end{minipage}
\end{figure}

\clearpage
\ecompanionsection{RRR-BRD Results}\label{appendix:CR-BRD}
We present the full results of the RRR-BRD algorithm using different initial strategy profiles in various problem settings. Specifically, we report results for selfish EBMC games with $|\mathcal{L}|=1$ and $|\mathcal{L}|\ge 2$, as well as KPG instances with interaction types (A), (B), and (C). The metrics and computational settings used in these experiments are described in Section~\ref{sec:6.1}.
\begin{table}[htbp]
\caption{RRR-BRD Results with Different Initial Strategy Profiles for Selfish EBMC Games $\mathcal{G}^{\text{Self}}_{|\mathcal{L}|=1}$.}
\centering
\small
\scalebox{0.8}{
\begin{tabular}{r r r r r r r r r r r r}
\toprule
& & \multicolumn{2}{c}{RRR-BRD ($\phi$)} & \multicolumn{2}{c}{Price of Stability} & \multicolumn{2}{c}{References from BZR} & \multicolumn{2}{c}{Time (secs)} & \multicolumn{2}{c}{Round}  \\
\cmidrule(lr){3-4} \cmidrule(lr){5-6} \cmidrule(lr){7-8} \cmidrule(lr){9-10} \cmidrule(lr){11-12} 
$n$ & BG & init:$\barxzero$ &   init:$\barxswtime{300}$ & $\widetilde{\text{OSW}}$ & $\widetilde{\text{POS}}$ & BZR($\phi$) & BZR(Bound) & init:$\barxzero$ &   init:$\barxswtime{300}$ & init:$\barxzero$ &   init:$\barxswtime{300}$  \\ \midrule
2  & 0.3 & 19.2   & 19.2   & 19.2   & 1.000 & 19.2   & 19.2   & 0.0  & 0.1   & 2   & 1   \\
   & 0.5 & 21.6   & 22.3   & 22.3   & 1.000 & 22.3   & 22.3   & 0.1  & 2.3   & 2   & 1   \\
   & 0.8 & 22.6   & 22.6   & 24.9   & 1.102 & 23.3   & 23.3   & 0.1  & 0.4   & 2   & 2   \\
3  & 0.3 & 22.0   & 27.6   & 27.6   & 1.000 & 27.6   & 27.6   & 0.2  & 2.4   & 2   & 1   \\
   & 0.5 & 38.8   & 42.1   & 42.1   & 1.000 & 42.1   & 42.1   & 0.2  & 6.9   & 2   & 1   \\
   & 0.8 & 53.1   & 53.1   & 53.1   & 1.000 & 53.1   & 53.1   & 0.1  & 0.2   & 2   & 1   \\
5  & 0.3 & 83.5   & 90.4   & 90.4   & 1.000 & 90.4   & 90.4   & 0.4  & 13.8  & 2   & 1   \\
   & 0.5 & 85.0   & 100.6  & 100.6  & 1.000 & 100.6  & 101.2  & 0.5  & 301.0 & 2   & 1   \\
   & 0.8 & 146.9  & 156.7  & 156.7  & 1.000 & 156.7  & 157.4  & 0.5  & 301.1 & 2   & 1   \\
8  & 0.3 & 130.1  & 130.1  & 149.9  & 1.152 & 134.1  & 134.2  & 1.6  & 5.3   & 2   & 3   \\
   & 0.5 & 169.1  & 252.2  & 252.2  & 1.000 & 252.2  & 260.2  & 1.6  & 302.8 & 2   & 1   \\
   & 0.8 & 269.9  & 341.4  & 381.3  & 1.117 & 349.2  & 353.4  & 1.4  & 232.6 & 2   & 3   \\
10 & 0.3 & 308.1  & 317.2  & 317.2  & 1.000 & 317.2  & 317.3  & 2.6  & 304.5 & 2   & 1   \\
   & 0.5 & 289.8  & 361.7  & 361.7  & 1.000 & 361.7  & 372.4  & 1.6  & 302.8 & 2   & 1   \\
   & 0.8 & 412.2  & 539.8  & 539.8  & 1.000 & 538.1  & 549.3  & 2.0  & 303.8 & 2   & 1   \\
15 & 0.3 & 630.4  & 725.4  & 725.4  & 1.000 & 725.4  & 745.9  & 4.7  & 307.8 & 2   & 1   \\
   & 0.5 & 576.9  & 576.9  & 857.5  & 1.487 & 708.4  & 730.1  & 6.1  & 23.7  & 2   & 3   \\
   & 0.8 & 886.3  & 1212.2 & 1219.7 & 1.006 & 1216.5 & 1249.2 & 5.7  & 311.3 & 2   & 3   \\
20 & 0.3 & 656.1  & 904.5  & 904.5  & 1.000 & 904.5  & 922.5  & 10.4 & 319.2 & 2   & 1   \\
   & 0.5 & 1089.0 & 1398.8 & 1529.1 & 1.093 & 1282.1 & 1472.8 & 11.1 & 319.8 & 2   & 1   \\
   & 0.8 & 1317.6 & 1701.4 & 2415.5 & 1.420 & 2024.4 & 2207.2 & 13.1 & 323.9 & 2   & 3   \\
25 & 0.3 & 1419.5 & 1765.0 & 1765.0 & 1.000 & 1765.0 & 1819.5 & 18.0 & 328.9 & 2   & 1   \\
   & 0.5 & 2140.7 & 2597.5 & 2597.7 & 1.000 & 2597.4 & 2767.9 & 19.0 & 334.7 & 2   & 2   \\
   & 0.8 & 2246.8 & 3294.0 & 3312.8 & 1.006 & 3307.4 & 3541.1 & 18.9 & 336.4 & 2   & 3   \\
30 & 0.3 & 1780.2 & 2448.2 & 2448.3 & 1.000 & 2448.1 & 2556.0 & 26.1 & 345.6 & 2   & 2   \\
   & 0.5 & 2704.9 & 3263.8 & 3263.8 & 1.000 & 3260.8 & 3482.4 & 29.8 & 349.1 & 2   & 1   \\
   & 0.8 & 3724.0 & 4919.7 & 4919.7 & 1.000 & 4910.4 & 5399.1 & 28.1 & 346.4 & 2   & 1    \\ \bottomrule
\end{tabular}
}

% \vspace{3pt}
% \footnotesize{\textit{Note.} Objective values ($\phi$) are reported in thousands and rounded to the nearest tenth.}
\label{tab:EBMC_games_k=1_CR_BRD}
\end{table}

\begin{table}[htbp]
\caption{RRR-BRD Results with Different Initial Strategy Profiles for Selfish EBMC Games $\mathcal{G}^{\text{Self}}_{|\mathcal{L}|\ge 2}$.}
\centering
\small
\scalebox{0.8}{
\begin{tabular}{r r r r r r r r r r r r}
\toprule
& & \multicolumn{2}{c}{RRR-BRD ($\phi$)} & \multicolumn{2}{c}{Price of Stability} & \multicolumn{2}{c}{References from BZR} & \multicolumn{2}{c}{Time (secs)} & \multicolumn{2}{c}{Round}  \\
\cmidrule(lr){3-4} \cmidrule(lr){5-6} \cmidrule(lr){7-8} \cmidrule(lr){9-10} \cmidrule(lr){11-12} 
$n$ & BG & init:$\barxzero$ &   init:$\barxswtime{300}$ & $\widetilde{\text{OSW}}$ & $\widetilde{\text{POS}}$ & BZR($\phi$) & BZR(Bound) & init:$\barxzero$ &   init:$\barxswtime{300}$ & init:$\barxzero$ &   init:$\barxswtime{300}$  \\ \midrule
2  & 0.3 & 51.4    & 51.4    & 56.3    & 1.096 & 51.4    & 51.4    & 0.4    & 1.6    & 3   & 3   \\
   & 0.5 & 83.8    & 83.8    & 86.6    & 1.033 & 83.8    & 86.0    & 0.8    & 58.8   & 3   & 2   \\
   & 0.8 & 92.1    & 91.4    & 95.2    & 1.034 & 92.8    & 94.3    & 1.0    & 5.6    & 3   & 2   \\
3  & 0.3 & 115.0   & 115.0   & 139.7   & 1.215 & 115.0   & 115.0   & 0.7    & 12.1   & 4   & 4   \\
   & 0.5 & 178.6   & 178.3   & 200.3   & 1.122 & 178.6   & 196.7   & 0.7    & 301.5  & 4   & 4   \\
   & 0.8 & 252.2   & 252.2   & 253.7   & 1.006 & 252.2   & 257.8   & 0.8    & 301.8  & 4   & 6   \\
5  & 0.3 & 380.0   & 380.0   & 399.8   & 1.052 & 380.0   & 380.0   & 1.7    & 303.5  & 3   & 3   \\
   & 0.5 & 557.6   & 557.6   & 577.3   & 1.035 & 557.6   & 618.5   & 1.7    & 303.1  & 3   & 3   \\
   & 0.8 & 650.1   & 650.1   & 661.9   & 1.018 & 650.1   & 677.5   & 2.1    & 303.9  & 4   & 4   \\
8  & 0.3 & 862.1   & 862.1   & 995.8   & 1.155 & 862.1   & 980.4   & 5.2    & 309.4  & 4   & 4   \\
   & 0.5 & 1214.5  & 1214.5  & 1387.9  & 1.143 & 1214.5  & 1535.1  & 5.6    & 309.2  & 4   & 4   \\
   & 0.8 & 1661.5  & 1661.5  & 1721.0  & 1.036 & 1661.5  & 1771.3  & 6.1    & 309.3  & 4   & 6   \\
10 & 0.3 & 1265.2  & 1265.2  & 1521.6  & 1.203 & 1265.2  & 1526.5  & 8.0    & 319.6  & 3   & 13  \\
   & 0.5 & 2084.2  & 2084.2  & 2253.9  & 1.081 & 2083.2  & 2635.7  & 8.7    & 315.5  & 4   & 5   \\
   & 0.8 & 2612.3  & 2612.3  & 2685.2  & 1.028 & 2612.3  & 2757.3  & 9.4    & 318.6  & 5   & 9   \\
15 & 0.3 & 2916.4  & 2916.4  & 3440.2  & 1.180 & 2916.4  & 3523.7  & 19.5   & 329.0  & 4   & 4   \\
   & 0.5 & 4476.8  & 4476.8  & 5030.1  & 1.124 & 4476.8  & 5750.7  & 29.3   & 337.9  & 6   & 5   \\
   & 0.8 & -       & -       & 6206.3  & -     & -       & 6428.7  & 402.9  & 687.6  & 200 & 200 \\
20 & 0.3 & 5373.6  & 5373.6  & 6284.2  & 1.169 & 5373.6  & 6483.1  & 44.9   & 366.4  & 5   & 4   \\
   & 0.5 & -       & -       & 9019.4  & -     & -       & 11452.5 & 874.9  & 1195.1 & 200 & 200 \\
   & 0.8 & 10149.3 & 10149.7 & 10510.8 & 1.036 & 10149.7 & 10858.2 & 45.7   & 377.6  & 4   & 6   \\
25 & 0.3 & 8023.9  & 8023.9  & 9502.9  & 1.184 & 8023.9  & 17369.6 & 51.8   & 394.3  & 3   & 4   \\
   & 0.5 & 12640.7 & 12640.7 & 13936.0 & 1.102 & 12640.7 & 17424.1 & 77.7   & 417.1  & 4   & 4   \\
   & 0.8 & -       & -       & 16806.1 & -     & -       & 17521.5 & 1789.4 & 1949.1 & 200 & 200 \\
30 & 0.3 & 11151.4 & 11151.4 & 13520.2 & 1.212 & 11151.4 & 24217.3 & 119.6  & 460.2  & 5   & 4   \\
   & 0.5 & 17342.4 & 17342.4 & 16969.0 & 0.978 & 17342.4 & 25103.4 & 146.5  & 503.8  & 6   & 6   \\
   & 0.8 & -       & -       & 24180.9 & -     & -       & 25028.6 & 2910.7 & 3240.5 & 200 & 200 \\ \bottomrule
\end{tabular}
}

% \vspace{3pt}
% \footnotesize{\textit{Note.} Objective values ($\phi$) are reported in thousands and rounded to the nearest tenth.}
\label{tab:EBMC_games_k=2_CR_BRD}
\end{table}

\begin{table}[htbp]
\caption{RRR-BRD Results with Different Initial Strategy Profiles for Type A KPG.}
\centering
\small
\scalebox{0.8}{
\begin{tabular}{r r r r r r r r r r r r}
\toprule
& & \multicolumn{2}{c}{RRR-BRD ($\phi$)} & \multicolumn{2}{c}{Price of Stability} & \multicolumn{2}{c}{References from BZR} & \multicolumn{2}{c}{Time (secs)} & \multicolumn{2}{c}{Round}  \\
\cmidrule(lr){3-4} \cmidrule(lr){5-6} \cmidrule(lr){7-8} \cmidrule(lr){9-10} \cmidrule(lr){11-12} 
$n$ & BG & init:$\barxzero$ &   init:$\barxswtime{300}$ & $\widetilde{\text{OSW}}$ & $\widetilde{\text{POS}}$ & BZR($\phi$) & BZR(Bound) & init:$\barxzero$ &   init:$\barxswtime{300}$ & init:$\barxzero$ &   init:$\barxswtime{300}$  \\ \midrule
2  & 0.2 & 18.4   & 18.5   & 18.7   & 1.011 & 18.5   & 18.5   & 0.1  & 0.6   & 3  & 2 \\
   & 0.5 & 17.6   & 18.3   & 18.3   & 1.002 & 18.3   & 18.3   & 0.0  & 0.4   & 3  & 2 \\
   & 0.8 & 8.1    & 8.2    & 8.4    & 1.022 & 8.3    & 8.3    & 0.0  & 0.4   & 3  & 2 \\
3  & 0.2 & 37.2   & 37.5   & 37.8   & 1.010 & 37.6   & 37.6   & 0.3  & 3.1   & 4  & 3 \\
   & 0.5 & 31.8   & 32.4   & 32.8   & 1.011 & 32.7   & 32.7   & 0.1  & 2.0   & 3  & 3 \\
   & 0.8 & 14.5   & 14.8   & 15.2   & 1.027 & 14.8   & 14.8   & 0.0  & 0.4   & 4  & 2 \\
5  & 0.2 & 129.2  & 131.3  & 131.3  & 1.000 & 131.3  & 131.3  & 0.9  & 47.0  & 4  & 1 \\
   & 0.5 & 59.6   & 60.0   & 61.4   & 1.024 & 60.6   & 61.0   & 0.8  & 33.5  & 7  & 3 \\
   & 0.8 & 27.0   & 27.2   & 29.3   & 1.079 & 27.3   & 27.3   & 0.1  & 2.0   & 4  & 3 \\
8  & 0.2 & 289.8  & 294.8  & 295.1  & 1.001 & 295.1  & 295.4  & 4.0  & 301.1 & 8  & 2 \\
   & 0.5 & 110.5  & 112.1  & 116.7  & 1.041 & 112.4  & 114.0  & 0.9  & 235.1 & 6  & 5 \\
   & 0.8 & 85.6   & 86.6   & 87.8   & 1.013 & 86.9   & 87.6   & 0.3  & 25.8  & 7  & 4 \\
10 & 0.2 & 500.5  & 504.6  & 508.7  & 1.008 & 505.5  & 506.1  & 6.1  & 302.5 & 8  & 3 \\
   & 0.5 & 255.0  & 260.4  & 262.8  & 1.009 & 260.9  & 263.0  & 1.0  & 302.3 & 7  & 4 \\
   & 0.8 & 139.5  & 141.6  & 143.9  & 1.016 & 141.2  & 142.9  & 0.5  & 301.3 & 7  & 5 \\
15 & 0.2 & 1050.2 & 1057.4 & 1058.8 & 1.001 & 1058.1 & 1060.8 & 6.5  & 302.8 & 8  & 3 \\
   & 0.5 & 583.1  & 599.7  & 604.2  & 1.007 & 600.4  & 605.2  & 1.9  & 304.1 & 9  & 6 \\
   & 0.8 & 246.4  & 254.2  & 257.7  & 1.014 & 255.3  & 258.3  & 0.8  & 302.4 & 8  & 6 \\
20 & 0.2 & 1663.5 & 1688.0 & 1689.9 & 1.001 & 1689.5 & 1693.9 & 8.6  & 303.2 & 10 & 4 \\
   & 0.5 & 1103.4 & 1120.6 & 1126.2 & 1.005 & 1120.4 & 1129.0 & 2.2  & 302.8 & 13 & 7 \\
   & 0.8 & 598.6  & 610.2  & 612.8  & 1.004 & 610.8  & 615.2  & 1.2  & 302.6 & 15 & 4 \\
25 & 0.2 & 2597.6 & 2626.4 & 2628.1 & 1.001 & 2627.4 & 2637.1 & 8.3  & 306.1 & 7  & 3 \\
   & 0.5 & 1789.9 & 1825.8 & 1833.3 & 1.004 & 1827.5 & 1838.4 & 3.7  & 307.2 & 17 & 8 \\
   & 0.8 & 690.1  & 706.2  & 711.5  & 1.008 & 706.6  & 714.5  & 1.4  & 303.8 & 11 & 5 \\
30 & 0.2 & 3311.2 & 3334.8 & 3351.2 & 1.005 & 3331.6 & 3356.8 & 15.1 & 307.4 & 10 & 4 \\
   & 0.5 & 2490.0 & 2528.8 & 2535.3 & 1.003 & 2528.7 & 2544.7 & 3.3  & 305.2 & 14 & 6 \\
   & 0.8 & 1070.2 & 1095.9 & 1104.6 & 1.008 & 1095.7 & 1109.7 & 1.4  & 304.4 & 8  & 6          \\ \bottomrule
\end{tabular}
}

% \vspace{3pt}
% \footnotesize{\textit{Note.} Objective values ($\phi$) are reported in thousands and rounded to the nearest tenth.}
\label{tab:KPG_A_CR_BRD}
\end{table}

\begin{table}[htbp]
\caption{RRR-BRD Results with Different Initial Strategy Profiles for Type B KPG.}
\centering
\small
\scalebox{0.8}{
\begin{tabular}{r r r r r r r r r r r r}
\toprule
& & \multicolumn{2}{c}{RRR-BRD ($\phi$)} & \multicolumn{2}{c}{Price of Stability} & \multicolumn{2}{c}{References from BZR} & \multicolumn{2}{c}{Time (secs)} & \multicolumn{2}{c}{Round}  \\
\cmidrule(lr){3-4} \cmidrule(lr){5-6} \cmidrule(lr){7-8} \cmidrule(lr){9-10} \cmidrule(lr){11-12} 
$n$ & BG & init:$\barxzero$ &   init:$\barxswtime{300}$ & $\widetilde{\text{OSW}}$ & $\widetilde{\text{POS}}$ & BZR($\phi$) & BZR(Bound) & init:$\barxzero$ &   init:$\barxswtime{300}$ & init:$\barxzero$ &   init:$\barxswtime{300}$  \\ \midrule
2  & 0.2 & 8.0    & 8.2    & 8.3    & 1.018 & 8.2    & 8.2    & 0.0  & 0.1   & 3  & 2  \\
   & 0.5 & 14.1   & 14.2   & 14.4   & 1.016 & 14.2   & 14.2   & 0.2  & 0.3   & 3  & 3  \\
   & 0.8 & 18.8   & 19.0   & 19.1   & 1.005 & 19.1   & 19.1   & 0.2  & 0.3   & 3  & 3  \\
3  & 0.2 & 14.8   & 15.4   & 15.9   & 1.035 & 15.6   & 15.6   & 0.0  & 0.3   & 5  & 3  \\
   & 0.5 & 28.3   & 28.7   & 29.4   & 1.025 & 28.9   & 29.1   & 0.1  & 2.8   & 5  & 4  \\
   & 0.8 & 39.0   & 39.8   & 40.3   & 1.012 & 40.0   & 40.1   & 0.9  & 2.6   & 4  & 2  \\
5  & 0.2 & 39.6   & 39.6   & 40.9   & 1.033 & 40.4   & 40.8   & 0.2  & 5.8   & 6  & 5  \\
   & 0.5 & 76.7   & 77.9   & 78.8   & 1.012 & 78.1   & 78.8   & 0.6  & 19.5  & 9  & 5  \\
   & 0.8 & 106.9  & 107.7  & 108.7  & 1.009 & 108.3  & 108.6  & 4.0  & 24.9  & 10 & 3  \\
8  & 0.2 & 87.7   & 91.1   & 92.6   & 1.017 & 91.6   & 92.7   & 0.4  & 253.4 & 5  & 4  \\
   & 0.5 & 185.2  & 187.5  & 189.9  & 1.013 & 188.6  & 189.9  & 1.8  & 184.4 & 13 & 7  \\
   & 0.8 & 267.6  & 272.3  & 273.1  & 1.003 & 272.8  & 273.2  & 4.5  & 302.5 & 6  & 4  \\
10 & 0.2 & 132.5  & 135.4  & 136.9  & 1.011 & 136.0  & 137.0  & 0.1  & 300.6 & 10 & 3  \\
   & 0.5 & 281.0  & 286.8  & 289.0  & 1.008 & 287.6  & 289.4  & 0.7  & 302.0 & 7  & 4  \\
   & 0.8 & 423.2  & 426.1  & 427.8  & 1.004 & 426.9  & 428.3  & 5.6  & 302.9 & 9  & 4  \\
15 & 0.2 & 279.2  & 285.8  & 289.5  & 1.013 & 287.0  & 290.5  & 0.6  & 302.5 & 7  & 7  \\
   & 0.5 & 627.1  & 641.3  & 644.0  & 1.004 & 642.7  & 645.4  & 1.1  & 302.1 & 9  & 6  \\
   & 0.8 & 954.7  & 963.8  & 965.2  & 1.001 & 963.8  & 966.5  & 6.6  & 303.1 & 9  & 3  \\
20 & 0.2 & 479.9  & 486.6  & 491.2  & 1.009 & 488.7  & 493.1  & 0.7  & 301.9 & 10 & 4  \\
   & 0.5 & 1108.2 & 1121.6 & 1125.5 & 1.004 & 1123.8 & 1129.0 & 1.0  & 302.0 & 9  & 4  \\
   & 0.8 & 1661.3 & 1671.9 & 1672.9 & 1.001 & 1670.7 & 1677.3 & 8.4  & 302.7 & 11 & 2  \\
25 & 0.2 & 753.5  & 759.8  & 763.5  & 1.005 & 761.4  & 767.5  & 0.9  & 303.4 & 9  & 4  \\
   & 0.5 & 1696.0 & 1717.7 & 1723.7 & 1.004 & 1719.9 & 1728.6 & 4.0  & 304.2 & 20 & 8  \\
   & 0.8 & 2586.8 & 2608.5 & 2614.3 & 1.002 & 2610.0 & 2622.9 & 12.1 & 306.5 & 13 & 5  \\
30 & 0.2 & 1055.2 & 1085.3 & 1088.8 & 1.003 & 1085.2 & 1094.7 & 2.3  & 305.8 & 12 & 14 \\
   & 0.5 & 2409.3 & 2460.4 & 2465.8 & 1.002 & 2460.3 & 2473.3 & 3.8  & 305.1 & 13 & 5  \\
   & 0.8 & 3729.8 & 3745.0 & 3750.8 & 1.002 & 3749.0 & 3764.0 & 17.0 & 310.2 & 11 & 6             \\ \bottomrule
\end{tabular}
}

% \vspace{3pt}
% \footnotesize{\textit{Note.} Objective values ($\phi$) are reported in thousands and rounded to the nearest tenth.}
\label{tab:KPG_B_CR_BRD}
\end{table}

\begin{table}[htbp]
\caption{RRR-BRD Results with Different Initial Strategy Profiles for Type C KPG.}
\centering
\small
\scalebox{0.8}{
\begin{tabular}{r r r r r r r r r r r r}
\toprule
& & \multicolumn{2}{c}{RRR-BRD ($\phi$)} & \multicolumn{2}{c}{Price of Stability} & \multicolumn{2}{c}{References from BZR} & \multicolumn{2}{c}{Time (secs)} & \multicolumn{2}{c}{Round}  \\
\cmidrule(lr){3-4} \cmidrule(lr){5-6} \cmidrule(lr){7-8} \cmidrule(lr){9-10} \cmidrule(lr){11-12} 
$n$ & BG & init:$\barxzero$ &   init:$\barxswtime{300}$ & $\widetilde{\text{OSW}}$ & $\widetilde{\text{POS}}$ & BZR($\phi$) & BZR(Bound) & init:$\barxzero$ &   init:$\barxswtime{300}$ & init:$\barxzero$ &   init:$\barxswtime{300}$  \\ \midrule
2  & 0.2 & 4.7  & 4.7  & 4.7  & 1.007 & 4.7  & 4.7  & 0.0  & 0.1   & 3   & 2   \\
   & 0.5 & 7.8  & 7.8  & 7.8  & 1.007 & 7.8  & 7.8  & 0.0  & 0.1   & 3   & 2   \\
   & 0.8 & 8.4  & 8.4  & 8.4  & 1.008 & 8.4  & 8.4  & 0.1  & 0.1   & 3   & 2   \\
3  & 0.2 & 7.1  & 7.1  & 7.2  & 1.012 & 7.1  & 7.1  & 0.0  & 0.3   & 4   & 3   \\
   & 0.5 & 10.3 & 10.3 & 10.4 & 1.011 & 10.3 & 10.3 & 0.1  & 0.3   & 4   & 2   \\
   & 0.8 & 11.5 & 11.6 & 11.8 & 1.024 & 11.6 & 11.6 & 0.1  & 0.1   & 5   & 3   \\
5  & 0.2 & 10.1 & 10.1 & 10.4 & 1.025 & 10.3 & 10.3 & 0.1  & 0.7   & 3   & 3   \\
   & 0.5 & 14.2 & 14.1 & 14.6 & 1.028 & 14.3 & 14.4 & 0.5  & 1.6   & 4   & 3   \\
   & 0.8 & 13.6 & 13.7 & 15.0 & 1.095 & 13.7 & 13.7 & 0.3  & 0.4   & 4   & 3   \\
8  & 0.2 & 15.6 & 15.6 & 16.2 & 1.034 & 15.8 & 16.0 & 0.1  & 11.4  & 4   & 3   \\
   & 0.5 & 15.6 & 15.6 & 17.6 & 1.122 & 15.8 & 16.5 & 0.5  & 1.9   & 6   & 4   \\
   & 0.8 & 15.6 & 15.5 & 18.5 & 1.192 & 15.7 & 16.6 & 0.2  & 1.3   & 5   & 4   \\
10 & 0.2 & 16.3 & 16.4 & 17.5 & 1.070 & 16.7 & 17.3 & 0.1  & 13.1  & 6   & 10  \\
   & 0.5 & 16.7 & 16.7 & 20.3 & 1.212 & 16.9 & 18.3 & 6.2  & 16.4  & 85  & 126 \\
   & 0.8 & 16.1 & 16.1 & 20.3 & 1.262 & 16.3 & 18.0 & 0.3  & 4.5   & 6   & 5   \\
15 & 0.2 & 19.6 & 19.6 & 21.8 & 1.114 & 19.8 & 21.5 & 0.5  & 224.2 & 8   & 17  \\
   & 0.5 & 15.9 & 16.1 & 23.0 & 1.435 & 16.5 & 19.8 & 0.4  & 72.6  & 10  & 7   \\
   & 0.8 & 16.3 & 16.4 & 23.0 & 1.398 & 16.8 & 19.8 & 0.3  & 42.8  & 9   & 8   \\
20 & 0.2 & 20.5 & 20.7 & 24.5 & 1.185 & 20.9 & 24.0 & 4.3  & 304.7 & 77  & 54  \\
   & 0.5 & -    & -    & 24.7 & -     & -    & 22.0 & 6.4  & 308.5 & 200 & 200 \\
   & 0.8 & 16.8 & 16.8 & 25.5 & 1.519 & 17.1 & 23.1 & 5.0  & 303.7 & 131 & 50  \\
25 & 0.2 & 20.6 & 20.4 & 26.2 & 1.272 & 20.7 & 26.4 & 15.1 & 303.9 & 196 & 20  \\
   & 0.5 & -    & -    & 25.7 & -     & -    & 24.9 & 10.2 & 313.0 & 200 & 200 \\
   & 0.8 & -    & -    & 26.4 & -     & -    & 25.4 & 11.5 & 315.3 & 200 & 200 \\
30 & 0.2 & 20.6 & -    & 27.7 & 1.350 & 20.5 & 40.8 & 17.7 & 320.2 & 196 & 200 \\
   & 0.5 & 15.9 & 15.7 & 27.3 & 1.720 & 16.3 & 33.0 & 4.6  & 305.5 & 74  & 33  \\
   & 0.8 & 15.6 & -    & 27.6 & 1.772 & 15.8 & 30.4 & 2.9  & 323.8 & 78  & 200         \\ \bottomrule
\end{tabular}
}

% \vspace{3pt}
% \footnotesize{\textit{Note.} Objective values ($\phi$) are reported in thousands and rounded to the nearest tenth.}
\label{tab:KPG_C_CR_BRD}
\end{table}

\clearpage
\ecompanionsection{BZR Full Results}\label{appendix:full_BZR}
We present the full BZR results of the EBMC $|\mathcal{L}|\ge 2$ games, KPG types (A) and (C). The metrics and computational settings used in these experiments are described in Section~\ref{sec:6.1}. 
\vspaceminus{12pt}
\begin{table}[htbp]
\caption{Comparison of BZR and ZR for Selfish EBMC Games $\mathcal{G}^{\text{Self}}_{|\mathcal{L}|\ge 2}$.}
\centering
\small
\scalebox{0.73}{
\begin{tabular}{rrrrrrrrrrrrrrrrrr}
\toprule
&& \multicolumn{2}{c}{\# PNEs Found} & \multicolumn{2}{c}{Best PNE ($\phi$)} & \multicolumn{2}{c}{Dual Bounds} & \multicolumn{2}{c}{Price of Stability} & \multicolumn{4}{c}{Time (secs)} & \multicolumn{2}{c}{\# EI Cuts}  \\
\cmidrule(lr){3-4} \cmidrule(lr){5-6} \cmidrule(lr){7-8} \cmidrule(lr){9-10} \cmidrule(lr){11-14} \cmidrule(lr){15-16}
$n$ & BG & $\#_\text{pne(ZR)}$ & $\#_\text{pne(BZR)}$ & ZR &  BZR  & ZR & BZR & $\widetilde{\text{OSW}}$ & $\widetilde{\text{POS}}$ & T(ZR$^{\text{1st}}$) & T(ZR) & T(BZR$^{\text{1st}}$) & T(BZR) &  $\#_\text{cuts(ZR)}$ & $\#_\text{cuts(BZR)}$  \\ \midrule
2  & 0.3 & 1 & 1          & 51.4  & 51.4             & 51.4             & 51.4             & 56.3    & 1.096 & 9.7    & 9.8    & 0.8   & 12.8   & 12  & 19  \\
   & 0.5 & 0 & \textbf{1} & -     & \textbf{83.8}    & \textbf{85.7}    & 86.0             & 86.6    & 1.033 & -      & 1800.3 & 1.4   & 1800.1 & 11  & 20  \\
   & 0.8 & 0 & \textbf{6} & -     & \textbf{92.8}    & \textbf{94.1}    & 94.3             & 95.2    & 1.027 & -      & 1800.1 & 2.1   & 1800.1 & 14  & 26  \\
3  & 0.3 & 1 & 1          & 115.0 & 115.0            & 115.0            & 115.0            & 139.7   & 1.215 & 1423.4 & 1423.7 & 1.9   & 1040.1 & 18  & 29  \\
   & 0.5 & 0 & \textbf{2} & -     & \textbf{178.6}   & 197.9            & \textbf{196.7}   & 200.3   & 1.122 & -      & 1800.4 & 2.3   & 1800.5 & 15  & 29  \\
   & 0.8 & 0 & \textbf{1} & -     & \textbf{252.2}   & \textbf{257.7}   & 257.8            & 253.7   & 1.006 & -      & 1800.3 & 1.9   & 1800.4 & 15  & 32  \\
5  & 0.3 & 1 & 1          & 380.0 & 380.0            & 380.0            & 380.0            & 399.8   & 1.052 & 489.2  & 665.9  & 5.7   & 642.9  & 28  & 49  \\
   & 0.5 & 0 & \textbf{1} & -     & \textbf{557.6}   & 626.7            & \textbf{618.5}   & 577.3   & 1.035 & -      & 1800.3 & 5.6   & 1800.0 & 25  & 43  \\
   & 0.8 & 0 & \textbf{1} & -     & \textbf{650.1}   & \textbf{671.1}   & 677.5            & 661.9   & 1.018 & -      & 1800.7 & 5.4   & 1800.5 & 25  & 57  \\
8  & 0.3 & 0 & \textbf{1} & -     & \textbf{862.1}   & \textbf{975.9}   & 980.4            & 995.8   & 1.155 & -      & 1800.1 & 14.8  & 1800.1 & 40  & 70  \\
   & 0.5 & 0 & \textbf{1} & -     & \textbf{1214.5}  & \textbf{1527.4}  & 1535.1           & 1387.9  & 1.143 & -      & 1800.1 & 14.4  & 1800.1 & 40  & 75  \\
   & 0.8 & 0 & \textbf{1} & -     & \textbf{1661.5}  & \textbf{1770.3}  & 1771.3           & 1721.0  & 1.036 & -      & 1800.1 & 14.8  & 1800.2 & 40  & 78  \\
10 & 0.3 & 0 & \textbf{1} & -     & \textbf{1265.2}  & \textbf{1517.8}  & 1526.5           & 1521.6  & 1.203 & -      & 1800.1 & 24.9  & 1800.1 & 50  & 95  \\
   & 0.5 & 0 & \textbf{1} & -     & \textbf{2083.2}  & \textbf{2635.2}  & 2635.7           & 2253.9  & 1.082 & -      & 1800.1 & 23.2  & 1800.1 & 50  & 93  \\
   & 0.8 & 0 & \textbf{1} & -     & \textbf{2612.3}  & \textbf{2755.8}  & 2757.3           & 2685.2  & 1.028 & -      & 1800.3 & 25.7  & 1800.3 & 50  & 100 \\
15 & 0.3 & 0 & \textbf{1} & -     & \textbf{2916.4}  & \textbf{3501.5}  & 3523.7           & 3440.2  & 1.180 & -      & 1800.3 & 56.7  & 1800.1 & 75  & 131 \\
   & 0.5 & 0 & \textbf{1} & -     & \textbf{4476.8}  & \textbf{5728.7}  & 5750.7           & 5030.1  & 1.124 & -      & 1800.3 & 58.6  & 1800.2 & 75  & 137 \\
   & 0.8 & 0 & 0          & -     & -                & \textbf{6423.5}  & 6428.7           & 6206.3  & -     & -      & 1800.3 & -     & 1800.3 & 75  & 150 \\
20 & 0.3 & 0 & \textbf{1} & -     & \textbf{5373.6}  & 6483.6           & \textbf{6483.1}  & 6284.2  & 1.169 & -      & 1800.4 & 118.5 & 1800.3 & 100 & 178 \\
   & 0.5 & 0 & 0          & -     & -                & \textbf{10639.6} & 11452.5          & 9019.4  & -     & -      & 1800.4 & -     & 1800.4 & 100 & 200 \\
   & 0.8 & 0 & \textbf{1} & -     & \textbf{10149.7} & \textbf{10856.7} & 10858.2          & 10510.8 & 1.036 & -      & 1800.4 & 117.4 & 1800.2 & 99  & 190 \\
25 & 0.3 & 0 & \textbf{1} & -     & \textbf{8023.9}  & \textbf{9787.5}  & 17369.6          & 9502.9  & 1.184 & -      & 1800.4 & 166.9 & 1800.8 & 125 & 210 \\
   & 0.5 & 0 & \textbf{1} & -     & \textbf{12640.7} & 17424.1          & 17424.1          & 13936.0 & 1.102 & -      & 1800.6 & 176.5 & 1800.5 & 125 & 236 \\
   & 0.8 & 0 & 0          & -     & -                & 17512.8          & \textbf{17521.5} & 16806.1 & -     & -      & 1800.4 & -     & 1800.5 & 125 & 175 \\
30 & 0.3 & 0 & \textbf{1} & -     & \textbf{11151.4} & \textbf{13751.2} & 24217.3          & 13520.2 & 1.212 & -      & 1800.6 & 284.1 & 1801.1 & 150 & 267 \\
   & 0.5 & 0 & \textbf{1} & -     & \textbf{17342.4} & 25103.4          & 25103.4          & 16969.0 & 0.978 & -      & 1800.8 & 288.5 & 1800.9 & 150 & 272 \\
   & 0.8 & 0 & 0          & -     & -                & \textbf{25019.9} & 25028.6          & 24180.9 & -     & -      & 1800.6 & -     & 1800.7 & 150 & 180

            \\ \bottomrule
\end{tabular}
}

\vspace{3pt}
\footnotesize{\textit{Note.} 
Boldfaced values indicate better solutions found within the time limit, tighter dual bounds, or more PNEs.}
\label{tab:EBMCP_games_2}
\end{table}

\begin{table}[htbp]
\caption{Comparison of BZR and ZR for Type A KPG.}
\centering
\small
\scalebox{0.73}{
\begin{tabular}{rrrrrrrrrrrrrrrrrr}
\toprule
&& \multicolumn{2}{c}{\# PNEs Found} & \multicolumn{2}{c}{Best PNE ($\phi$)} & \multicolumn{2}{c}{Dual Bounds} & \multicolumn{2}{c}{Price of Stability} & \multicolumn{4}{c}{Time (secs)} & \multicolumn{2}{c}{\# EI Cuts}  \\
\cmidrule(lr){3-4} \cmidrule(lr){5-6} \cmidrule(lr){7-8} \cmidrule(lr){9-10} \cmidrule(lr){11-14} \cmidrule(lr){15-16}
$n$ & BG & $\#_\text{pne(ZR)}$ & $\#_\text{pne(BZR)}$ & ZR &  BZR  & ZR & BZR & $\widetilde{\text{OSW}}$ & $\widetilde{\text{POS}}$  & T(ZR$^{\text{1st}}$) & T(ZR) & T(BZR$^{\text{1st}}$) & T(BZR) &  $\#_\text{cuts(ZR)}$ & $\#_\text{cuts(BZR)}$  \\ \midrule
2  & 0.2 & 1 & \textbf{3}   & 18.5  & 18.5            & 18.5   & 18.5            & 18.7   & 1.011 & 1.1    & 1.7    & 0.1  & 2.5    & 2031 & 2040 \\
   & 0.5 & 1 & \textbf{9}   & 18.3  & 18.3            & 18.3   & 18.3            & 18.3   & 1.002 & 1.1    & 1.5    & 0.1  & 1.2    & 155  & 116  \\
   & 0.8 & 1 & \textbf{10}  & 8.3   & 8.3             & 8.3    & 8.3             & 8.4    & 1.007 & 2.7    & 3.1    & 0.1  & 2.3    & 671  & 657  \\
3  & 0.2 & 5 & \textbf{9}   & 37.6  & 37.6            & 37.6   & 37.6            & 37.8   & 1.006 & 12.0   & 14.8   & 1.1  & 9.6    & 1058 & 1009 \\
   & 0.5 & 2 & \textbf{131} & 32.7  & 32.7            & 32.7   & 32.7            & 32.8   & 1.003 & 958.7  & 1646.6 & 0.2  & 174.2  & 3394 & 805  \\
   & 0.8 & 0 & \textbf{61}  & -     & \textbf{14.8}   & 14.9   & \textbf{14.8}   & 15.2   & 1.025 & -      & 1800.2 & 0.2  & 339.1  & 2739 & 1167 \\
5  & 0.2 & 3 & \textbf{8}   & 131.3 & 131.3           & 131.3  & 131.3           & 131.3  & 1.000 & 7.7    & 81.5   & 1.8  & 47.9   & 58   & 78   \\
   & 0.5 & 0 & \textbf{93}  & -     & \textbf{60.6}   & 61.0   & 61.0            & 61.4   & 1.014 & -      & 1800.3 & 0.9  & 1800.1 & 2892 & 1730 \\
   & 0.8 & 1 & \textbf{18}  & 27.3  & 27.3            & 27.4   & \textbf{27.3}   & 29.3   & 1.072 & 1784.9 & 1800.1 & 0.5  & 183.8  & 3982 & 3857 \\
8  & 0.2 & 0 & \textbf{17}  & -     & \textbf{295.1}  & 295.6  & \textbf{295.4}  & 295.1  & 1.000 & -      & 1800.7 & 5.0  & 1800.3 & 1657 & 200  \\
   & 0.5 & 0 & \textbf{35}  & -     & \textbf{112.4}  & 114.2  & \textbf{114.0}  & 116.7  & 1.038 & -      & 1800.4 & 1.5  & 1800.3 & 4833 & 4680 \\
   & 0.8 & 0 & \textbf{49}  & -     & \textbf{86.9}   & 87.7   & \textbf{87.6}   & 87.8   & 1.010 & -      & 1800.5 & 0.6  & 1800.9 & 571  & 649  \\
10 & 0.2 & 0 & \textbf{8}   & -     & \textbf{505.5}  & 506.5  & \textbf{506.1}  & 508.7  & 1.006 & -      & 1800.3 & 6.2  & 1800.6 & 2604 & 2157 \\
   & 0.5 & 0 & \textbf{36}  & -     & \textbf{260.9}  & 263.1  & \textbf{263.0}  & 262.8  & 1.007 & -      & 1800.9 & 2.6  & 1801.0 & 583  & 608  \\
   & 0.8 & 0 & \textbf{54}  & -     & \textbf{141.2}  & 143.0  & \textbf{142.9}  & 143.9  & 1.019 & -      & 1800.8 & 1.1  & 1800.6 & 1584 & 1653 \\
15 & 0.2 & 0 & \textbf{13}  & -     & \textbf{1058.1} & 1061.3 & \textbf{1060.8} & 1058.8 & 1.001 & -      & 1800.5 & 12.9 & 1801.5 & 344  & 254  \\
   & 0.5 & 0 & \textbf{31}  & -     & \textbf{600.4}  & 605.3  & \textbf{605.2}  & 604.2  & 1.006 & -      & 1800.8 & 11.3 & 1801.5 & 461  & 755  \\
   & 0.8 & 0 & \textbf{35}  & -     & \textbf{255.3}  & 258.3  & 258.3           & 257.7  & 1.009 & -      & 1800.7 & 9.1  & 1801.6 & 325  & 822  \\
20 & 0.2 & 0 & \textbf{13}  & -     & \textbf{1689.5} & 1694.4 & \textbf{1693.9} & 1689.9 & 1.000 & -      & 1801.6 & 12.9 & 1800.6 & 279  & 372  \\
   & 0.5 & 0 & \textbf{15}  & -     & \textbf{1120.4} & 1129.0 & 1129.0          & 1126.2 & 1.005 & -      & 1800.7 & 3.8  & 1800.8 & 215  & 514  \\
   & 0.8 & 0 & \textbf{14}  & -     & \textbf{610.8}  & 615.5  & \textbf{615.2}  & 612.8  & 1.003 & -      & 1800.6 & 3.0  & 1800.8 & 348  & 395  \\
25 & 0.2 & 0 & \textbf{8}   & -     & \textbf{2627.4} & 2637.5 & \textbf{2637.1} & 2628.1 & 1.000 & -      & 1800.9 & 17.3 & 1801.1 & 309  & 355  \\
   & 0.5 & 0 & \textbf{10}  & -     & \textbf{1827.5} & 1838.6 & \textbf{1838.4} & 1833.3 & 1.003 & -      & 1800.7 & 5.6  & 1801.0 & 290  & 447  \\
   & 0.8 & 0 & \textbf{12}  & -     & \textbf{706.6}  & 714.5  & 714.5           & 711.5  & 1.007 & -      & 1801.4 & 3.9  & 1800.9 & 332  & 547  \\
30 & 0.2 & 0 & \textbf{7}   & -     & \textbf{3331.6} & 3356.9 & \textbf{3356.8} & 3351.2 & 1.006 & -      & 1800.7 & 26.3 & 1802.9 & 197  & 397  \\
   & 0.5 & 0 & \textbf{9}   & -     & \textbf{2528.7} & 2545.1 & \textbf{2544.7} & 2535.3 & 1.003 & -      & 1800.8 & 8.4  & 1801.3 & 357  & 475  \\
   & 0.8 & 0 & \textbf{11}  & -     & \textbf{1095.7} & 1109.9 & \textbf{1109.7} & 1104.6 & 1.008 & -      & 1800.9 & 4.7  & 1801.0 & 312  & 586  
   \\ \bottomrule
\end{tabular}
}

\vspace{3pt}
\footnotesize{\textit{Note.} 
Boldfaced values indicate better solutions found within the time limit, tighter dual bounds, or more PNEs.}
\label{tab:KPG_A}
\end{table}

\begin{table}[htbp]
\caption{Comparison of BZR and ZR for Type C KPG.}
\centering
\small
\scalebox{0.73}{
\begin{tabular}{rrrrrrrrrrrrrrrrrr}
\toprule
&& \multicolumn{2}{c}{\# PNEs Found} & \multicolumn{2}{c}{Best PNE ($\phi$)} & \multicolumn{2}{c}{Dual Bounds} & \multicolumn{2}{c}{Price of Stability} & \multicolumn{4}{c}{Time (secs)} & \multicolumn{2}{c}{\# EI Cuts}  \\
\cmidrule(lr){3-4} \cmidrule(lr){5-6} \cmidrule(lr){7-8} \cmidrule(lr){9-10} \cmidrule(lr){11-14} \cmidrule(lr){15-16}
$n$ & BG & $\#_\text{pne(ZR)}$ & $\#_\text{pne(BZR)}$ & ZR &  BZR  & ZR & BZR & $\widetilde{\text{OSW}}$ & $\widetilde{\text{POS}}$  & T(ZR$^{\text{1st}}$) & T(ZR) & T(BZR$^{\text{1st}}$) & T(BZR) &  $\#_\text{cuts(ZR)}$ & $\#_\text{cuts(BZR)}$  \\ \midrule
2  & 0.2 & 1 & 1            & 4.7  & 4.7           & 4.7           & 4.7           & 4.7  & 1.007 & 0.4   & 0.6    & 0.1  & 0.5    & 4075 & 4085 \\
   & 0.5 & 3 & \textbf{14}  & 7.8  & 7.8           & 7.8           & 7.8           & 7.8  & 1.007 & 2.3   & 3.3    & 0.1  & 3.7    & 4159 & 4196 \\
   & 0.8 & 1 & \textbf{2}   & 8.4  & 8.4           & 8.4           & 8.4           & 8.4  & 1.008 & 0.6   & 0.8    & 0.1  & 0.9    & 3567 & 3583 \\
3  & 0.2 & 1 & \textbf{3}   & 7.1  & 7.1           & 7.1           & 7.1           & 7.2  & 1.012 & 3.6   & 3.9    & 0.1  & 1.8    & 4385 & 4407 \\
   & 0.5 & 2 & \textbf{5}   & 10.3 & 10.3          & 10.3          & 10.3          & 10.4 & 1.011 & 39.1  & 51.2   & 0.3  & 12.3   & 4442 & 4451 \\
   & 0.8 & 2 & \textbf{4}   & 11.6 & 11.6          & 11.6          & 11.6          & 11.8 & 1.024 & 1.3   & 1.4    & 0.1  & 1.6    & 5629 & 5647 \\
5  & 0.2 & 0 & \textbf{14}  & -    & \textbf{10.3} & 10.3          & 10.3          & 10.4 & 1.012 & -     & 1800.4 & 0.3  & 243.4  & 4198 & 4274 \\
   & 0.5 & 0 & \textbf{20}  & -    & \textbf{14.3} & 14.5          & \textbf{14.4} & 14.6 & 1.023 & -     & 1800.6 & 0.5  & 1800.5 & 4986 & 5144 \\
   & 0.8 & 2 & \textbf{33}  & 13.7 & 13.7          & 13.7          & 13.7          & 15.0 & 1.092 & 235.0 & 446.0  & 0.8  & 177.6  & 4119 & 4227 \\
8  & 0.2 & 0 & \textbf{19}  & -    & \textbf{15.8} & 16.1          & \textbf{16.0} & 16.2 & 1.024 & -     & 1800.8 & 0.4  & 1801.0 & 2713 & 2856 \\
   & 0.5 & 0 & \textbf{34}  & -    & \textbf{15.8} & 16.8          & \textbf{16.5} & 17.6 & 1.108 & -     & 1800.9 & 1.0  & 1800.5 & 2267 & 2612 \\
   & 0.8 & 0 & \textbf{146} & -    & \textbf{15.7} & 16.8          & \textbf{16.6} & 18.5 & 1.183 & -     & 1800.5 & 0.7  & 1800.5 & 3033 & 4072 \\
10 & 0.2 & 0 & \textbf{19}  & -    & \textbf{16.7} & 17.4          & \textbf{17.3} & 17.5 & 1.051 & -     & 1800.5 & 0.7  & 1800.4 & 1559 & 1749 \\
   & 0.5 & 0 & \textbf{49}  & -    & \textbf{16.9} & 18.7          & \textbf{18.3} & 20.3 & 1.196 & -     & 1800.7 & 6.0  & 1800.6 & 1564 & 2377 \\
   & 0.8 & 0 & \textbf{156} & -    & \textbf{16.3} & 18.6          & \textbf{18.0} & 20.3 & 1.244 & -     & 1800.5 & 1.1  & 1800.8 & 1775 & 3841 \\
15 & 0.2 & 0 & \textbf{18}  & -    & \textbf{19.8} & 21.7          & \textbf{21.5} & 21.8 & 1.104 & -     & 1800.5 & 22.3 & 1800.4 & 1608 & 1831 \\
   & 0.5 & 0 & \textbf{39}  & -    & \textbf{16.5} & 20.3          & \textbf{19.8} & 23.0 & 1.397 & -     & 1800.3 & 21.7 & 1800.3 & 1651 & 2302 \\
   & 0.8 & 0 & \textbf{37}  & -    & \textbf{16.8} & 20.2          & \textbf{19.8} & 23.0 & 1.366 & -     & 1800.3 & 21.4 & 1800.4 & 1633 & 2362 \\
20 & 0.2 & 0 & \textbf{11}  & -    & \textbf{20.9} & 24.1          & \textbf{24.0} & 24.5 & 1.170 & -     & 1800.6 & 4.4  & 1800.4 & 120  & 340  \\
   & 0.5 & 0 & 0            & -    & -             & \textbf{21.8} & 22.0          & 24.7 & -     & -     & 1800.2 & -    & 1800.2 & 160  & 240  \\
   & 0.8 & 0 & \textbf{13}  & -    & \textbf{17.1} & \textbf{22.7} & 23.1          & 25.5 & 1.492 & -     & 1800.2 & 3.7  & 1800.3 & 200  & 352  \\
25 & 0.2 & 0 & \textbf{7}   & -    & \textbf{20.7} & \textbf{25.9} & 26.4          & 26.2 & 1.267 & -     & 1800.3 & 10.1 & 1800.4 & 150  & 360  \\
   & 0.5 & 0 & 0            & -    & -             & \textbf{24.7} & 24.9          & 25.7 & -     & -     & 1800.3 & -    & 1800.6 & 175  & 250  \\
   & 0.8 & 0 & 0            & -    & -             & 25.5          & \textbf{25.4} & 26.4 & -     & -     & 1800.3 & -    & 1800.5 & 175  & 250  \\
30 & 0.2 & 0 & \textbf{2}   & -    & \textbf{20.5} & \textbf{29.3} & 40.8          & 27.7 & 1.355 & -     & 1800.5 & 14.5 & 1800.8 & 150  & 526  \\
   & 0.5 & 0 & \textbf{11}  & -    & \textbf{16.3} & \textbf{27.5} & 33.0          & 27.3 & 1.675 & -     & 1800.5 & 8.7  & 1806.9 & 180  & 719  \\
   & 0.8 & 0 & \textbf{18}  & -    & \textbf{15.8} & \textbf{27.9} & 30.4          & 27.6 & 1.747 & -     & 1800.4 & 5.6  & 1801.5 & 210  & 891 
          \\ \bottomrule
\end{tabular}
}

\vspace{3pt}
\footnotesize{\textit{Note.} 
Boldfaced values indicate better solutions found within the time limit, tighter dual bounds, or more PNEs.}
\label{tab:KPG_C}
\end{table}

\clearpage
\ecompanionsection{Approximate-PNE} \label{appendix:approximate_PNE}

This appendix reports \emph{approximate} equilibrium certificates for the
instances in which our algorithms did not report an \emph{exact} pure Nash
equilibrium (PNE) within the prescribed computational limits. The goal is not
to claim that a PNE exists in these instances, but rather to quantify the stability of the best-observed strategy profiles against unilateral deviations.

For each ``no-PNE-reported'' instance (both EBMC games and KPGs), we run
RRR--BRD with a single restart and a fixed round cap:
$L=1$ and $R=20$. Let $\x^{(r)}$ denote the
strategy profile at the \emph{end} of round $r\in\{1,\dots,20\}$ (i.e., after all
players have played once in that round under a randomized play order). We store
these end-of-round profiles, yielding at most 20 candidate profiles per
instance.

For any candidate profile $\x^{(r)}$, we compute a multiplicative approximation
factor as follows. For each player $i\in N$, we solve the best-response problem
against $\x^{-i}$ and obtain a best response
$\hat{\x}^i\in\arg\max_{\x^i\in\mathcal{X}_i} u_i(\x^i,\x^{-i})$. Assuming
$u_i(\x^i,\x^{-i})>0$ for all feasible profiles, we define
$$
\alpha_i(\x^{(r)}) ~:=~
\frac{u_i(\hat{\x}^i,(\x^{(r)})^{-i})}{u_i(\x^{(r),i},(\x^{(r)})^{-i})},
\qquad
\alpha(\x^{(r)}) ~:=~ \max_{i\in N} \alpha_i(\x^{(r)}).
$$
Then $\x^{(r)}$ is an $\alpha(\x^{(r)})$-approximate PNE in the multiplicative
sense, and $\alpha(\x^{(r)})=1$ if and only if $\x^{(r)}$ is an exact PNE.
Among the 20 end-of-round candidates, we identify the profile with the smallest
approximation factor,
\[
\alpha^\star ~:=~ \min_{r=1,\dots,20} \alpha(\x^{(r)}),
\]
and report: (i) the best round $r^\star$ attaining $\alpha^\star$, (ii) the time
taken for this RRR--BRD run, and (iii) the social welfare (global objective
value) evaluated at $\x^{(r^\star)}$.

Tables~\ref{tab:approx_EBMC} and~\ref{tab:approx_KPG} summarize these results
for EBMC games and KPGs, respectively.

\begin{table}[h!]
\caption{Approximate PNE for the instances without a PNE in EBMC Games}
\label{tab:approx_EBMC}
\centering
\scalebox{0.80}{
\begin{tabular}{rrrrrrr}
\toprule
Type & $n$ & BG & Approx BRD ($\phi$) & Best $\alpha$ & Time (secs) & Best Round \\
\midrule
$|\mathcal{L}|\ge 2$ & 15 & 0.8 & 6043.8  & 1.000003 & 106.3 & 11 \\
$|\mathcal{L}|\ge 2$ &20 & 0.5 & 8292.9  & 1.000003 & 210.4 & 3  \\
$|\mathcal{L}|\ge 2$ &25 & 0.8 & 16255.9  & 1.000002 & 393.8 & 19  \\
$|\mathcal{L}|\ge 2$ &30 & 0.8 & 23509.4 & 1.000004 & 639.2 & 8 \\
\bottomrule
\end{tabular}
}

% \vspace{3pt}
% \footnotesize{\textit{Note.} Objective values ($\phi$) are reported in thousands and rounded to the nearest tenth.}

\end{table}

\begin{table}[h!]
\caption{Approximate PNE for the instances without a PNE in KPGs}
\label{tab:approx_KPG}
\centering
\scalebox{0.80}{
\begin{tabular}{rrrrrrr}
\toprule
Type & $n$ & BG & Approx BRD ($\phi$) & Best $\alpha$ & Time (secs) & Best Round \\
\midrule
C & 20 & 0.5 & 15.2 & 1.003140 & 0.9  & 9 \\
C & 25 & 0.5 & 14.5 & 1.003591 & 1.1 & 19 \\
C & 25 & 0.8 & 15.9 & 1.005714 & 1.2  & 20 \\
\bottomrule
\end{tabular}
}

% \vspace{3pt}
% \footnotesize{\textit{Note.} Objective values ($\phi$) are reported in thousands and rounded to the nearest tenth.}

\end{table}

\clearpage
\ecompanionsection{County-level Utility Values}\label{appendix:county-level}
Table~\ref{tab:county-level_full} presents the complete set of county-level utility values under different strategy profiles, as introduced in Table~\ref{tab:county-level} of the main manuscript.

\begin{table}[h!]
\caption{County-level Utility Values Using Different Strategy Profiles.}
\label{tab:county-level_full}
\centering
\scalebox{0.75}{
\begin{tabular}{lrrrr@{\hspace{0.5cm}}|lrrrr@{\hspace{0.5cm}}}
\toprule
County & $u_i^{\text{Self}}(\bar{\x}_{\text{ng}})$ & $u_i^{\text{Self}}(\hat{\x}_{\text{pne}}^1)$ & $u_i^{\text{Self}}(\hat{\x}_{\text{pne}}^*)$ & $u_i^{\text{Self}}(\barxswtime{3600})$ & County & $u_i^{\text{Self}}(\bar{\x}_{\text{ng}})$ & $u_i^{\text{Self}}(\hat{\x}_{\text{pne}}^1)$ & $u_i^{\text{Self}}(\hat{\x}_{\text{pne}}^*)$ & $u_i^{\text{Self}}(\barxswtime{3600})$ \\ \midrule
Aitkin            & 23746 & 24956 & 24982 & 29013 & Martin          & 24            & 24            & 24            & 24            \\
Anoka             & 11292 & 12466 & 12495 & 15723 & Mcleod          & 7194          & 7479          & 7503          & 8933          \\
Becker            & 24219 & 26489 & 26497 & 31712 & Meeker          & 8866          & 9423          & 9428          & 9964          \\
Beltrami          & 18720 & 20049 & 20061 & 23756 & Mille lacs      & 1778          & 1818          & 1818          & 1886          \\
Benton            & 1571  & 1688  & 1688  & 1779  & Morrison        & 5243          & 5460          & 5466          & 6184          \\
Big Stone         & 4555  & 4827  & 4827  & 5664  & Mower           & 43            & 43            & 43            & 48            \\
Blue Earth        & 4712  & 4924  & 4932  & 5439  & Murray          & 3481          & 3576          & 3590          & 3906          \\
Brown             & 2338  & 2436  & 2436  & 2805  & Nicollet        & 32            & 32            & 32            & 32            \\
Carlton           & 5393  & 5808  & 5813  & 6910  & Nobles          & 2328          & 2396          & 2396          & 2520          \\
Carver            & 6664  & 7306  & 7331  & 8211  & Norman          & 8             & 8             & 8             & 8             \\
Cass              & 40151 & 42450 & 42500 & 48926 & Olmsted         & 587           & 596           & 598           & 852           \\
Chippewa          & 60    & 60    & 60    & 60    & Otter tail      & 43479         & 47839         & 47863         & 58200         \\
Chisago           & 5463  & 5923  & 5930  & 6827  & Pennington      & 22            & 22            & 22            & 22            \\
Clay              & 2075  & 2481  & 2485  & 2893  & Pine            & 6061          & 6313          & 6321          & 7088          \\
Clearwater        & 8484  & 9210  & 9235  & 10484 & Pipestone       & 38            & 73            & 73            & 78            \\
Cook              & 28157 & 29949 & 30031 & 34442 & Polk            & 5584          & 6031          & 6037          & 6906          \\
Cottonwood        & 2713  & 2824  & 2824  & 3212  & Pope            & 10157         & 11020         & 11032         & 13728         \\
Crow Wing         & 38044 & 39727 & 39789 & 46254 & Ramsey          & 6995          & 7726          & 7761          & 9372          \\
Dakota            & 6958  & 7400  & 7406  & 8664  & Redwood         & 177           & 177           & 177           & 177           \\
Dodge             & 7     & 7     & 7     & 7     & Renville        & 837           & 866           & 871           & 1056          \\
Douglas           & 16827 & 18127 & 18136 & 22158 & Rice            & 4496          & 4839          & 4847          & 5076          \\
Faribault         & 251   & 251   & 251   & 251   & Rock            & 0             & 0             & 0             & 0             \\
Freeborn          & 1172  & 1172  & 1172  & 1172  & Roseau          & 234           & 234           & 234           & 234           \\
Goodhue           & 770   & 845   & 845   & 879   & Saint Louis     & 78691         & 81719         & 81770         & 91172         \\
Grant             & 6635  & 7389  & 7391  & 8817  & Scott           & 4983          & 5249          & 5253          & 6381          \\
Hennepin          & 21136 & 22442 & 22474 & 27024 & Sherburne       & 5077          & 5557          & 5564          & 6698          \\
Hubbard           & 22780 & 24426 & 24444 & 29543 & \textbf{Sibley} & \textbf{2390} & \textbf{2388} & \textbf{2388} & \textbf{2429} \\
Isanti            & 3635  & 3876  & 3884  & 4227  & Stearns         & 20200         & 21544         & 21554         & 26076         \\
Itasca            & 51214 & 53588 & 53640 & 61367 & Steele          & 372           & 372           & 372           & 372           \\
Jackson           & 3061  & 3167  & 3171  & 3474  & Stevens         & 2938          & 3340          & 3382          & 4142          \\
Kanabec           & 3697  & 3917  & 3927  & 4634  & Swift           & 1287          & 1461          & 1461          & 1778          \\
Kandiyohi         & 15551 & 16438 & 16450 & 18723 & Todd            & 9108          & 9832          & 9842          & 11661         \\
Kittson           & 254   & 254   & 254   & 254   & Traverse        & 1427          & 1466          & 1466          & 1523          \\
Koochiching       & 2217  & 2278  & 2284  & 2314  & Wabasha         & 331           & 331           & 331           & 331           \\
Lac Qui Parle     & 1331  & 1376  & 1376  & 1405  & Wadena          & 1834          & 1950          & 1950          & 2289          \\
Lake              & 30194 & 31584 & 31640 & 36702 & Waseca          & 2940          & 3074          & 3080          & 3089          \\
Lake of the woods & 1009  & 1009  & 1009  & 1009  & Washington      & 11252         & 11946         & 11968         & 15845         \\
Le Sueur          & 6804  & 7135  & 7157  & 7929  & Watonwan        & 946           & 955           & 955           & 1114          \\
Lincoln           & 4221  & 4292  & 4292  & 4501  & Wilkin          & 94            & 102           & 102           & 128           \\
Lyon              & 2752  & 2925  & 2939  & 3314  & \textbf{Winona} & \textbf{1561} & \textbf{1579} & \textbf{1579} & \textbf{1444} \\
Mahnomen          & 4177  & 4424  & 4428  & 6012  & Wright          & 22047         & 23782         & 23805         & 29464         \\
Marshall          & 4585  & 5076  & 5078  & 6081  & Yellow Medicine & 829           & 858           & 858           & 938    \\
\bottomrule
\end{tabular}
}

\vspace{3pt}
\footnotesize{\textit{Note.} Counties, where the inequalities regarding utility value comparisons do not hold, are highlighted in bold.}

\end{table}

\end{document}